\documentclass[reprint,notitlepage,amsmath,amssymb,aps,prl,onecolumn]{article}

\usepackage[utf8]{inputenc}
\usepackage{subcaption}
\usepackage[labelformat=parens,labelsep=quad,skip=3pt,font=small,labelfont=bf]{caption}
\usepackage{graphicx}
\usepackage{float}
\usepackage[export]{adjustbox}
\usepackage{dcolumn}
\usepackage{bm}
\usepackage{color}
\usepackage{overpic}
\usepackage{psfrag}
\usepackage{amsmath}
\usepackage{amsfonts}
\usepackage[title]{appendix}
\usepackage[square,numbers]{natbib}
\usepackage[dvipsnames]{xcolor}

\usepackage[symbol]{footmisc}

\usepackage{authblk}

\providecommand\xx{\boldsymbol{x}}
\providecommand\ww{\boldsymbol{\omega}}
\providecommand\Dbar{\bar{D}}
\providecommand\uu{\boldsymbol{u}}
\providecommand\nn{\textbf{n}}
\providecommand\ee{\textbf{e}}
\renewcommand\d{\partial}

\newcommand\numberthis{\addtocounter{equation}{1}\tag{\theequation}}

\usepackage[normalem]{ulem}

\begin{document}

\renewcommand*{\thefootnote}{\fnsymbol{footnote}}

\title{Symmetry breaking of azimuthal waves: Slow-flow dynamics on the Bloch Sphere}
\author{Abel Faure-Beaulieu\footnote{abelf@ethz.ch}\,}
\author{Nicolas Noiray\footnote{noirayn@ethz.ch}}
\affil{\textit{CAPS Laboratory,
	Department of Mechanical and Process Engineering,
	ETH Z\"urich}}

\renewcommand*{\thefootnote}{\arabic{footnote}}

\date{}
\maketitle


\begin{abstract}
	Depending on the reflectional and rotational symmetries of annular combustors for aeroengines and gas turbines, self-sustained azimuthal thermoacoustic eigenmodes can be  standing, spinning or mix of these two types of waves. These thermoacoustic limit cycles are unwanted because the resulting  intense acoustic fields induce high-cycle fatigue of the combustor components. 
	This paper presents a new theoretical framework for describing, in an idealized annular combustor, the dynamics of the slow-flow variables, which define the state of an eigenmode, i.e. if the latter is standing, spinning or mixed. The acoustic pressure is  expressed as a hypercomplex field and this ansatz is inserted into a one dimensional wave equation that describes the thermoacoustics of a thin annulus. Slow-flow averaging of this wave equation is performed by adapting the classic Krylov-Bogoliubov method to the quaternion field in order to derive a system of coupled first order differential equations for the four slow-flow variables, i.e. the amplitude, the nature angle, the preferential direction and the temporal phase of the azimuthal thermoacoustic mode. The state of the mode can be conveniently depicted by using the first three slow-flow variables as spherical coordinates for a Bloch sphere representation. Stochastic forcing from the turbulence in  annular combustors is also accounted for. This new analytical model describes both rotational and reflectional explicit symmetry breaking bifurcations induced by the non-uniform distribution of thermoacoustic sources along the annulus circumference and by the presence of a mean swirl. 
\end{abstract}

\section{Introduction}\label{intro}
One of the environmental challenges for civil aircraft transportation is the necessity to significantly reduce emissions of nitrogen oxides NO$_\text{x}$, which cause toxic pollution in the vicinity of airports and contributes to the formation of ozone in the high atmosphere. A technical solution for this problem is to burn lean fuel-air mixtures made as homogeneous as possible, which prevents near-stoichiometric hot-spots where NO$_\text{x}$ are produced, and which has also the advantage of reducing soot formation and unburned hydrocarbons. These flow conditions can be obtained by injecting the fuel in the compressed air well upstream the flame, as it is done in the combustors of modern gas turbines for power generation. The problem is that robust anchoring of these lean premixed flames over a wide range of operating conditions is extremely challenging to achieve in compact aeroengine combustors. In addition, the thermoacoustic stability is significantly reduced with these flames, because they are more sensitive to flow perturbations. In particular, they exhibit strong heat release rate response to acoustically-induced modulation of mixture composition or velocity field. Thermoacoustic instabilities are induced by the constructive linear and nonlinear coupling between flames and combustion chamber acoustics. They yield high amplitude limit cycles which reduce the lifetime of the hot gas path parts because of the induced vibrations, or in the worst case destroy them during a sudden catastrophic event. The phenomenon is known and studied since more than a century, in particular with the early works of Rayleigh \cite{rayleigh78} showing the importance of the phase between the oscillations of heat release $\dot{Q}'$ and of the acoustic pressure $p_a$ upon the thermoacoustic oscillation in the stability of the system. His famous criterion for instability is
\begin{align}\label{eq_Rayleigh}
\int_\mathcal{T}{\int_\mathcal{V}{p_a(\xx,t)\dot{Q}'(\xx,t)\,d\xx \,dt}}>0,
\end{align}
where $\mathcal{T}$ is the acoustic period and $\mathcal{V}$ the  volume of the combustor. This criterion results from the acoustic energy balance and it is a necessary, but not sufficient condition to get an instability \cite{nicoud_2005}.
A strong research effort aims at understanding the fundamental phenomena associated with thermoacoustic instabilities. In this context, numerous studies have been conducted about thermoacoustic instabilities in annular combustor geometries which are very common for land-based gas turbines and aeroengines. Annular combustion chambers for aeroengines are more compact and light-weighted, and produce more uniform temperature distribution at the turbine inlet than can-annular combustors. A fascinating aspect of thermoacoustic instabilities in annular combustion chambers is that they display pairs of azimuthal eigenmodes that yield standing or spinning thermoacoustic waves depending on the system symmetries. The measurements of Krebs \textit{et al.}  \cite{krebs02} and Noiray \textit{et al.}  \cite{noiray13} on  large gas turbines for power generation featuring annular combustors show the existence of spinning waves propagating at the speed of sound in the azimuthal direction, standing  modes with fixed or slowly varying nodal line direction, and mixed modes resulting from the combination of spinning and standing waves. However, the root causes of these observations are not fully understood, although thermoacoustics of annular combustors has been for several years the topic of intense research \cite{oconnor15a}.  \\ 
In addition to the data from practical gas turbine combustors, numerical and experimental studies in academia shed light on the nonlinear dynamics of azimuthal thermoacoustic eigenmodes and can be used to put the theoretical findings of the present paper into perspective. Regarding numerical simulations, Wolf \textit{et al.} \cite{wolf12} performed compressible reactive Large Eddy Simulations (LES) of a whole annular helicopter-engine combustor. In this numerical work, the observed  azimuthal mode is first standing, then mixed, and then standing again, and the angular position of the standing modes drifts at the azimuthal velocity of the mean flow. Nevertheless, the simulated physical time is not long enough to draw  conclusions about the predominance of spinning or standing modes. Regarding experiments, one can for instance refer to the work performed using academic annular combustors at atmospheric pressure -- effects of equivalence ratio \cite{worthproci2017}, presence of baffles \cite{dawsonproci2015}, turbulent bluff-body flames without swirl \cite{mazur_proci2019}, with laminar flames \cite{prieur17,bourgouinproci15} -- and at elevated pressure \cite{fanaca10,zahn17}.\\
The present work falls in the category of low-order modelling approaches, where the annular combustor is approximated by a one dimensional (1D) waveguide. In previous studies using this 1D simplified description, the acoustic field has been projected onto orthogonal basis formed by pairs of degenerate standing modes \cite{schuermans06,noiray13,noiray11}  or spinning modes \cite{hummel17b}, which we now briefly explain. Regarding the former modelling strategy, Noiray \textit{et al.} \cite{noiray13,noiray11} proposed a nonlinear theoretical model of the spinning/standing/mixed mode dynamics by projecting the azimuthal thermoacoustic eigenmodes on orthogonal standing waves. This projection is truncated to keep only one degenerate pair of azimuthal standing modes of amplitudes $A$ and $B$ sharing the same wave number, which gives a compact description of the modal dynamics when there is one dominant azimuthal thermoacoustic mode in the combustor. This ansatz for the acoustic field is injected in a wave equation with  nonlinear thermoacoustic feedback. The wave equation is averaged in space and time to get slow-time system of first order differential equations for the amplitudes $A$ and $B$ and phase difference $\phi$ of the two standing modes used for projecting the acoustic field. With a simple cubic nonlinearity to describe the flame response, it is shown that sufficiently strong spatial non-uniformities in the thermoacoustic coupling along the circumference of the annular combustor yields standing modes, whereas a perfectly symmetric configuration leads to spinning modes. In \cite{schuermans06,noiray11}, these theoretical findings are confirmed by time-domain simulations thanks to a reduced-order network model of the thermoacoustic system, including experimentally measured flame transfer functions and three-dimensional (3D) azimuthal modes computed with a finite-element Helmholtz solver. Hummel \textit{et al.} \cite{hummel17,hummel17b} proposed to use a different 1D nonlinear formalism where the azimuthal mode is described as the sum of  counterclockwise  and a clockwise-spinning waves. Performing spatial averaging and time averaging of the wave equation, they also conclude that limit cycle spinning modes are stable for uniformly distributed thermoacoustic feedback, while standing modes are unstable. \\ 
It should be emphasized that the low-order 1D model presented in this work predicts the nonlinear dynamics of azimuthal thermoacoustic modes in an \textit{idealized} annular chamber, but it cannot be used to predict linear thermoacoustic stability or limit cycle amplitudes of a \textit{real-world} annular combustor. Such predictions can be achieved using reduced-order thermoacoustic network models which include modelled or measured flame transfer functions and acoustic propagation in 3D complex geometries -- e.g. \cite{bauerheim14,bauerheim15,rouwenhorst17,li19} for linear description of the annular thermoacoustics, with stability analysis and investigation of symmetry breaking, \cite{mensah16} for computationally efficient numerical methods to solve the associated nonlinear eigenvalue problem and \cite{mensah19} to perform sensitivity analysis to perturbations in the heat release rate distribution , \cite{Laerajegtp2017,Yang_JSV19} for frequency domain description of the thermoacoustics in annular combustors with measured and modelled nonlinear flame feedback, \cite{noiray11,bothien15} for state-space description allowing for time and frequency domain analysis.\\
Coming back to the 1D description adopted in this work,
previous 1D studies from Ghirardo \textit{et al.} \cite{ghirardo15,ghirardo16} use a general formulation with flame describing function for both static and dynamic nonlinearities. They show that static nonlinearities associated with non-monotonic describing function gains can ex{\tiny }plain the occurrence of stable standing modes in perfectly symmetric annular combustors. Another study from Ghirardo \textit{et al.} \cite{ghirardo13} shows that a heat release model depending on both the axial and azimuthal acoustic velocities is another possible explanation for stable standing modes in rotationally symmetric chambers. In that regard, one can also refer to the interesting experimental studies dealing with the response of flames to transverse acoustic forcing \cite{hauser_jegtp2011,oconnor15,lespinasse13,saurabhcnf2017,smith18}.\\
Most of the theoretical findings from 1D models that are based on projection of the acoustic field onto pairs of orthogonal standing or spinning modes have not yet been validated with well-controlled experiments and these two modelling approaches have shortcomings. First, the nature of the acoustic oscillations in the annular chamber (spinning, standing or mixed) is not a phase space variable but a derived quantity. Second, the phase space is ill-posed for both approaches, because when the amplitude of one of the modes   vanishes, the phase difference between these modes becomes undetermined. These two shortcomings have been raised by Ghirardo and Bothien \cite{ghirardo18} who proposed an alternative quaternion-based projection of the acoustic field, which results in a well-posed phase space. We show in the present paper that an additional limitation of the aforementioned projections is overcome by using the hypercomplex  ansatz from \cite{ghirardo18}. Indeed, it allows to describe \textit{both} the reflectional symmetry breaking when a mean flow exists in the annulus and the rotational symmetry breaking when the thermoacoustic sources are not uniformly distributed along the annulus.\\
In this context, the present work aims at i) describing the nonlinear stochastic dynamics of azimuthal thermoacoustic modes in a simplified annular chamber with a new theoretical model, in order to address shortcomings of the models available in the literature, and ii) explaining the different types of symmetry breaking bifurcation that are observed in these systems with this unified framework.
Our starting point is the quaternion formalism introduced in \cite{ghirardo18} to represent azimuthal waves. In section \ref{sec2}, we derive an azimuthal wave equation for the acoustic pressure including mean flow effects and thermoacoustic sources. In section \ref{sec3}, we briefly introduce some properties of the quaternion algebra and the quaternion representation of the acoustic pressure field in a thin annular combustor. In section \ref{sec4}, the quaternion formalism is introduced into the wave equation which is spatially and temporally averaged in order to obtain a dynamical system for the slow-flow variables. The attractors of the symmetric system are discussed in section \ref{sec5}. Sections \ref{sec6} to \ref{sec9} respectively deal with situations where the nonlinear flame response is delayed, where turbulent forcing is present, and where the symmetry of the thermoacoustic system is broken by a non-uniform distribution of the thermoacoustic feedback from the flames around the annulus and by the presence of a mean swirl.


\section{Thermoacoustic wave equation in thin annulus with mean swirl }\label{sec2}
\subsection{1D thermoacoustic wave equation}

\begin{figure}
	\centering
	\includegraphics[width=.5\textwidth]{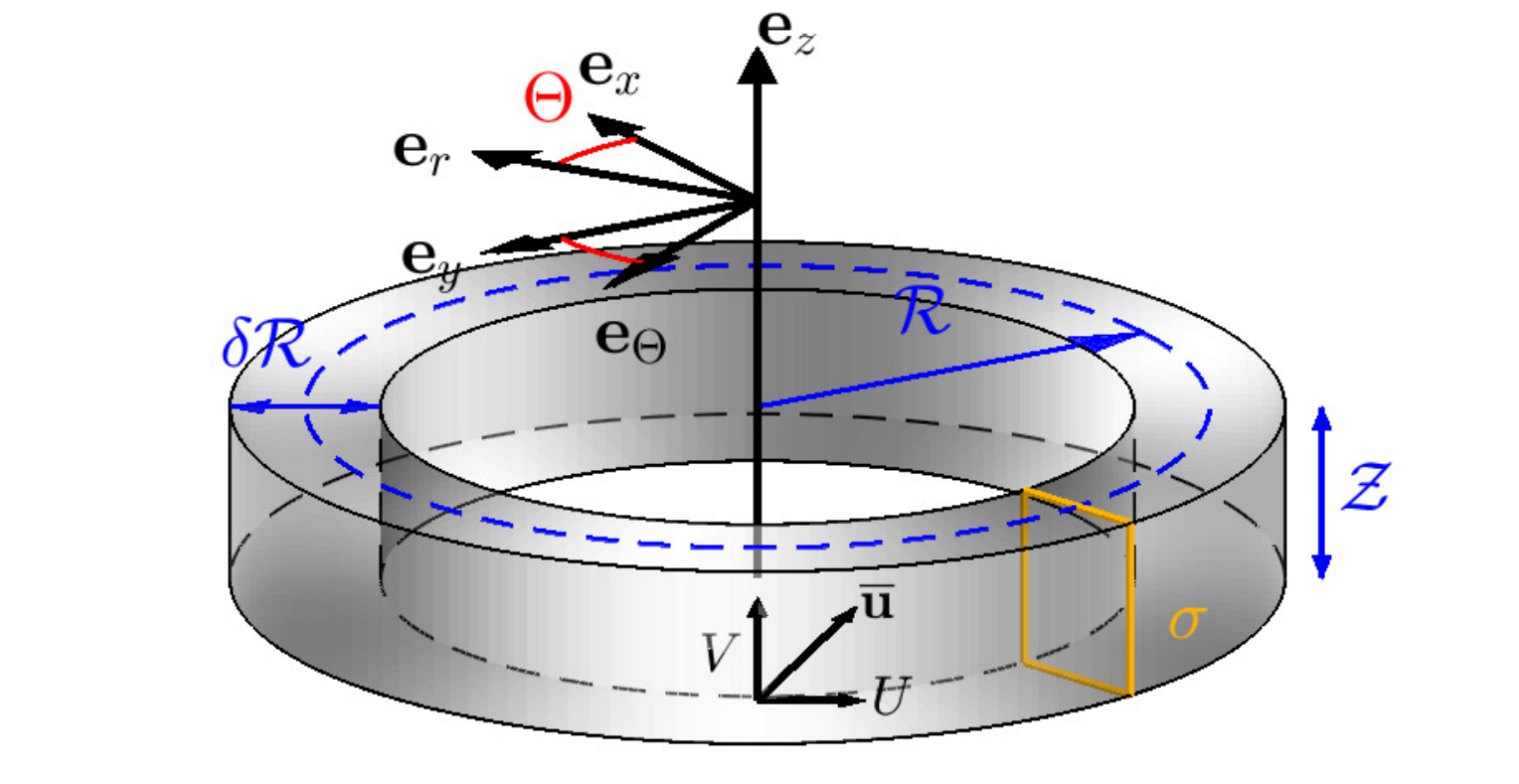}
	\caption{Simplified geometry of a thin annular combustor. $\sigma$ is a poloidal cross-section on which the wave equation is averaged.}
	\label{fig:chamber}
\end{figure}

In this part, we present a model for thermoacoustic instabilities in an idealized annular combustor. The combustion chamber is modelled as a thin annulus of mean radius $\mathcal{R}$, of thickness $\delta \mathcal{R}$, with $\mathcal{R}\gg\delta \mathcal{R}$, and of length $\mathcal{Z}$. The geometry considered for the annular chamber is given in Fig. \ref{fig:chamber}. The response of the flame is modelled with a 1D distribution of heat release rate $\dot{Q}(t,\Theta)$, with $t$ the time, and $\Theta$ the azimuthal coordinate. The chamber features a uniform mean flow $\bar{\uu}=U \ee_\Theta + V\ee_z$, corresponding to a simple swirling motion. The Mach number of this mean flow is low, therefore quadratic terms in the Mach number are neglected. The temperature and entropy gradients are not considered, as well as the entropy fluctuations and unsteady conduction phenomena. With these assumptions, the thermoacoustic wave equation for the linearised pressure perturbations  $p'(\xx,t)$ is:
\begin{align}\label{eq:wavepSimp}
\Box_{\bar{\uu}}\,p' \:=\: \dfrac{\gamma - 1}{c^2}\left(\dfrac{\partial \dot{Q}'}{\partial t} +\frac{U}{\mathcal{R}}\frac{\partial \dot{Q}'}{\partial\Theta}\right).
\end{align}
where $\xx$ is the position vector, $c$ the speed of sound, $\gamma$ the heat capacities ratio, $\dot{Q}'$ refers to the linearised perturbations of the heat release rate, and  $$\Box_{\bar{\uu}}\equiv\dfrac{1}{c^2}\left(\dfrac{\partial^2}{\partial t^2} + 2\bar{\uu}\cdot\nabla\dfrac{\partial}{\partial t} \right)- \nabla^2$$ 
the convected wave operator. The details of the derivation of this equation are provided in the appendix \ref{apC}. The strong simplifications leading to Eq. \eqref{eq:wavepSimp} prevent any quantitative modeling of the thermoacoustic stability in a practical configuration. However, as explained in the introduction, our aim is not to establish a model for quantitative prediction of the thermoacoustic dynamics in a real configuration, but to derive a model which governs the thermoacoustic dynamics in an \textit{idealized} annular chamber with uniform temperature distribution and without vorticity and entropy disturbances.

In order to obtain a 1D equation for $p'$, Eq. \eqref{eq:wavepSimp} is averaged over a poloidal cross section $\sigma=\mathcal{Z}\, \delta\mathcal{R}$. The averaging operator is denoted $\langle\cdot\rangle_\sigma$. For instance, the acoustic pressure $p_a(\Theta,t)$ associated to pure azimuthal thermoacoustic modes can be approximated by
\begin{align}
\langle p'\rangle_\sigma= \dfrac{1}{ \mathcal{Z}\, \delta\mathcal{R}}\int_0^{\mathcal{Z} }\int_{\mathcal{R}-\delta\mathcal{R}/2}^{\mathcal{R}+\delta\mathcal{R}/2} p'(r,\Theta,z,t) \,dr dz=\dfrac{1}{\sigma}\int_\sigma p'(r,\Theta,z,t) d\sigma.
\end{align}
This averaging is performed on eq. \eqref{eq:wavepSimp}. Focusing first on the left-hand-side of this equation, one can write
\begin{align}\label{eq:SWp}
\langle\Box_{\bar{\uu}}\,p'\rangle_\sigma= \dfrac{1}{c^2}\dfrac{\partial^2 p_a}{\partial t^2} + 2\dfrac{U}{c^2}\dfrac{\partial^2 p_a}{\partial \Theta\partial t} - \langle\nabla^2 p' \rangle_\sigma.
\end{align}
The last term can be rewritten as:
\begin{align}\label{eq:Slaplacian}
\langle\nabla^2 p'\rangle_\sigma = \dfrac{1}{\sigma}\int_{\sigma} \nabla^2_{r,z} p'\; d\sigma + \dfrac{1}{\sigma}\int_{\sigma} \dfrac{1}{r^2}\dfrac{\partial^2 p'}{\partial \Theta^2} \; d\sigma
\end{align}
where $\nabla^2_{r,z}$ is the 2D Laplace operator in the slice $S$. The divergence theorem gives:
\begin{align}\label{eq:ostrogradsky}
\dfrac{1}{\sigma}\int_{\sigma} \nabla^2_{r,z} p'\; d\sigma = \dfrac{1}{\sigma}\int_{l} \nabla_{r,z} p'\cdot\nn\; dl 
\end{align} 
where $l$ is the contour of $\sigma$ and $\nn$ is the outwards normal.
Considering the specific acoustic impedance $Z_b(\xx,\omega)$ of the annulus boundary, with $\omega=2\pi f$ being the angular frequency,  the local boundary condition for the acoustic pressure is
\begin{align}\label{eq:BCfreq}
\nabla_{r,z} \hat{p}_a\cdot\nn = -\dfrac{i\omega}{Z_b c}\hat{p}_a ,
\end{align}
where $\hat{p}_a$ is the acoustic pressure in the frequency domain.
The boundary delimiting annular combustion chambers comprise the outlets of the burners, the hard walls of the inner and outer liners and the high-Mach turbine inlet. The latter two exhibit very high resistance and negligible reactance, and the  impedance of the burners defines the acoustic coupling between the chamber and the compressor-outlet plenum \cite{evesque02}. In this work, we present an analytical model for the dynamics of pure azimuthal thermoacoustic modes in combustion chambers, and we focus on situations where the acoustic coupling with the plenum can be neglected. We thus neglect the reactance and consider a purely resistive specific impedance, i.e. $Z_b\in\mathbb{R}$, at the annulus boundary. Consequently, eq. (\ref{eq:BCfreq}) can be rewritten in the time domain as 
\begin{align}\label{eq:BCtime}
\nabla_{r,z} p_a\cdot\nn = -\dfrac{1}{Z_b \,c}\frac{\partial p_a}{\partial t},
\end{align}
We assume that the integrated acoustic losses at the annulus boundary can be modelled by a simple term involving an effective, uniform positive specific resistance, i.e. $Z_b(\xx,\omega)= Z_e>0$, and we obtain from (\ref{eq:ostrogradsky}) and  (\ref{eq:BCtime})
\begin{align}
\dfrac{1}{\sigma}\int_\sigma \nabla^2_{r,z} p'\; d\sigma \approx -\frac{\alpha}{c^2}\dfrac{\partial p_a}{\partial t},
\end{align}
with the linear damping coefficient $\alpha=lc/(\sigma Z_e)$
which sets the acoustic energy losses at the annulus boundaries.
In the last term of eq. (\ref{eq:Slaplacian}), $1/r^2$ can be replaced by $1/\mathcal{R}^2$ and factorized out of the integral because the thickness of the chamber is small compared to the radius $\mathcal{R}$. Equation (\ref{eq:Slaplacian}) becomes:
\begin{align}\label{eq:SlaplacianSimp}
\langle\nabla^2 p'\rangle_\sigma \approx -\frac{\alpha}{c^2}\dfrac{\partial p_a}{\partial t} + \dfrac{1}{\mathcal{R}^2}\dfrac{\partial p_a}{\partial \Theta^2}
\end{align}
and the spatial averaging applied to eq. (\ref{eq:wavepSimp}) gives the following 1D convected wave equation with damping and unsteady heat release rate source:
\begin{align}\label{eq:azWavEq}
\dfrac{\partial^2 p_a}{\partial t^2} + 2\,\dfrac{U}{\mathcal{R}}\dfrac{\partial^2 p_a }{\partial \Theta \partial t} + \alpha\dfrac{\partial p_a}{\partial t} - \dfrac{c^2}{\mathcal{R}^2} \dfrac{\partial^2 p_a}{\partial \Theta^2}
\:=\: (\gamma - 1)\:\left<\dfrac{\Dbar \dot{Q}'}{D t}\right>_\sigma \: .
\end{align}

\subsection{Thermoacoustic feedback}\label{sec24}
In the previous section, we averaged the left-hand-side of eq. \eqref{eq:wavepSimp} over a poloidal cross section $\sigma$ of the annular combustion chamber. Now, we focus on the averaging and the modelling of the right-hand-side, which contains the heat release rate source term. Thermoacoustic instabilities result from the  two-way coupling between the acoustic field and the flames. One of the many coupling mechanisms is as follows: in the case of an acoustically-stiff fuel injection, an acoustically-modulated air mass flow through the burner will induce equivalence ratio modulations. The latter are convected toward the flame front, reach it after a convective delay and induce heat release rate oscillations $\dot{Q}'$. Depending on the delay, these oscillations, which constitute a source term of the wave equation, destructively or constructively interact with the acoustic field which produced them. In the following paragraph, we will use a minimal model to account for the heat release rate response $\dot{Q}'$ to the acoustic field $p_a$. Considering the 1D convected wave equation in frequency domain
\begin{multline}\label{eq:azWavEqFr}
(i\omega)^2 \hat{p}_a + 2i\omega\,\dfrac{U}{\mathcal{R}}\dfrac{\d \hat{p}_a }{\d \Theta} +i\omega\alpha \,\hat{p}_{a} - \dfrac{c^2}{\mathcal{R}^2} \dfrac{\d^2 \hat{p}_a}{\d \Theta^2}\\
\:=\: (\gamma - 1)\left\{i\omega + \dfrac{U}{\mathcal{R}}\dfrac{\d}{\d \Theta}\right\}\left<\hat{\dot{Q}}\right>_\sigma,
\end{multline}  
we would like to express the unsteady heat release as a function of the acoustic  pressure $\hat{p}_a$. Before elaborating further on this model, it is important to mention that, for gas turbine applications, although the acoustic pressure $p_a$ is \emph{directly} involved in the Rayleigh criterion given in eq. \eqref{eq_Rayleigh}, its influence on the heat release rate oscillations from the flames is \emph{indirect}. Indeed, turbulent propagating flames are significantly affected by fluctuations of the reactants velocity field and fluctuations of the local equivalence ratio, but very weakly by the acoustic pressure. Considering that the former perturbations are caused by $\hat{p}_a$ at the burners via their impedance $\rho c \,Z_\text{bu}(\omega)=\hat{p}_a(\xx_\text{bu},\omega)/\hat{u}_a(\xx_\text{bu},\omega)$, one can nonetheless express the coherent heat-release-rate as function of $\hat{p}_a$. Usually, this transfer function varies smoothly with the angular frequency and it can be approximated by a complex constant over a frequency range that spans the width of the thermoacoustic mode peak: 
\begin{align}
(\gamma - 1) \left<\hat{\dot{Q}}\right>_\sigma   = (\beta + i\upsilon)\hat{p}_a,
\end{align}
with  $(\beta,\upsilon)\in\mathbb{R}^2$.
In the time domain, it gives $(\gamma - 1)\langle\dot{Q}'\rangle_\sigma = \beta p_a + (\upsilon/\omega)\d p_a/\d t$. 
Since the time derivative of $\langle\dot{Q}'\rangle_\sigma $ acts as a source term in the wave equation, the term $\beta p_a$ will act as a linear damping term, which will make the system more stable if $\beta<0$ and less stable, if not linearly unstable, if $\beta>0$. The second term  affects the oscillation frequency. In gas turbine combustors, it usually results in a slight shift of the frequency of the thermoacoustic eigenmode from the one of the corresponding acoustic eigenmode -- the latter being a solution of the homogeneous problem associated with eq. \eqref{eq:azWavEqFr}. For the sake of simplicity, this term is omitted in the present model and we just write that $(\gamma - 1)\langle\dot{Q}'\rangle_\sigma = \beta p_a$. 
This model leads to unphysical unbounded acoustic pressure levels when the system is thermoacoustically unstable. In gas turbine combustors, when the acoustic level rises, the flame response to acoustic perturbations is the main nonlinear element in the thermoacoustic feedback loop, while acoustic propagation remains mostly linear.  The response at the fundamental frequency saturates, with a redistribution of the energy to the higher order harmonics in the power spectral density of the spatially-integrated heat release rate. This differs from thermoacoustic instabilities in rocket motors where nonlinear acoustics governs the limit cycle amplitude \cite{yang_1990}. A minimal model to account for saturation of the flame response in the case of supercritical Hopf bifurcations \cite{noiray11,noiray13,noiray_proci2017} consists in adding a 3$^\text{rd}$ order term to the relationship linking $\langle\dot{Q}'\rangle_\sigma$ and  $p_a$:
\begin{align}\label{eq:heatRelease}
(\gamma-1)\langle\dot{Q}'\rangle_\sigma=\beta p_a - \kappa p_a^3.
\end{align}

In this expression, $\beta$ and $\kappa$ can be chosen constant in space  in the case of a uniform azimuthal distribution of  heat release rate, or, as it is done in \cite{noiray11}, they can depend on $\Theta$ to account for asymmetries that are usually present in  real combustors. In this paper, $\kappa$ is assumed constant in the volume in order to simplify the derivations, and $\beta$ is expressed as a Fourier series with real coefficients:
\begin{align}\label{eq:fourierBeta}
\beta(\Theta) = \beta_0\left\{1\: +\: \sum\limits_{m=1}^{\infty} c_m \cos[m(\Theta-\Theta_{m})]\right\}
\end{align}
The angles $\Theta_m$ are chosen so that all the $c_m$ are positive.\\
The simple cubic nonlinearity used in the expression \eqref{eq:heatRelease} is not meant to give a predictive description of the flame response: The thermoacoustic bifurcation diagram is defined by the details of the nonlinear component of the flame response to acoustic perturbations in a given combustor. Nevertheless, for flames exhibiting any sigmoid-type nonlinear response to acoustic perturbations, the topology of stochastic super-critical Hopf bifurcations will not differ from the one obtained with a simple cubic nonlinearity, at least in the vicinity of the bifurcation point (see for instance Figs. 2, 4 and 8 in \cite{noiray17}, where the cubic nonlinearity is compared to an arctangent nonlinearity). 
In the case of sub-critical Hopf bifurcations, the topology of the bifurcation diagram is changed (see Figs. 9 and 11 in \cite{noiray17}) and can only be described by using a more complex flame describing functions \cite{ghirardo16}. Additionally,  expression \eqref{eq:heatRelease} cannot be used to represent dynamic nonlinearities involving an amplitude-dependent delay of the flame response to acoustic perturbation, unlike the general flame describing function considered in \cite{ghirardo16}.
From now on, the subscript $(\cdot)_a$ will be omitted and  $p$ approximates the acoustic pressure of  azimuthal eigenmodes at a given azimuthal position. 
One finally obtains the 1D thermoacoustic wave equation in thin annulus with mean swirl:
\begin{multline}\label{eq:thEq}
\dfrac{\partial^2 p}{\partial t^2} + \dfrac{2U}{\mathcal{R}}\dfrac{\partial^2 p }{\partial \Theta \partial t} +  \alpha \dfrac{\partial p}{\partial t} - \dfrac{c^2}{\mathcal{R}^2} \dfrac{\partial^2 p}{\partial \Theta^2}
\\[10pt]
\quad\:=\: \beta(\Theta)\dfrac{\partial p}{\partial t}+\dfrac{U}{\mathcal{R}}\dfrac{\partial [\beta(\Theta)p]}{\partial \Theta} -3\kappa p^2 \left(\dfrac{\partial p}{\partial t} +\dfrac{U}{\mathcal{R}}\dfrac{\partial p}{\partial \Theta} \right).
\end{multline}

\section{  Quaternion-based acoustic field projection }\label{sec3}

The wave equation \eqref{eq:thEq} involves several different time scales: A time scale associated to acoustic oscillation, a time scale associated to the thermoacoustic growth rate, and a time scale associated to convective phenomena due to the presence of the mean azimuthal flow. In practical applications, the  latter two are slow compared to the first one. The acoustic time scale will be referred to as \textit{fast time scale}, and the other two, as \textit{slow time scales}.
In the remainder of the paper, approximate solutions of eq. \eqref{eq:thEq} will be derived using time-scale separation in order to describe the thermoacoustic instability on the slow time scales only. To that end, a quaternion-based acoustic pressure will be used in order to unambiguously describe the state of azimuthal eigenmodes in presence of spatial asymmetries of the thermoacoustic feedback as well as in presence of a mean swirling flow.  This quaternion-based projection of the acoustic field has recently been proposed in \cite{ghirardo18}. It can be linked to the concept of analytical signal $\tilde{x}(t)=A(t)\exp(i(\omega t + \varphi(t)))$ for  monovariate signals $x(t)$ oscillating at pulsation $\omega$,  with $A(t)$ representing the real-valued instantaneous amplitude of the oscillation, and $\varphi(t)$ the real valued phase drift. The oscillations being quasi-harmonic, $A(t)$ and $\varphi(t)$ have to be slow compared to the oscillation frequency: This imposes the conditions $\dot{\varphi}\ll\omega$ and $\dot{A}\ll\omega A$. $A$ and $\varphi$ constitute the \textit{complex embedding} of the signal and they describe its evolution on time scales that are large compared to an acoustic period.  When the evolution of $x$ is governed by a second order nonlinear oscillator equation, the method of Krylov-Bogoliubov \cite{sanders85} allows to separate the slow scales from the fast oscillations and to obtain first order differential equations for $A$ and $\varphi$. The concept of \textit{quaternion embedding} was introduced in \cite{lebihan14} and further developed in \cite{flamant17} in order to represent a bivariate (or complex valued) signal in a similar way. The quaternion embedding can be seen as an equivalent of the analytical signal for a bivariate signal. In the case of almost harmonic perturbations, the quaternion embedding will be defined by a slow amplitude $A$, a slow phase drift $\varphi$ and two additional slow variables $\chi$ and $\theta$ characterizing the spatial dynamics: $\chi$ will affect the ellipticity of the oscillations and $\theta$, their direction. Ghirardo and Bothien showed in \cite{ghirardo18} that this quaternion formalism is perfectly suited to represent azimuthal waves in annular combustion chambers. They emphasize on the fact that this representation overcomes the problem of ill-posedness of the phase in  previously adopted projections on standing modes e.g. \cite{schuermans06,noiray13} or on counter-rotating modes \cite{hummel17b}. In the quaternion representation, it is always possible to properly define a temporal phase, making this representation very suited to model wave propagation phenomena in annular or cylindrical elements, and we will show that another advantage of the quaternion embedding is that it captures both rotational and reflectional symmetry breaking bifurcations.

\subsection{Basic properties of the quaternions}\label{sec31}
This section presents very succinctly some properties of the quaternion algebra, which will be referred to as $\mathbb{H}$.
This non-commutative algebra features 3 imaginary units $i$, $j$, $k$, such as $i^2=j^2=k^2=ijk=-1$. It follows that $ij=k$, $jk=i$, $ki=j$ and that, for instance $ji=-k$.
Every quaternion can be written uniquely in its cartesian form:
\begin{align}
\forall\: h \in \mathbb{H}\quad \exists\: (a,b,c,d)\in\mathbb{R}^4 \quad h = a + b i + c j + d k,
\end{align}
with $a$ the real part of the quaternion $h$, which will be noted $\Re(h)$. The $i$,$j$,$k$-imaginary parts $b$, $c$ and $d$ are respectively designed as $\Im_i(h)$,$\Im_j(h)$,$\Im_k(h)$. The modulus is defined as:
\begin{align}
\forall\: (a,b,c,d)\in\mathbb{R}^4 \quad \lvert a + b i + c j + d k \rvert = \sqrt{a^2+b^2+c^2+d^2}.
\end{align}
A unitary quaternion is a quaternion of modulus 1. The pure imaginary units $i$, $j$, $k$ are unitary quaternions. The quaternion conjugate (q.c.) is:
\begin{align}
\forall\: (a,b,c,d)\in\mathbb{R}^4 \quad ( a + b i + c j + d k )^* = a - b i - c j - d k .
\end{align}
So we have:
\begin{align}
\forall\: h\in\mathbb{H} \quad \Re(h) = \dfrac{1}{2}(h+h^*).
\end{align}
An other useful property is:
\begin{align}\label{eq:conjCom}
\forall\: (x,y)\in\mathbb{H}^2 \quad (xy)^* = (y^*x^*).
\end{align}
We also see that the sub-algebras $\mathbb{R}+i\mathbb{R}$, $\mathbb{R}+j\mathbb{R}$, $\mathbb{R}+k\mathbb{R}$ are commutative.
For all $x \in \mathbb{R}$ and $\mu\in\{i,j,k\}$, we have the following Euler decomposition:
\begin{align}
e^{\mu x} = \cos(x) + \mu\sin(x),
\end{align}
which is a unitary quaternion. Any quaternion $h$ can be written in its exponential form:
\begin{align}
h = A e^{i\theta}e^{-k \chi}e^{j\phi},
\end{align}
with $A\in\mathbb{R}$ the modulus  and $(\theta,\chi,\phi)\in]-\pi,\pi]\times[-\pi/4,\pi/4]\times]-\pi,\pi]$ the phase triplet of $h$ (see \cite{flamant17}).

\subsection{Hypercomplex acoustic field}\label{sec32}

A quaternion-domain pressure field $\tilde{p}$ is associated to the real-valued pressure field as
\begin{align}\label{eq:preal}
p = \Re(\tilde{p})\,=\dfrac{\tilde{p}+\tilde{p}^*}{2}
\end{align}
Due to the axisymmetric character of the geometry considered in this work, $p(t,\Theta)$ and $\tilde{p}(t,\Theta)$ are $2\pi$-periodic in $\Theta$.
\\
For  a $2\pi$-periodic function $f(\Theta):\mathbb{R}\mapsto\mathbb{H}$, a quaternion Fourier series can be expressed  by writing the Fourier series of each of the quaternion components. It can be shown that
\begin{align}\label{eq:ft}
f+f^*=f_0 + \sum\limits_{n=1}^{\infty} e^{-in\Theta}K_n + K^*_n e^{+in\Theta},
\end{align}
where $(K_n)_{n\in\mathbb{N}^*}$ are quaternion coefficients. Using this result and \eqref{eq:preal}, $p(\Theta,t)$ can be expanded as a sum of azimuthal modes:
\begin{align}
p(\Theta,t) = \eta_0(t)+ \dfrac{1}{2}\sum\limits_{n=1}^{\infty}(e^{-in\Theta}\eta_n(t) + \eta^*_n(t) e^{+in\Theta})
\end{align}
This work focuses on situations where the acoustic field is governed by the $n^\text{th}$ azimuthal eigenmode. It does not include situations where several eigenmodes nonlinearly interact, such as  the intriguing synchronization phenomenon involving azimuthal and longitudinal thermoacoustic modes \cite{moeck18}. One can therefore continue with the following approximation:
\begin{align}\label{eq:pQuat}
p (\Theta,t)\approx \, \dfrac{1}{2}(e^{-in\Theta}\eta_n(t) + \eta_n^*(t) e^{+in\Theta})\,= \Re\left[e^{-in\Theta}\eta_n(t)\right].
\end{align}
Without loss of generality, the hypercomplex time signal $\eta_n$ can be written with the quaternion exponential form initially proposed by B\"{u}low and Sommer \cite{bulow01} and later modified by Flamant \textit{et al.} \cite{flamant17}:
\begin{align}
\eta_n(t)=A(t)e^{in\theta(t)}e^{-k\chi(t)}e^{j \phi(t)}
\end{align}
In the present work, it is assumed that the sources distribution along the annulus induces only weak damping or amplification of the acoustic energy.  Consequently, $\eta_n$   exhibits quasi-sinusoidal oscillations at a pulsation that is close to the eigenfrequency $\omega_n=2\pi c/\lambda_n=nc/R$ of the associated homogeneous Helmholtz equation with zero mean swirl, and we can write it in the form:
\begin{align}\label{etaExp}
\eta_n(t)=A(t)e^{in\theta(t)}e^{-k\chi(t)}e^{j(\omega_n t + \varphi(t))}.
\end{align}
In that expression, $A$, $\theta$, $\chi$, $\varphi$ are real valued functions whose time variations are slow compared to the fast oscillations at pulsation $\omega_n$, with $(A,\theta,\chi,\phi)\in\mathbb{R}^+\times]-\pi,\pi]\times[-\pi/4,\pi/4]\times]-\pi,\pi]$. The exponential notation for the hypercomplex pressure is thus 
\begin{align}\label{eq:pH}
\tilde{p}(\Theta,t)=A(t)e^{in(\theta(t)-\Theta)}e^{-k\chi(t)}e^{j(\omega_n t + \varphi(t))}.
\end{align}

\subsection{Physical meaning of the slow variables}\label{sec33}

Using the slow-flow variables $A$, $\theta$, $\chi$, $\varphi$ is particularly convenient for differentiating standing, clockwise and counterclockwise spinning modes over time scales that are long compared to the acoustic period.
The convention used here is the same as the one presented in \cite{ghirardo18}: 
\\
The \textit{nature angle} $\chi$ characterizes the dynamic nature of the mode: it defines whether the mode is standing ($\chi=0$) or  spinning at the speed of sound in the clockwise $(\chi=-\pi/4)$ or counterclockwise $(\chi=\pi/4$) direction. Intermediate values of $\chi$ correspond to mixed modes, which can be written as the sum of a spinning and a standing mode. $\chi>0$ means that the spinning component spins counterclockwise (CCW), $\chi<0$ means it spins clockwise (CW).\\
The angle $\theta$ is the \textit{preferential direction} because it indicates where the acoustic amplitude is the largest. For a standing mode, $\theta$ is the anti-nodal direction of the acoustic pressure oscillations. For mixed modes, $\theta$ is the direction of the standing component. For pure spinning modes, there is no preferential direction and $\theta$ acts as a temporal phase angle. When $\theta$ vary in time, the preferential direction of the mode drifts slowly. This has to be distinguished from the phenomenon of spinning wave: in the case of standing modes with a slowly drifting preferential direction, the modes do not propagate in the azimuthal direction and the rotation velocity is small compared to the speed of sound. As discussed in \cite{ghirardo18}, the range of $\theta$ could be limited to $[0,\pi[$ instead of $]-\pi,\pi]$, because $\theta$ and $\theta+\pi \mod 2\pi$  represent exactly the same state. However, the range  $[0,\pi[$ is not chosen since it leads to potentially misleading phase jumps  in  $\theta$ time traces.\\
The angle $\varphi$ is the slowly varying \textit{temporal phase} of the wave. A slow linear drift of $\varphi$ with time can be interpreted as a small frequency shift.\\
Finally, $A$ gives the amplitude of the acoustic pressure oscillations. In the case of a standing mode, $A$ is the amplitude of $p$ in the anti-nodal direction. For a pure spinning mode, the amplitude of the Riemann invariant is $A/\sqrt{2}$.
To illustrate this quaternion-based modal projection, one considers the real part  of eq. \eqref{eq:pH}, which provides the expression for the real pressure:
\begin{align}\label{eq:pRe}
p(\Theta,t) =  \,& A \cos(n(\Theta - \theta))\cos(\chi) \cos(\omega_n t + \varphi) \nonumber \\
&+ A \sin(n(\Theta - \theta)) \sin(\chi) \sin(\omega_n t + \varphi)
\end{align}
From this expression, it becomes clear that a pure standing wave is obtained when $\chi(t)=\theta(t)=0$. If the slow variable $\theta$ varies over time, the nodal lines of the mode correspondingly slowly drifts, but the oscillation nevertheless falls in the standing wave category ($\chi=0$) as opposed to the pure spinning modes ($\chi=\pm \pi/4$) for which the nodal lines spin at the speed of sound. Equation \eqref{eq:pRe} can be rewritten as a sum of two waves spinning respectively counterclockwise and clockwise:
\begin{align}\label{eq:pReSpin}
p (\Theta,t) = \,& A^{+} \cos[-n(\Theta-\theta) + \omega_nt + \varphi] \nonumber \\
&\:+\: A^{-} \cos[n(\Theta-\theta) + \omega_nt + \varphi]
\end{align}
where the counterclockwise mode amplitude $A^+$ and the clockwise mode amplitude $A^-$ are defined as:
\begin{align}
A^{+} = \dfrac{\cos(\chi)+\sin(\chi)}{2}\,A\quad\text{and}\quad   A^{-} = \dfrac{\cos(\chi)- \sin(\chi)}{2}\,A
\end{align}
For a pure counterclockwise spinning mode ($\chi=\pi/4$), $A^{-}$ is zero, while for a pure clockwise mode  ($\chi=-\pi/4$), $A^{+}$ is zero. In these cases, slow time variations of $\theta$ and $\varphi$ result in  slow changes of the instantaneous frequency.

\subsection{Representation on the Bloch sphere}\label{sec34}

\begin{figure}[H]
	\centering
	\includegraphics[width=0.7\textwidth]{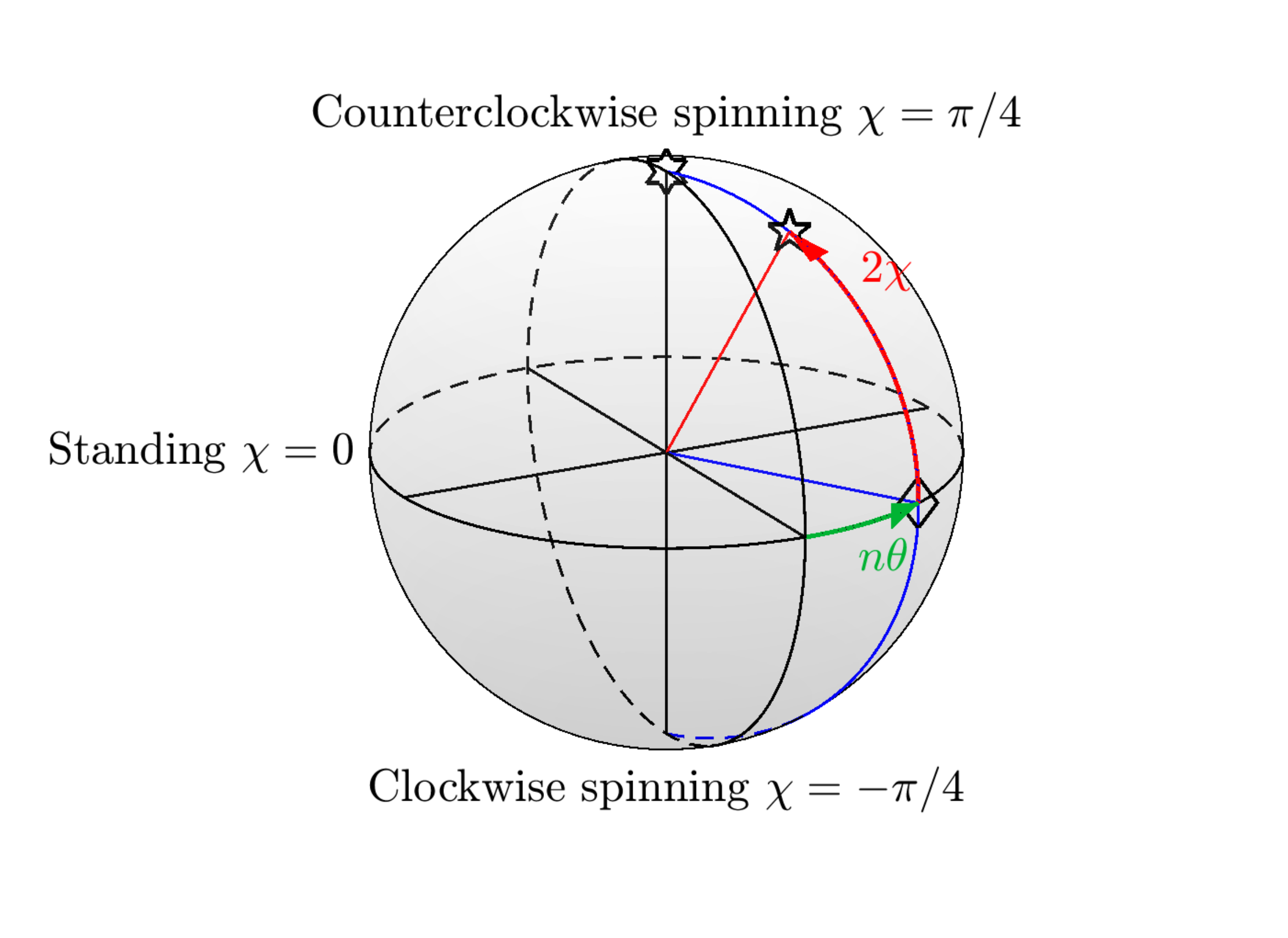}
	\caption{Representation of the mode state on the Bloch sphere. The markers correspond to the 3 cases displayed on Fig. \ref{fig:mode_types}}
	\label{fig:bloch}
\end{figure}

\begin{figure}[H]
	\centering
	\begin{subfigure}{.3\linewidth}
		\centering
		\includegraphics[width=\linewidth]{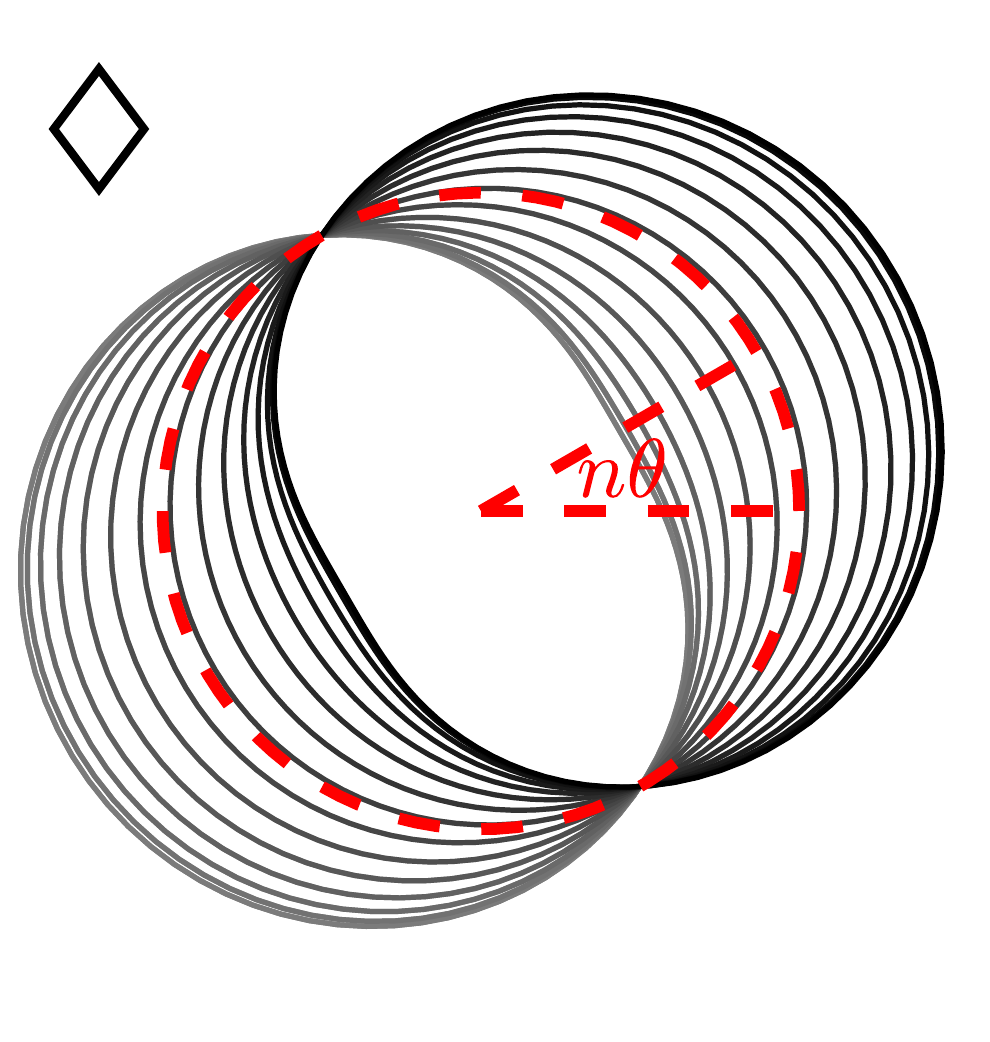}
		\caption{standing mode}
		\label{fig:stmode}
	\end{subfigure}%
	\begin{subfigure}{.3\linewidth}
		\centering
		\includegraphics[width=\linewidth]{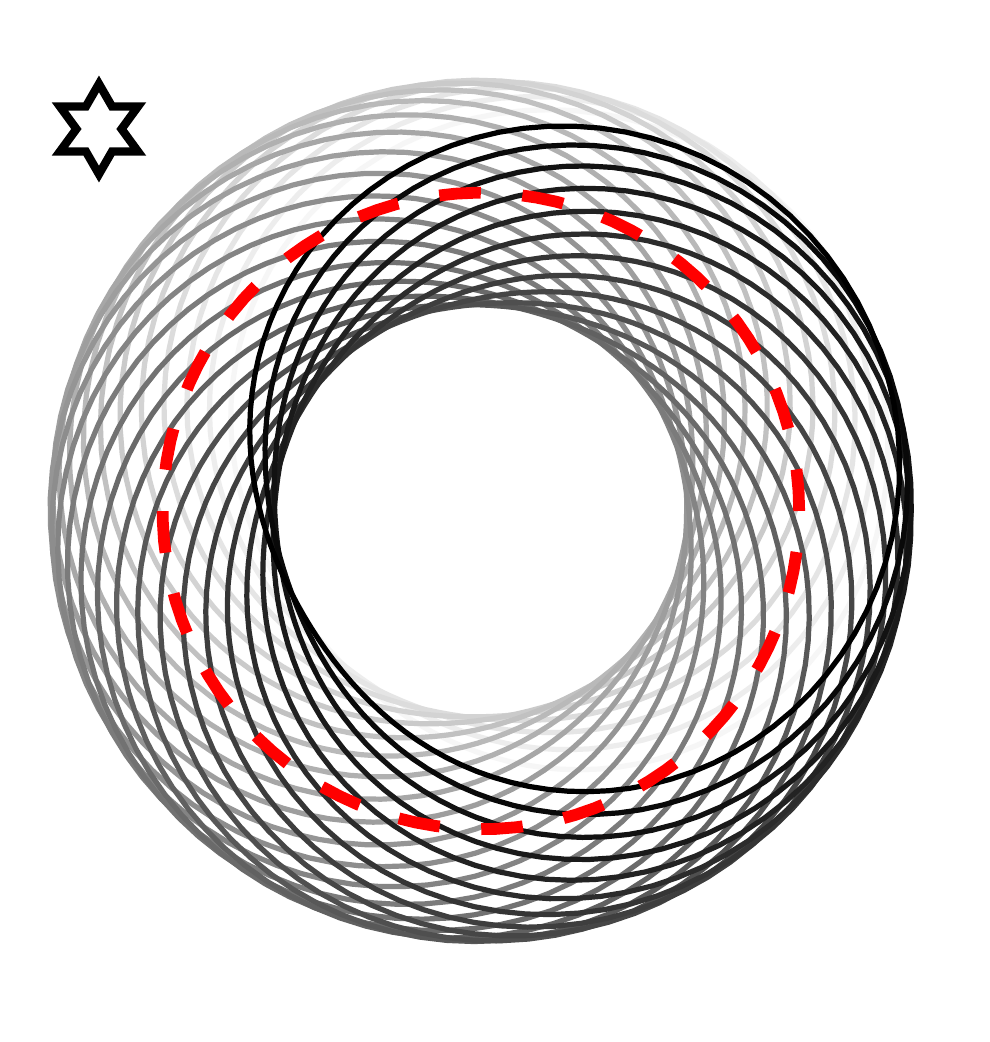}
		\caption{spinning mode}
		\label{fig:spmode}
	\end{subfigure}%
	\begin{subfigure}{.3\linewidth}
		\centering
		\includegraphics[width=\linewidth]{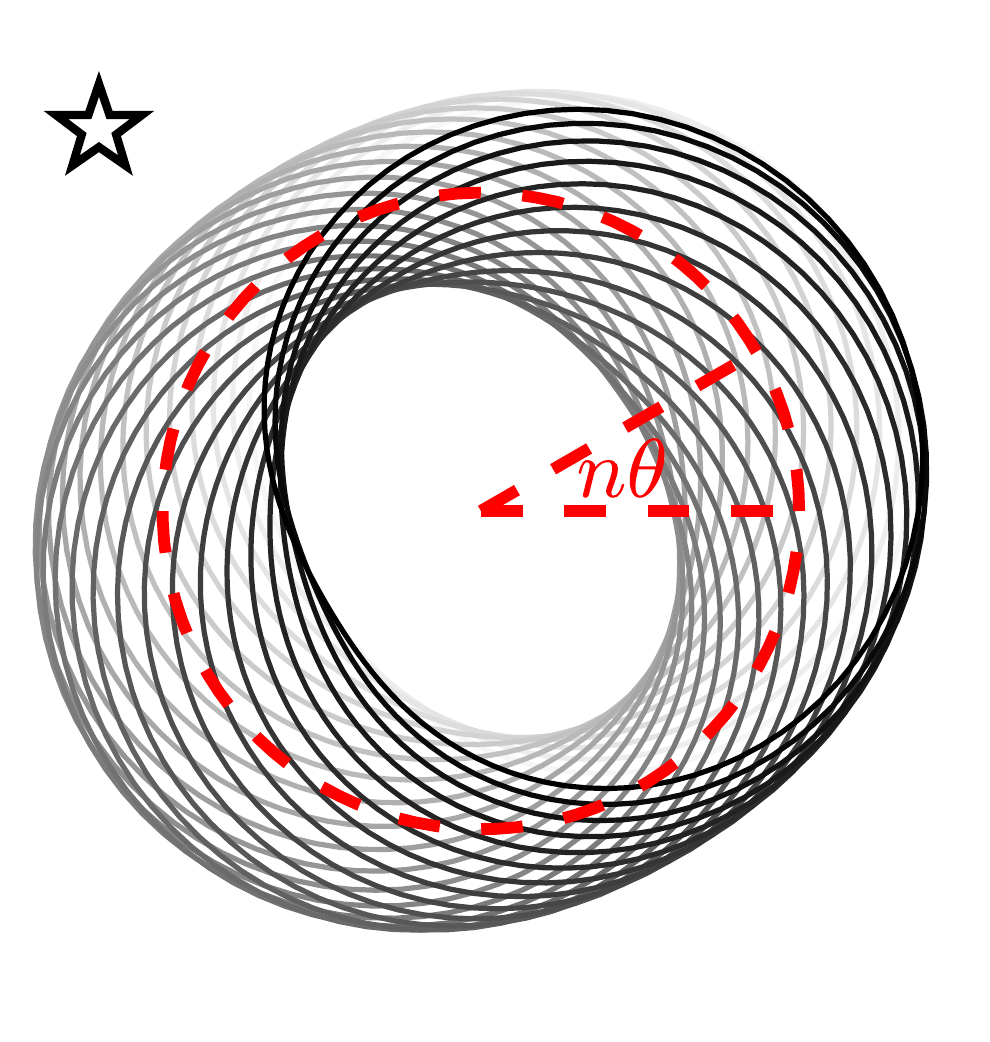}
		\caption{mixed mode}
		\label{fig:mixmode}
	\end{subfigure}%
	\begin{subfigure}{.07\linewidth}
		\centering
		\includegraphics[width=\linewidth]{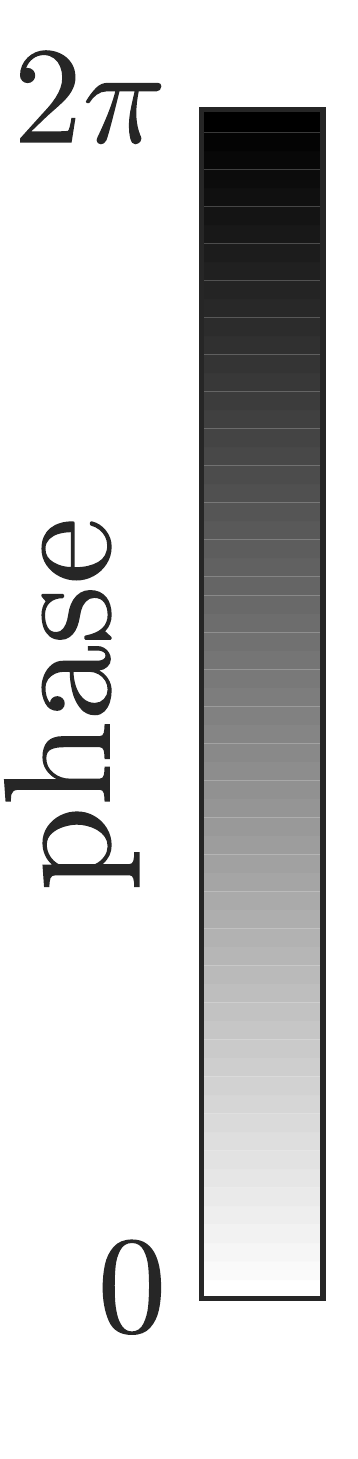}
	\end{subfigure}%
	\caption{Representation of the acoustic pressure in the chamber for three different states of the 1$^{st}$ azimuthal mode (n=1), each of them being described by a point on the Bloch sphere: (\ref{fig:stmode}) standing mode with $\chi=0$ and $\theta=\pi/6$; (\ref{fig:spmode}) counterclockwise spinning mode with $\chi=\pi/4$; (\ref{fig:mixmode}) counterclockwise mixed mode with $\chi=\pi/6$ and $\theta = \pi/6$. }
	\label{fig:mode_types}
\end{figure}

As brightly proposed by Ghirardo and Bothien \cite{ghirardo18}, an appropriate way to represent graphically the state of an azimuthal mode and the trajectories of its slow evolution in time is the Bloch sphere, with the following  spherical coordinates $\{A,n\theta,2\chi\}$. As shown in Fig. \ref{fig:bloch}, for constant amplitude A, the three-vector $\left\{A,n\theta,2\chi\right\}$ points on a sphere whose equator corresponds to pure standing modes and the two poles, to pure spinning modes, for which the preferential angle $\theta$ is not meaningless since it acts as a temporal phase. This spherical coordinate system involves twice the nature angle $\chi$ because it is convenient to represent spinning modes $\chi=\pm\pi/4$ at the poles. Another specificity of the chosen representation is that $\left[A,n\theta,2\chi\right]$ and $\left[A,n\theta+\pi,2\chi\right]$ indicate the same state. It is also important to mention that this representation of the azimuthal mode state is incomplete in the sense that it does not provide information about slow variation of the temporal phase.
Examples of states for the first azimuthal eigenmode are given in Fig. (\ref{fig:mode_types}).

\section{Derivation of the slow-flow equations}\label{sec4}

In this section, the quaternion formalism of section \ref{sec3}.\ref{sec32} is inserted into the wave equation with thermoacoustic feedback. The objective of the section is to propose a method for obtaining equations that describe the dynamics at slow time scales. The proposed method can be seen as an adaption of the Krylov-Bogoliubov method for bivariate signals: It provides a system of first order differential equations for the four variables $A,\chi,\theta,\varphi$, describing the quaternion embedding on the slow time scales.
We begin the derivation with the thermoacoustic wave equation (\ref{eq:thEq}), into which we replace $p$  with the quaternion exponential notation \eqref{eq:pQuat} introduced in the previous part, and where the subscript $(\cdot)_n$ of $\eta_n$ has been dropped:
\begin{equation}\label{eq:thEqQuat}
\begin{split}
e^{-in\Theta} &\left[\ddot{\eta}  + (\alpha  - i\,2M\omega_n)\dot{\eta} + \omega_n^2\eta \right] \\
&+\left[\ddot{\eta}^* + \dot{\eta}^* (\alpha+i\,2M\omega_n) + \omega_n^2\eta^* \right]e^{+in\Theta}\\
&\quad\quad\quad 
\:=\: \left[\beta(\Theta) - \frac{3\kappa}{4} \left(e^{-in\Theta}\eta + \eta^*e^{+in\Theta}\right)^2\right]\\
&\quad\quad\quad\quad\quad
\times\left(e^{-in\Theta}\left[\dot{\eta} - i\,M\omega_n\eta\right]+ \left[\dot{\eta}^* + i\,M \omega_n\eta^*\right]e^{+in\Theta} \right)  \\
&\quad\quad\quad\quad\quad\quad\quad\quad\quad\quad\quad\quad\quad 
+ \dfrac{M\omega_n}{n}\dfrac{\d\beta(\Theta)}{\d \Theta}\left(e^{-in\Theta}\eta + \eta^*e^{+in\Theta}\right).
\end{split}
\end{equation}\\
In this equation, the eigenfrequency $\omega_n=nc/\mathcal{R}$ and the azimuthal Mach number $M=U/c$ have been introduced. The former corresponds to the wavelength of the $n$-th pure azimuthal mode of a thin annulus of radius $\mathcal{R}$.
The latter is positive when the swirl is counterclockwise and negative if the swirl is clockwise.
\\
The equation (\ref{eq:thEqQuat}) can also be rewritten
\begin{equation}\label{eq:thEqQuat2}
\begin{split}
e^{-in\Theta} \left[\ddot{\eta}  + (\alpha  - i\,2M\omega_n)\dot{\eta} + \omega_n^2\eta \right] + &\,\text{q.c.}\\
\:=\: \left[\beta(\Theta) - \frac{3\kappa}{4} \left(e^{-in\Theta}\eta + \,\text{q.c.}\right)^2\right]&\left(e^{-in\Theta}\left[\dot{\eta} - i\,M\omega_n\eta\right]+ \,\text{q.c.}\right)  \\
&   + \dfrac{M\omega_n}{n}\dfrac{\d\beta(\Theta)}{\d \Theta}\left(e^{-in\Theta}\eta + \,\text{q.c.}\right),
\end{split}
\end{equation}
where q.c. stands for `quaternion conjugate'.

\subsection{Spatial averaging}\label{sec42}
To remove the dependency in $\Theta$, we use the following averaging operator in the azimuthal direction:
\begin{align}\label{eq:volAvg}
\langle\,\cdot\,\rangle_\Theta \:=\: \dfrac{1}{2\pi }\int_0^{2\pi} e^{in\Theta}\:(\,\cdot\,) \: d\Theta
\end{align}
We also introduce the following projection $\mathcal{C}$ from $\mathbb{H}$ to $\mathbb{C}$:
\begin{align}\label{eq:projC}
\forall \; (a,b,c,d)\in\mathbb{R}^4  \quad \mathcal{C}(a + b i + c j + d k) \; \equiv \; a + b i,
\end{align}
which will be a convenient notation for the following steps.
When azimuthal averaging is applied to the left hand side (LHS) of the thermoacoustic equation (\ref{eq:thEqQuat}), we obtain\footnote{For all functions $h:\mathbb{R}\rightarrow\mathbb{H}$ with $h(t)=a(t)+ib(t)+jc(t)+kd(t)$,  we have $(1/2\pi)\int_0^{2\pi} e^{in\Theta}h e^{in\Theta}d\Theta = cj+dk$. Then $\langle e^{-in\Theta}h +h^* e^{in\Theta}\rangle_\Theta = h-cj-dk = a+bi = \mathcal{C}(h)$. }
\begin{align}
\langle\text{LHS}\rangle_\Theta\:=\:\mathcal{C}\left(\ddot{\eta}+\left( \alpha- i\,2M\omega_n\right)\dot{\eta} + \omega_n^2\eta \right).
\end{align}
The azimuthally averaged LHS does not have  $j$-imaginary and $k$-imaginary parts,
and we used the fact that $\mathcal{C}$ commute with the time derivative.\\
To apply the spatial averaging operator to the linear part of the thermoacoustic term, we use the Fourier series of $\beta(\Theta)$ given in eq. (\ref{eq:fourierBeta}) and we show that only the terms of order 0 and $2n$ have a non-zero contribution, which is in agreement with the study of Noiray, Bothien and Schuermans \cite{noiray11}.
We have:
\begin{align}\label{eq:AzAvbp}
\left<\beta \left(e^{-in\Theta}\dot{\eta} + \,\text{q.c.}\right)\right>_\Theta\:  = \,\beta_0\,\Big( \mathcal{C}(\dot{\eta}) \:+\: \dfrac{c_{2n}}{2}\,e^{2in\Theta_{2n}}\,\mathcal{C}(\dot{\eta})^* \Big)
\end{align}
and
\begin{align}\label{eq:AzAvbdp}
\left<M\omega_n \beta(\Theta) \left(-i\,e^{-in\Theta}\eta + \,\text{q.c.}\right) \right>_\Theta = M\omega_n\,\beta_0\,i\, \Big(-\mathcal{C}(\eta) \:+\:  \dfrac{c_{2n}}{2}\,e^{2in\Theta_{2n}}\,\mathcal{C}(\eta)^*\Big)
\end{align}
and
\begin{align}\label{eq:AzAvdbp}
\left<\dfrac{M\omega_n}{n}\dfrac{\d \beta(\Theta)}{\d \Theta} \left(e^{-in\Theta}\eta + \,\text{q.c.}\right) \right>_\Theta = 
\;-\; M\,\omega_n\, \beta_0\, i\, c_{2n}\,e^{2in\Theta_{2n}}\, \mathcal{C}(\eta)^* . 
\end{align}
Since the azimuthal origin can be chosen arbitrarily, we set it so that $\Theta_{2n}=0$.
The locations $\Theta = 0 \mod(\pi/n)$ corresponds therefore to the angles where the Fourier component of order $2n$ of $\beta$ reaches its maximum.
The averaged terms  \eqref{eq:AzAvbp}, \eqref{eq:AzAvbdp}, \eqref{eq:AzAvdbp} terms also have zero $j$- and $k$-imaginary parts.
The nonlinear terms coming from the saturation model are:
\begin{align}
\text{N.L.}\:=\:\frac{3\kappa}{4} \left(e^{-in\Theta}\eta + \,\text{q.c.}\right)^2\left(e^{-in\Theta}\dot{\eta}- M\omega_n e^{-in\Theta}i\eta \:\: +\:\: \text{q.c.}\right).
\end{align}
Replacing $\eta$ with:
\begin{align}
\eta=A(t)e^{in\theta(t)}e^{-k\chi(t)}e^{j(\omega_n t + \varphi(t))}
\end{align}
and $\dot{\eta}$ being written as its quaternion Cartesian decomposition:
\begin{align}
\dot{\eta} = a(t)i + b(t)i + c(t)j + d(t)k
\end{align}
we show that the volume averaging of the nonlinear terms gives an expression with zero $j$- and $k$- imaginary parts and depending only on $t$, $A$, $\chi$, $\theta$, $\varphi$ and the two first parts of $\dot{\eta}$, $a(t)$ and $b(t)$. The global expression of the spatially averaged terms has been obtained with a computer algebra software is too long to be reproduced in this paper. The averaged nonlinear terms will be noted $f_{\text{nl}}\left(A,\chi,\theta,\varphi,\mathcal{C}(\dot{\eta})\right)$.
We define $\xi(t)=\mathcal{C}(\eta(t))\in \mathbb{C}$ and we write the azimuthally averaged equation:
\begin{multline}\label{eq:waveComplex}
\ddot{\xi} + (\alpha-\beta_0)\dot{\xi} - 2i M\omega_n\dot{\xi} + \omega_n^2\xi 
\\
\:=\beta_0\left[\dfrac{c_{2n}}{2}\dot{\xi}^* - iM\omega_n \left( \xi + \dfrac{c_{2n}}{2}\xi^* \right) \right]  + f_{\text{nl}}\left(A,\chi,\theta,\varphi,\dot{\xi}\right)
\end{multline}
We remind that the system is governed by one fast time scale, of characteristic time $1/\omega_n$, and two slow time scales, one associated to the growth rate of thermoacoustic instabilities, with a characteristic time $1/(\beta_0-\alpha)$ and one associated to the convective phenomena due to the azimuthal mean flow, whose characteristic time is $1/(M\omega_n)$. In Eq. \eqref{eq:waveComplex}, the terms $\ddot{\xi}$ and $\omega_n^2\xi$ are both of the same order of magnitude $A\omega_n^2$ and are the dominant terms. The order of magnitude of the terms $(\alpha-\beta_0)\dot{\xi}$ and $2i M\omega_n\dot{\xi}$  are respectively   $A\omega_n(\beta_0-\alpha)$ and $AM\omega_n^2$. In the right-hand side, the term $i\beta_0M\omega_n\xi$ involves a product of $\beta_0$ and $M\omega_n$ and is of the order of $\beta_0M\omega_n A$. In the next steps of the derivation we will remove the second order terms $\mathcal{O}\left[M^2\omega_n^2 A;\, (\beta_0-\alpha)^2A;\,(\beta_0-\alpha)M\omega_n A\right]$. Under these conditions, keeping the term $\mathcal{O}\left(\beta_0 M\omega_n A\right)$ can be justified only in the case where $\beta_0\gg\beta_0-\alpha$, which means that the damping $\alpha$ of the chamber is large compared to the global growth rate $\beta_0-\alpha$  of the instability.

\subsection{Slow-flow averaging}\label{sec43}

The time derivative of $\xi$ writes:
\begin{align*}
\dot{\xi}=\mathcal{C}(\dot{\eta}) \:=\: \mathcal{C}\Big(\dot{A}\, e^{in\theta}e^{-k\chi}e^{j(\omega_n t + \varphi)}
\\
+\: A\, in\dot{\theta} e^{in\theta}e^{-k\chi}e^{j(\omega_n t + \varphi)}
\\
-\: A\,\dot{\chi}\, e^{in\theta} k \,e^{-k\chi}e^{j(\omega_n t + \varphi)}
\\
+\: A\,(\omega_n+\dot{\varphi})\, e^{in\theta} e^{-k\chi}e^{j(\omega_n t + \varphi)}j \Big). \numberthis \label{eq:D1}
\end{align*}
In this expression, the factor of $A\omega_n$ is large compared to the terms in $\dot{A}$, $A\dot{\theta}$, $A\dot{\chi}$, $A\dot{\varphi}$ since the acoustic oscillations are fast compared to the slow variables evolution. In a way similar to the Krylov-Bogoliubov method \cite{sanders85}, $\dot{\xi}$ will be sought under the form:
\begin{align}\label{eq:D1sf}
\dot{\xi} \:=\: \mathcal{C}\Big( A\,\omega_n\, e^{in\theta} e^{-k\chi}e^{j(\omega_n t + \varphi)}j \Big).
\end{align}

This imposes the following condition:
\begin{align*}\label{eq:slowFlowCond}
0\:=\: \mathcal{C}\Big(\dot{A}\, e^{in\theta}e^{-k\chi}e^{j(\omega_n t + \varphi)}
\\
+\: A\, in\dot{\theta} e^{in\theta}e^{-k\chi}e^{j(\omega_n t + \varphi)}
\\
-\: A\,\dot{\chi}\, e^{in\theta} k \,e^{-k\chi}e^{j(\omega_n t + \varphi)}
\\
+\: A\,\dot{\varphi}\, e^{in\theta} e^{-k\chi}e^{j(\omega_n t + \varphi)}j \Big).\numberthis
\end{align*}
The second derivative is then obtained by differentiating (\ref{eq:D1sf}):
\begin{align*}
\ddot{\xi} \:=\: \mathcal{C}\Big( \dot{A}\,\omega_n\, e^{in\theta}e^{-k\chi}e^{j(\omega_n t + \varphi)}j
\\
+\: A\, \omega_n\, in\dot{\theta} e^{in\theta}e^{-k\chi}e^{j(\omega_n t + \varphi)}j
\\
-\: A\, \omega_n\dot{\chi}\, e^{in\theta} k \,e^{-k\chi}e^{j(\omega_n t + \varphi)}j
\\
-\: A\, \omega_n(\omega_n+\dot{\varphi})\, e^{in\theta} e^{-k\chi}e^{j(\omega_n t + \varphi)}\Big). \numberthis \label{eq:D2sf}
\end{align*}
Using Eq. \eqref{eq:D1sf} implies that the expression  $\ddot{\xi}$ does not involve the second derivatives of the slow variables ($\ddot{A},\ddot{\chi},\ddot{\theta},\ddot{\varphi}$) and the products of their first derivatives (e.g $\dot{A}\dot{\varphi}$). Consequently, the terms of order 2 in the slow time scales $1/(\beta_0-\alpha)$ and $1/(M\omega_n)$ will not appear in the final system of equations.
Equations (\ref{eq:D1sf}) and (\ref{eq:D2sf}) are injected in the spatially averaged equation (\ref{eq:waveComplex}) and the identification of the real and the $i$-imaginary parts gives two equations in $A$,$\chi$,$\theta$,$\varphi$ and their first time derivatives.
The real and $i$-imaginary parts of the condition (\ref{eq:slowFlowCond}) provides two additional equations in $A$,$\chi$,$\theta$,$\varphi$ and their first derivatives. The four equations are linear equations in $\dot{A}$, $\dot{\chi}$, $\dot{\theta}$, $\dot{\varphi}$ with non constant coefficients. The system can be inverted and written under the form:
\begin{align}\label{eq:sysGen}
\dot{Y}=F_\text{osc}(Y,t)
\end{align}
where  $Y=\left(A, \chi, \theta, \varphi\right)^T$
and $F_\text{osc}(Y,t)$ a nonlinear function $\mathbb{R}^5\mapsto\mathbb{R}^4$. The subscript 'osc' means that the function contains fast oscillatory terms. Indeed, for a given $Y$, $F_\text{osc}(Y,\cdot)$ is a periodic function of pulsation $\omega_n$. The system's expression is relatively complex and we show only here the equations for the simplest configuration where the asymmetries and the swirl are set to 0 ($c_{2n}=0$ and $M=0$). 
\begin{align}
\begin{cases}
\;\dot{A} \:=\: \dfrac{1}{2}(\beta_0-\alpha)A \,-\, \dfrac{3\kappa}{64}\left[5+\cos(4\chi)\right]A^3 
\\[10pt]
\quad\quad+ \:A\cos(2\chi)\left[\dfrac{\alpha-\beta_0}{2}\cos(2(\varphi+\omega_nt)) + \dfrac{9A^2\kappa}{32}\cos(4(\varphi+\omega_nt))\right]
\\[15pt]
\; \dot{\chi} \:=\: \dfrac{3\kappa}{64}A^2\sin{4\chi}
\\[10pt]
\quad-\:\sin(2\chi)\left[\dfrac{\alpha-\beta_0}{2}\cos(2(\varphi+\omega_nt)) + \dfrac{3A^2\kappa}{32}\cos(2(\varphi+\omega_nt)) + \dfrac{9A^2\kappa}{32}\cos(4(\varphi+\omega_nt))\right]
\\[15pt]
\; n\dot{\theta} \:=\: \tan(2\chi)\left[\dfrac{\alpha-\beta_0}{2}\sin(2(\varphi+\omega_nt)) +\dfrac{3A^2\kappa}{16}\sin(2(\varphi+\omega_nt
))\right.
\\[10pt]
\quad\quad\left.+\; \dfrac{9A^2\kappa\cos(2\chi)}{32}\sin(4(\varphi+\omega_nt))\right]
\\[15pt]
\dot{\varphi}  \:=\: -\dfrac{1}{\cos(2\chi)}\left[\dfrac{\alpha-\beta_0}{2}\sin(2(\varphi+\omega_nt)) +\dfrac{3A^2\kappa}{8}\sin(2(\varphi+\omega_nt
))\right.
\\[10pt]
\quad\quad\left.+\; \dfrac{9A^2\kappa\cos(2\chi)}{32}\sin(4(\varphi+\omega_nt))\right]
\end{cases}
\end{align}
In this system, the right hand side of each equation is the sum of slow variables and fast oscillating terms which are weakly perturbed harmonic oscillations.

The time averaging operator $\langle\cdot\rangle_{\mathcal{T}}$ over the one acoustic period $\mathcal{T}=2\pi/\omega_n$ is applied on the system, 
giving a time-averaged system of form:
\begin{align}\label{eq:sysSF}
\dot{Y}=F_\text{slow}(Y)
\end{align} 
where $F_\text{slow}$ is the average of $F_\text{osc}$ over one oscillation. This new system is called the slow-flow system. The classical slow-flow averaging theorems of Krylov-Bogoliubov presented in \cite{sanders85,stratonovich67} show that for a weakly perturbed harmonic motion, if the perturbations have a characteristic time $\tau_\text{slow}$ large compared to the fast oscillations, then the solutions of eqs. (\ref{eq:sysSF}) and (\ref{eq:sysGen}) will differ of an order $1/\tau_\text{slow}$ on a time range $\tau_\text{slow}$. 
The system (\ref{eq:sysSF}) is a nonlinear dynamic system in $A$, $\chi$, $\theta$ and $\varphi$. The whole system with $M\neq0$ and $c_{2n}\neq0$ has the same properties as the one presented above and the time averaging can also be applied in this case.
\section{Uniform thermoacoustic distribution without mean swirl}\label{sec5}
\begin{figure}
	\centering
	\begin{subfigure}{.5\textwidth}
		\centering
		\includegraphics[width=\textwidth]{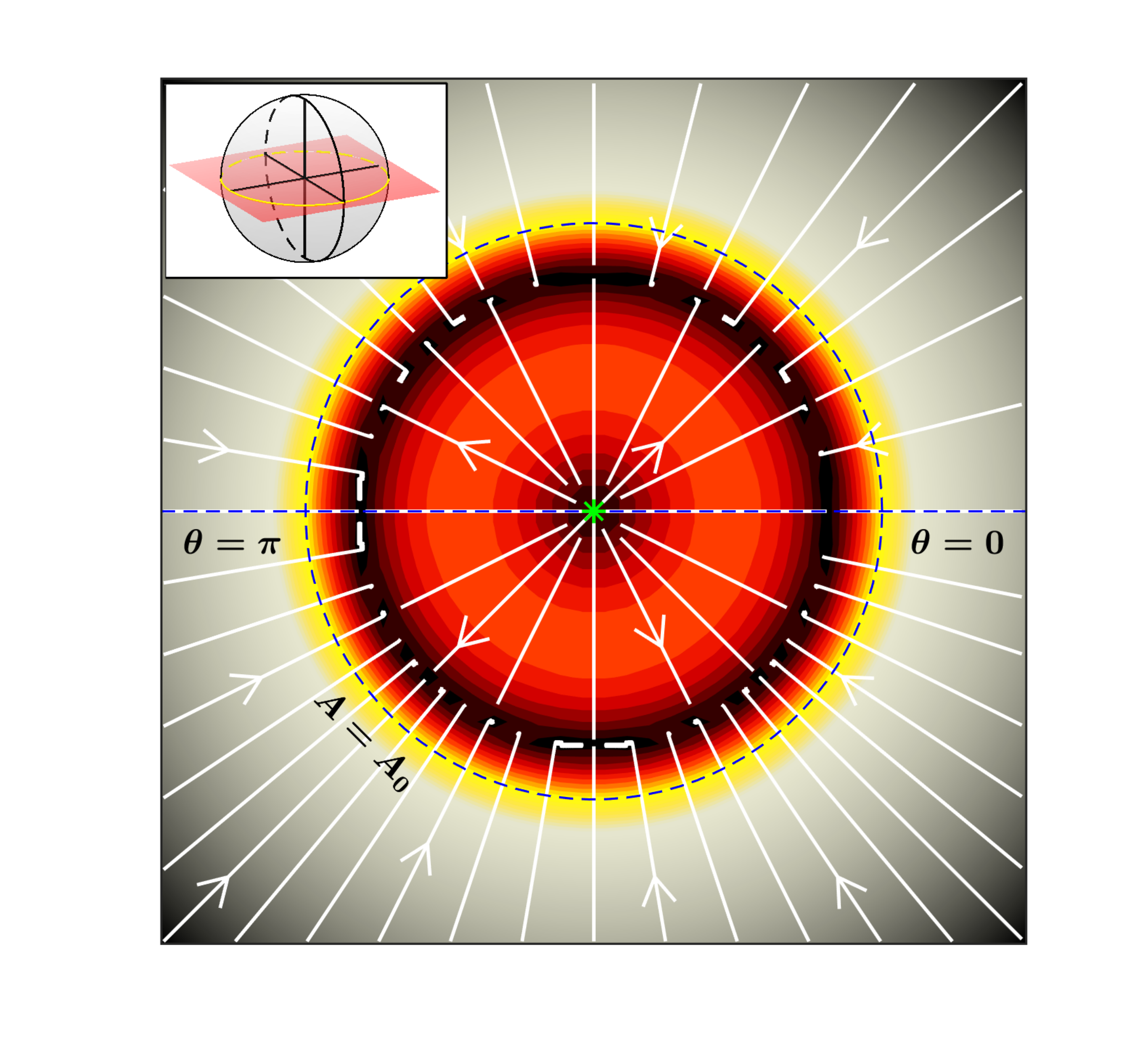}
		\caption{plane $\chi = 0$}
		\label{fig:basica}
	\end{subfigure}%
	\begin{subfigure}{.5\textwidth}
		\centering
		\includegraphics[width=\textwidth]{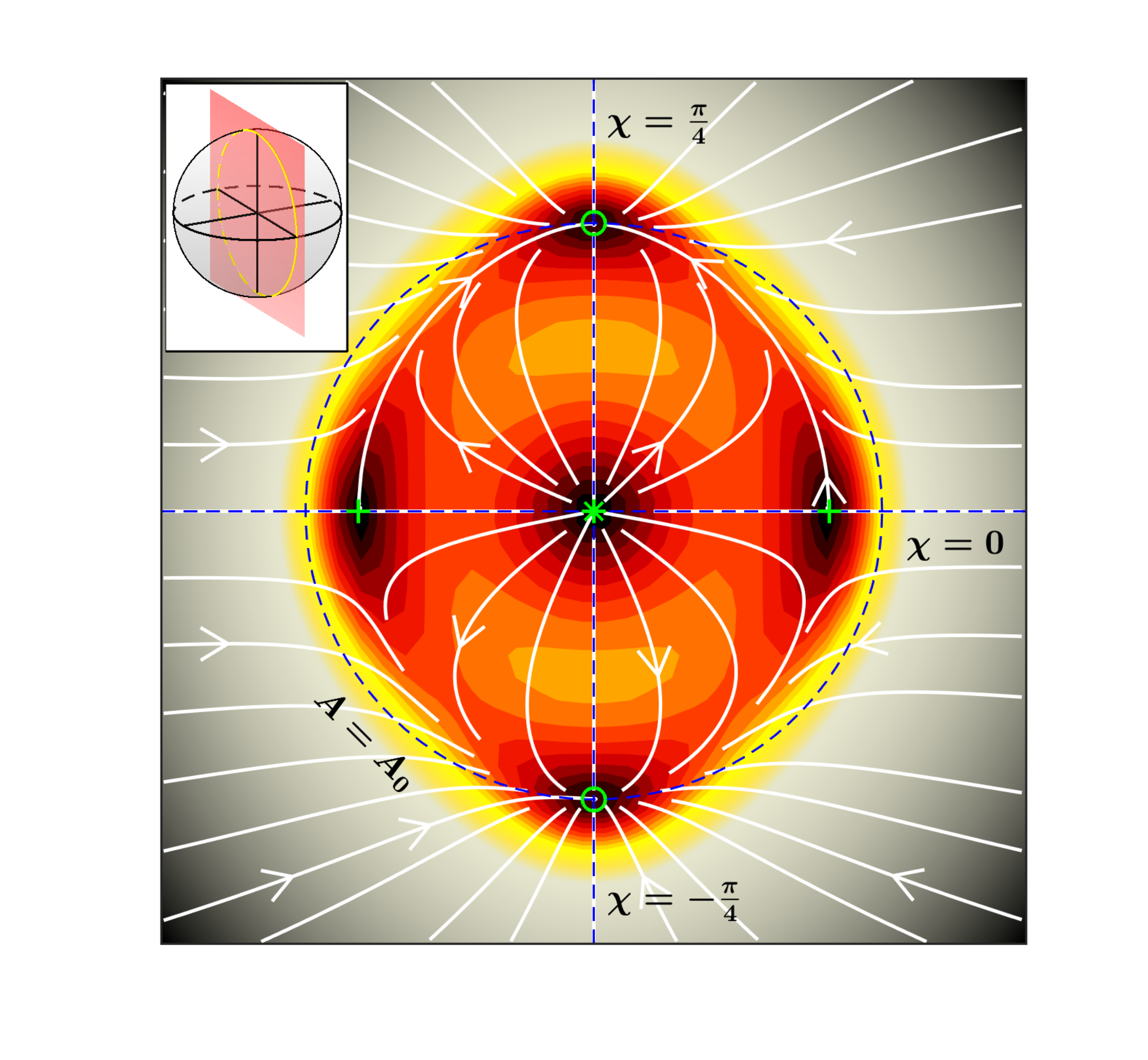}
		\caption{plane $\theta=0\mod\pi$}
		\label{fig:basicb}
	\end{subfigure}%
	\caption{Stream lines describing the evolution of the state of the linearly-unstable azimuthal-thermoacoustic-mode on the Bloch Sphere, for a uniform distribution of the thermoacoustic feedback, with no mean swirl. Parameters: $\beta_0$=160 rad.s$^{-1}$, $\alpha$=100 rad.s$^{-1}$, $\kappa$=0.5 rad.s$^{-1}$Pa$^{-2}$, $\omega_n$=1167 rad.s$^{-1}$. Symbols: $\circ$: attractors $*$: repellers, +: saddles. The color scale is given in Fig. \ref{fig:color}.}
	\label{fig:basic}
\end{figure}

\begin{figure}
	\centering
	\includegraphics[width=\textwidth]{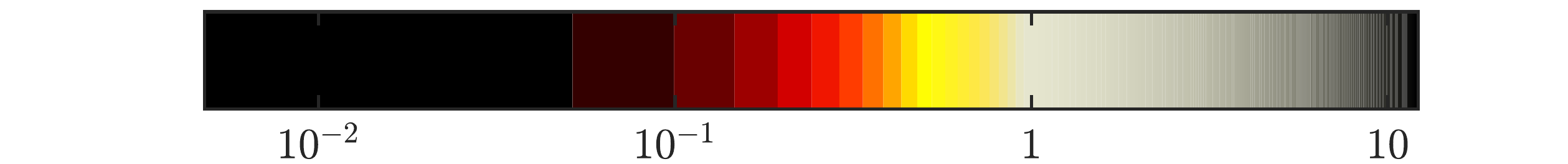}
	\caption{Color bar used for the contours of the streamline plots throughout the paper. The values correspond to the norm of the stream vectors normalized by $A_0\,(\beta_0-\alpha)$ with $A_0$ the amplitude of the stable spinning modes in the symmetric configuration.This bar is represented in logarithmic scale in order to  better visualise the state change speed near equilibrium points. }
	\label{fig:color}
\end{figure}

This section deals with the analysis of the coupled equations describing the dynamics of the slow-flow variables, when the thermoacoustic feedback is uniformly distributed along the annulus circumference, i.e. $\beta(\Theta)=\beta_0$, when the system is linearly unstable, i.e. $\beta_0>\alpha$, and when there is no mean swirl, i.e $M=0$. The effect of time-delayed thermoacoustic feedback and stochastic forcing will be investigated in  sections \ref{sec6} and \ref{sec7} respectively. The symmetry breaking induced by non-uniform distribution of the thermoacoustic coupling and by the presence of a mean swirl  will be scrutinized  in sections \ref{sec8} and \ref{sec9}.\\
The wave equation is invariant by azimuthal rotations around the chamber axis and reflections with respect to any axis of equation $\Theta=\text{constant}$.
The system of equations obtained with the method detailed in the section \ref{sec4} is:
\begin{align}\label{eq:sysSym}
\begin{cases}
\;\dot{A} \:=\: \nu A \,-\, \dfrac{3\kappa}{64}\left(5+\cos(4\chi)\right)A^3,
\\[10pt]
\; \dot{\chi} \:=\: \dfrac{3\kappa}{64}A^2\sin{4\chi},
\\[10pt]
\; n\dot{\theta} \:=\: 0,
\\[10pt]
\; \dot{\varphi}  \:=\: 0.
\end{cases}
\end{align}
In the amplitude equation, the linear growth rate $\nu=(\beta_0-\alpha)/2$, which is positive in the case of a linearly unstable situation, has been introduced.
The first two equations in $A$ and $\chi$ are independent of the two other variables $\theta$ and $\varphi$, which will be fixed by the initial condition and remain constant, and this system of two equations can be analysed separately. It
has four equilibrium points:
\begin{align}\label{eq:solSym}
\begin{array}{llll}
& A=0 &   & \\[10pt]
A=\dfrac{4}{3}\sqrt{\dfrac{2\nu}{\kappa}}& \quad\mbox{and} &\chi=0 &\mbox{(standing)}\\[10pt]
A=4\sqrt{\dfrac{\nu}{3\kappa}}\equiv A_0 & \quad\mbox{and} & \chi=\dfrac{\pi}{4}&\mbox{(spinning CCW)}\\[10pt]
A=4\sqrt{\dfrac{\nu}{3\kappa}} & \quad\mbox{and} & \chi=-\dfrac{\pi}{4} &\mbox{(spinning CW)}
\end{array}
\end{align}
A stability analysis is performed by deriving the eigenvalues of the Jacobian matrix associated with Eq. \eqref{eq:sysSym} at these fixed points. It shows that the solution $A=0$ is a repeller. The second equilibrium point, which corresponds to a standing mode, is a saddle point and is therefore unstable and repelling in the $\chi$ direction. The pure spinning modes in both directions are stable solutions. This result and the amplitudes predicted for the spinning and the standing modes reconstructed with the real expression of the pressure given in (\ref{eq:pRe}) are in agreement with the results of \cite{noiray11}.
Using the Bloch-Poincar\'e spherical representation introduced in the section \ref{sec34}, Fig. \ref{fig:basic} shows the 3D streamlines of the system for the coordinates $A$, $\chi$ and $\theta$. For more clarity, the figure shows only the dynamics in two slices of the Bloch sphere instead of showing the full 3D trajectories. The streamlines are obtained by computing the stream vector field in the slice and by using only the projected component of the vector on the slice. For this illustration, the numerical values of the parameters were arbitrarily set to  $\beta_0=160$ rad\,s$^{-1}$, $\alpha=100$ rad\,s$^{-1}$ (and therefore $\nu=(\beta_0-\alpha)/2=30$ rad\,s$^{-1}$), $\kappa=0.5$ rad\,s$^{-1}$\,Pa$^{-2}$ and $f_n=\omega_n/2\pi=1167/2\pi\simeq185$ Hz. These values satisfy the condition $\nu\ll \omega_n$, which is verified for thermoacoustic instabilities in practical gas turbine combustors, e.g. \cite{bothien15}. In the present case, $\nu/\omega_n\approx2.5\%$ and $\alpha/\omega_n\approx8.5\%$, which can be compared for instance to the experimental work  \cite{boujo16} where $\nu/\omega_n\approx 1\%$ and $\alpha/\omega_n\approx 5\%$. The color defines the norm of the stream vector and the colormap is shown in Fig. \ref{fig:color}. In particular, the dark red zones hues correspond to regions where the dynamics is slow and corresponds to the equilibrium points. Fig. \ref{fig:basica} shows that  for initial conditions corresponding to low-amplitude standing modes (in the plane $\chi=0$), the transient is characterised by a growth of the  amplitude $A$, while the direction of the anti-nodal line $n\theta$ does not change. A slowdown of the amplitude growth occurs when the mode approaches the saddle circle $A=(4/3)\times(2\nu/\kappa)^{1/2}$ of the equatorial plane $\chi=0$, from which the state of the azimuthal mode is repelled towards one of  the poles of the Bloch sphere, which correspond to pure spinning modes. This transient was discovered by Schuermans, Paschereit and Monkewitz, who performed numerical simulations of thermoacoustic instabilities in annular combustors by means of a reduced-order  network model \cite{schuermans06}. Afterwards, it was  further investigated analytically and numerically  by Noiray, Bothien and Schuermans in \cite{noiray11} and by Ghirardo and Juniper in \cite{ghirardo13}.\\
More generally, the dynamics of thermoacoustic azimuthal modes in such perfectly symmetrical configuration  corresponds to the general case of $O(2)$-symmetric supercritical Hopf bifurcations investigated by Crawford and Knobloch in their famous paper \cite{crawford91}. They show that the only equilibrium points are: the origin, the clockwise spinning wave, the counter-clocklockwise spinning wave, and the standing waves in every direction $\theta$. Depending on the system nonlinearities, either the spinning waves are stable and the standing waves are saddles, or the standing waves are stable and the spinning waves are saddles. They also show that the amplitudes of the two spinning waves are equal, and that when the spinning modes are stable, their amplitude is larger than the amplitude of the unstable standing modes.
\section{Time-delayed thermoacoustic feedback}\label{sec6}
Rayleigh \cite{rayleigh78} was the first to show the importance of the phase difference between the pressure oscillations and the unsteady heat release in the phenomenon of thermoacoustic instability. This phase difference defines whether the thermoacoustic feedback is constructive or destructive. In practical gas turbine and aeroengine combustors, the delay corresponding to the phase lag between acoustic field and flame heat release rate results from the finite-time convective propagation of acoustically-induced flame perturbations, such as equivalence ratio  or swirl number fluctuations. This time-delay is associated to the phase of the Flame Transfer Function (FTF).\\
The introduction of a time delay between the heat release and the pressure in thermoacoustic instabilities models has been applied in various studies, including annular configurations \cite{schuermans06}. In previous studies dealing with thermoacoustic instabilities in annular configurations, the time delay between the acoustic field and the heat release rate oscillations has been accounted for in different ways: i) by using the simple $n$-$\tau$ formulation (e.g. \cite{morgans_2007,bauerheim15}) and 
its state-space equivalent \cite{Laurent_2019}  for modelling the \emph{linear thermoacoustic problem in time and frequency domain}, ii) by using measured flame describing functions (e.g.  \cite{bourgouin14,ghirardo16}) for modelling the \emph{nonlinear thermoacoustic problem in frequency domain}, and iii) by modelling the flame response to acoustic perturbations with a distribution of time delays \cite{schuermans06}, or by using state-space representations of experimentally measured FTFs \cite{noiray11,bothien15} for simulating the \emph{nonlinear thermoacoustic problem in time domain}. \\ 
Recently, Ghirardo, Juniper and Bothien published a very interesting detailed study \cite{ghirardo17} on the effect of flame phase on thermoacoustic limit cycles in symmetric annular combustors without mean swirl and  without stochastic forcing. Notice that in this paper, time delay spread \cite{schuermans06}, which is always more or less present in practical systems, was not accounted for. In \cite{ghirardo17}, a Galerkin expansion based on the natural acoustic modes of the annular combustor  was performed, which differs from \cite{ghirardo18} and the present work where the quaternion-based acoustic field projection is adopted. The  convective phase lag is implemented in the present model by replacing the ``non-delayed'' thermoacoustic source term (\ref{eq:heatRelease}) by the following delayed model
\begin{align}\label{eq:Qtau}
Q(\Theta,t)=\beta p(\Theta,t-\tau) - \kappa p(\Theta,t-\tau)^3.
\end{align}
The thermoacoustic stability is very sensitive to small changes of the time delay  \cite{juniper18}. This delay $\tau$  typically does not exceed a few fast oscillations and it is small compared to the characteristic times of the slow-flow dynamics.
In this section, we consider the configuration with a delayed flame response ($\tau\neq0$) without swirling flow ($M=0$) and with a distribution of the heat release that does not exhibit 2$n$ component in its Fourier decomposition ($c_{2n}=0$).
The operations of averaging described in the section \ref{sec4} can be applied without modification. In addition, $A$, $\chi$, $\theta$ and $\phi$ are assumed constant over $\tau$ in order to perform the time averaging. In contrast with section \ref{sec8} and \ref{sec9}, neither the rotational nor the reflectional symmetry of the wave equation is broken and the results of \cite{crawford91} are applicable in this case. The stable solutions will therefore be either two counter-spinning waves, or standing waves.
In the present case, the system of slow-flow equations becomes:
\begin{align}\label{eq:sysSymTd}
\begin{cases}
\;\dot{A} \:=\: \dfrac{\beta_0 \cos(\omega_n\tau) - \alpha}{2} A \:-\: \dfrac{3\kappa}{64}\left(5+\cos(4\chi)\right)\cos(\omega_n\tau)A^3
\\[10pt]
\; \dot{\chi} \:=\: \dfrac{3\kappa}{64}A^3\sin(4\chi)\cos(\omega_n\tau)
\\[10pt]
\; n\dot{\theta}  \:=\: -\dfrac{3\kappa}{32}A^2\sin(2\chi)\sin(\omega_n\tau)
\\[10pt]
\; \dot{\varphi}  \:=\: \left(-\dfrac{\beta_0}{2} +  \dfrac{9 A^2 \kappa}{32}\right)\sin(\omega_n\tau).
\end{cases}
\end{align}
From the amplitude equation, the following condition leads to a linearly unstable thermoacoustic system:
\begin{align}\label{eq:rayleigh}
\beta_0 \cos(\omega_n\tau) - \alpha > 0.
\end{align}
It corresponds to the Rayleigh's criterion. In particular, when $\cos(\omega_n\tau)<0$, the system is linearly stable. The first two equations are similar to the system without time delay (\ref{eq:sysSym}), with $\beta_0$ replaced by  $\beta_0 \cos(\omega_n\tau) $ and $\kappa$ replaced by $\kappa\cos(\omega_n\tau)$. When the inequality (\ref{eq:rayleigh}) is fulfilled, the dynamics of the system is therefore the same, but the amplitude of the stable spinning modes is now:
\begin{align}
A_\text{sp} = \sqrt{\dfrac{8(\beta_0-\alpha/\cos(\omega_n\tau))}{3\kappa}}
\end{align}
At the limit cycle, the time derivative of the temporal phase
$\dot{\varphi}$ is constant $\left(\beta_0 - 3\alpha/\cos(\omega_n\tau)\right)\sin(\omega_n\tau)/4$, while the time derivative of the preferential direction $n\dot{\theta}$ converges exponentially fast to the constant value $-\left(\beta_0 - \alpha/\cos(\omega_n\tau)\right)\sin(\omega_n\tau)/4$ for the counterclockwise spinning solution and the opposite for the clockwise spinning solution. Consequently, one can express the evolution of these slow variables at the limit cycle as
\begin{align}\label{eq:affine}
\begin{cases}
n\theta \:\simeq\: \pm\dfrac{1}{4}\left(\beta_0 - \alpha/\cos(\omega_n\tau)\right)\sin(\omega_n\tau)\,t + n\theta_i,
\\
\varphi \:=\: \dfrac{1}{4}\left(\beta_0 - 3\alpha/\cos(\omega_n\tau)\right)\sin(\omega_n\tau)\,t + \varphi_i,
\end{cases}
\end{align}
with $\theta_i$ and $\varphi_i$ constants depending on the initial conditions.
For a pure CCW spinning mode, the amplitude $A^{-}$ of the CW component is zero. for a pure CW mode, the amplitude $A^{+}$ of the CCW component is zero. In the case of a pure CCW spinning mode, injecting (\ref{eq:affine}) in (\ref{eq:pReSpin}) gives:
\begin{align}\label{eq:pReCCW}
p \;=\; A^{+} \cos(-n(\Theta-\theta_i) + \omega_\text{th}t + \varphi_{i})
\end{align}
where the drift of frequency from the pure acoustic frequency $\omega_n$ to the thermoacoustic frequency $\omega_\text{th}$ comes from the linear drift of $\theta$ and $\varphi$. The expression of $\omega_\text{th}$ is:
\begin{align}\label{eq:freqDrift}
\omega_\text{th} = \omega_n - \alpha \tan(\omega_n\tau)/2
\end{align}
Notice that for pure spinning waves, there is no preferential direction and $\theta$ becomes a purely temporal phase angle.
The counterclockwise solution is:
\begin{align}\label{eq:pReCW}
p \;=\; A^{-} \cos(n(\Theta-\theta_{i}) + \omega_\text{th}t + \varphi_{i})
\end{align}
with the same definition of $\omega_\text{th}$. In the limit case where the system is neutrally stable ($\alpha=\beta_0\cos(\omega_n\tau)$), eq. \eqref{eq:freqDrift} becomes $$\omega_\text{th} = \omega_n - \beta_0 \sin(\omega_n\tau)/2$$. This expression is consistent with eq. (26b) of \cite{ghirardo17} under the condition that $\beta_0\ll\omega_n$, which is the case in the present study.
Various examples provided in \cite{ghirardo17} show that $\beta_0$ is usually  between one and a few times greater than $\alpha$. With Eq. \eqref{eq:rayleigh}, $\beta_0\cos(\omega_n\tau)$ and $\alpha$ are therefore of the same order of magnitude, then $\alpha \tan(\omega_n\tau)$ is of the order of magnitude of $\beta_0\sin(\omega_n\tau)$ which is small compared to $\omega_n$ since the pressure oscillations are short compared to the characteristic times of slow-flow dynamics.\\
\section{Stochastic forcing}\label{sec7}
Gas turbine combustors are subject to high level of turbulence. The flame's heat release rate is also affected by the turbulence and its fluctuations can therefore be decomposed as the sum of an acoustically coherent and a stochastic components. The latter acts as a random source in the wave equation. As done in previous studies \cite{lieuwen03,noiray13}, the contribution of stochastic noise is modelled by adding a random process in the equations. The stochastic forcing is represented by a Gaussian white noise process because it is significantly more convenient for not increasing the dimension of our system. For interested readers, the effect of noise colour on the thermoacoustic feedback has been considered in \cite{bonciolini18}  and it was shown that when one dominant mode governs the thermoacoustic dynamics, in general, white noise is a proxy which does not significantly affects the system dynamics and the shape of the basin of attraction.
\subsection{Stochastic averaging}
To account for the presence of noise in the system, a random function $\Xi(t,\Theta)$ is added in the wave equation:
\begin{multline}\label{eq:thEqNoise}
\dfrac{\partial^2 p}{\partial t^2} + \dfrac{2U}{\mathcal{R}}\dfrac{\partial^2 p }{\partial \Theta \partial t} +  \alpha \dfrac{\partial p}{\partial t} - \dfrac{c^2}{\mathcal{R}^2} \dfrac{\partial^2 p}{\partial \Theta^2}
\\[10pt]
\quad\:=\: \beta(\Theta)\dfrac{\partial p}{\partial t}+\dfrac{U}{\mathcal{R}}\dfrac{\partial [\beta(\Theta)p]}{\partial \Theta} -3\kappa p^2 \left(\dfrac{\partial p}{\partial t} +\dfrac{U}{\mathcal{R}}\dfrac{\partial p}{\partial \Theta} \right) + \Xi(t,\Theta)\,.
\end{multline}
To apply the spatial averaging on this new term, we write it as a Fourier series in $\Theta$:
\begin{equation}
\Xi(t,\Theta) = \Xi_0(t) + \sum_{m>0} \Xi_m(t)^*e^{im\Theta} + \Xi_m(t)e^{-im\Theta}\,.
\end{equation}
Therefore:
\begin{equation}
\left<\Xi(t,\Theta)\right>_\Theta \, = \,\Xi_n(t)
\end{equation}
The azimuthally averaged equation Eq. \eqref{eq:waveComplex} becomes:
\begin{multline}\label{eq:waveStoch}
\ddot{\xi} + (\alpha- 2i M\omega_n)\dot{\xi} + \omega_n^2\xi 
\\
\:=\beta_0\left[\dot{\xi}+\dfrac{c_{2n}}{2}\dot{\xi}^* - iM\omega_n \left( \xi + \dfrac{c_{2n}}{2}\xi^* \right) \right]  + f_{\text{nl}}\left(A,\chi,\theta,\varphi,\dot{\xi}\right)
+ 2\,\Xi_n(t).
\end{multline}
$\Xi_n(t)$ is complex and is written as $2\,\Xi_n=\zeta_1+i \zeta_2$ with $\zeta_1$ and $\zeta_2$ two real zero-mean gaussian random processes ($\langle \zeta_1 \rangle = 0 = \langle \zeta_2 \rangle$, where $\langle \cdot \rangle$ denotes the averaging on a time duration which is long compared to the characteristic fluctuation time of the noises $\zeta_1$ and $\zeta_2$, but small compared to an acoustic oscillation) , with infinitely small correlation time and identical intensity $\Gamma$: $ \langle \zeta_1(t_1)\zeta_1(t_2) \rangle = \Gamma\delta(t_1-t_2) = \langle \zeta_2(t_1)\zeta_2(t_2) \rangle$. Moreover, $\zeta_1$ and $\zeta_2$ are uncorrelated: $\forall\: t_1, t_2\quad \langle \zeta_1(t_1)\zeta_1(t_2) \rangle = 0$.
\\
Like in section \ref{sec43}, a 1$^{\text{st}}$ order dynamic system in the slow variables can be obtained from eq. \eqref{eq:waveStoch}:
\begin{align}\label{eq:sysGenSt}
\dot{Y}=F_{\text{tot}}(Y,t)
\end{align}
where $Y=(A,\chi,\theta,\varphi)^T=(Y_1, ... ,Y_4)^T$ and $F_\text{tot}(Y,t)=(F_{\text{tot}1}, ... , F_{\text{tot}4})^T(Y,t)$
The functions $F_{\text{tot}1...4}$ are real-valued stochastic processes and they include fast oscillating terms at even multiples of the natural eigenfrequency $\omega_n$ . The deterministic and stochastic parts in eq. \eqref{eq:sysGenSt} can be separated:
\begin{align}\label{eq:stoch}
\dot{Y} = F_{\text{osc}}(Y,t) + S_{\text{osc}}^{(1)}(Y,t)\zeta_1(t) + S_{\text{osc}}^{(2)}(Y,t)\zeta_2(t)
\end{align}
where $F_\text{osc}$, $S_{\text{osc}}^{(1)}$ and $S_{\text{osc}}^{(2)}$ are deterministic functions $\mathbb{R}^{5}\mapsto\mathbb{R}^4$. In particular, $F_\text{osc}$ has exactly the same expression as in the deterministic system \eqref{eq:sysGen}. The expressions for the components of $S_{\text{osc}}^{(1)}$ and $S_{\text{osc}}^{(2)}$ are given in appendix \ref{apB}.  In the book \cite{stratonovich67}, Stratonovich  
describes a methodology to get the slow-flow equations for such a system. The first step consists in applying Krylov-Bogoliubov averaging on the deterministic part in order to get rid of the fast oscillations in $F_\text{osc}$, giving the system

\begin{align}\label{eq:stochAv}
\dot{Y} = F_\text{slow}(Y) + S_{\text{osc}}^{(1)}(Y,t)\zeta_1(t) + S_{\text{osc}}^{(2)}(Y,t)\zeta_2(t)
\end{align}
which can be rewritten as: 
\begin{align}
\dot{Y} = G(Y,t)
\end{align}
The deterministic part $F_\text{slow}$ has the same expression as in the deterministic slow-flow system \eqref{eq:sysSF}.

The subscripts ``osc''  will be dropped from now on for the functions  $S^{(m)}_{\text{osc}}$, but it has to be kept in mind that they include fast oscillating terms. The subscript ``slow'' is also dropped for sake of readability. The subscripts 1,2,3,4 will be used to refer to the four components of $F_\text{slow}$, $G$, $S^{(1)}$ and $S^{(2)}$. To get the contributions of the stochastic part of Eq. \eqref{eq:stochAv} on the slow-flow dynamics, Stratonovitch \cite{stratonovich63} proposes to use the corresponding generalized Fokker-Planck equation to describe the temporal evolution of the joint probability density function $P(Y,t)$ of the system \eqref{eq:stochAv}:
\begin{align}
\frac{\partial P(Y,t)}{\partial t} = -  \sum_{i=1}^{4} \dfrac{\partial}{\partial Y_i}\left\{ \left(\langle G_i \rangle + \sum_{j=1}^{4}\left(\int_{-\infty}^{0} K\left[\dfrac{\partial G_i}{\partial Y_j},G_{j\tau} \right]\,d\tau\right) \right)P(Y,t) \right\}   
\\
+  \sum_{i,j} \dfrac{\partial^2}{\partial Y_i \partial Y_j}\left(P(Y,t) \int_{-\infty}^{0}K\left[G_i,G_{j\tau}\right]\,d\tau \right) \,.
\end{align}
This expression is valid under the assumption that the correlation time of the noise is small compared to the acoustic period, allowing to set the lower limit of the integrals to $-\infty$. The correlation function $K$ is defined as $K[f,g]=\langle f g\rangle - \langle f \rangle\langle g \rangle $. For a function $f(t)$, the notation $f_{\tau}$ designs the time-shifted function $t\mapsto f(t+\tau)$.
\\
Using the difference of time scales between the noise and the acoustic oscillations, the terms in the Fokker-Planck equation can be computed: 
\begin{flalign*}
\langle G_i \rangle &= F_i&
\\[10pt]
K\left[G_i,G_{j\tau}\right] &= S_i^{(1)} S_{j\tau}^{(1)} \langle \zeta_1\zeta_{1\tau}\rangle + S_i^{(2)} S_{j\tau}^{(2)} \langle \zeta_2\zeta_{2\tau}\rangle&
\\[10pt]
K\left[\dfrac{\partial G_i}{\partial Y_j},G_{j\tau}\right] &= \dfrac{\partial S_i^{(1)}}{\partial Y_j} S_{j\tau}^{(1)} \langle \zeta_1 \zeta_{1\tau}\rangle + \dfrac{\partial S_i^{(2)}}{\partial Y_j} S_{j\tau}^{(2)} \langle \zeta_2 \zeta_{2\tau}\rangle &
\end{flalign*}
The delay $\tau$ makes appear products of fast oscillating terms of phase $\omega_n t$ and $\omega_n (t-\tau)$. Using trigonometry expansion formulae, we can write:
\begin{align}
S_i^{(k)} S_{j\tau}^{(k)} = D_{Cij}^{(k)}\cos(\omega_n\tau) + D_{Sij}^{(k)}\sin(\omega_n\tau)\,,\quad k=1,2
\end{align}
where $D_{Cij}^{(k)}$ and $D_{Sij}^{(k)}$ are functions of $Y$ and $t$ only. Then:
\begin{align}
\int_{-\infty}^{0}K\left[F_i,F_{j\tau}\right]\,d\tau 
= 
\sum_{k=1}^2 \left[
D_{Cij}^{(k)}
\int_{-\infty}^0 \cos(\omega_n\tau) \langle \zeta_k \zeta_{k\tau}\rangle \,d\tau
+
D_{Sij}^{(k)}
\int_{-\infty}^0 \sin(\omega_n\tau)\langle \zeta_k \zeta_{k\tau}\rangle \,d\tau
\right]
\end{align}
Considering the the random forcing $\zeta_m$ are white noises, one can deduce that the first integral is proportional to the noise intensity\\ $\int_{-\infty}^0 \cos(\omega_n\tau) \langle \zeta_k \zeta_{k\tau}\rangle \,d\tau = \Gamma/2$ and the second vanishes:\\
$\int_{-\infty}^0 \sin(\omega_n\tau) \langle \zeta_k \zeta_{k\tau}\rangle \,d\tau = 0$. It remains 
\begin{align}
\int_{-\infty}^{0}K\left[G_i,G_{j\tau}\right]\,d\tau = \dfrac{\Gamma}{2}\left(D_{Cij}^{(1)}+D_{Cij}^{(2)}\right)\,.
\end{align}
The right hand side of this expression still contains fast coherent oscillations. Using the previously introduced time averaging operator over one acoustic period $\langle \cdot\rangle_\mathcal{T}$, the oscillations are separated from the slow dynamics:
\begin{align}
\int_{-\infty}^{0}K\left[G_i,G_{j\tau}\right]\,d\tau  = \dfrac{\Gamma}{2} \left(  \left<D_{Cij}^{(1)}\right>_\mathcal{T} + \left<D_{Cij}^{(2)}\right>_\mathcal{T}  \right) +\:\text{oscillatory terms}
\\
\equiv \dfrac{\Gamma}{2}D_{ij} +\:\text{oscillatory terms}\,.
\end{align}
$D$ is a 4$\times$4 matrix of functions of the slow variables only. We can show that $D$ is always positive definite and real, it therefore admits a Cholesky decomposition: $D=(1/2\omega^2)B^TB$ with $B$ a real upper triangular matrix. The expression for B is:
\begin{align}
B=\begin{bmatrix} 1 & 0 & 0 & 0 \\
0 & \dfrac{1}{A} & 0 & 0 \\
0 & 0 & \dfrac{1}{nA\cos(2\chi)} & -\dfrac{\tan(2\chi)}{A} \\
0 & 0 & 0 & \dfrac{1}{A}
\end{bmatrix}\:.
\end{align}
\\
The same procedure can be applied to deal with the terms $\int_{-\infty}^{0} K\left[\partial G_i/\partial Y_j,G_{j\tau} \right]\,d\tau$:
\begin{align}
\sum_{j=1}^{4}\left(\int_{-\infty}^{0} K\left[\dfrac{\partial G_i}{\partial Y_j},G_{j\tau} \right]\,d\tau\right) = \dfrac{\Gamma}{2}H_i+\:\text{oscillatory terms}
\end{align}
Where $H$ is a four components vector of functions of the slow variables only.
As explained by Stratonovich in \cite{stratonovich67}, the effect of these oscillations on the slow flow dynamics is a high order of perturbation and can be neglected. The final Fokker-Planck equation for the slow-flow is:
\begin{align}
\frac{\partial P(Y,t)}{\partial t} = - \left\{   \sum_{i=1}^{4} \dfrac{\partial}{\partial Y_i}\left[ \left(F_i(Y) + \dfrac{\Gamma}{2}H_i(Y) \right)P(Y,t) \right]    \right\} 
\\
+  \dfrac{1}{2}\sum_{i,j} \dfrac{\partial^2}{\partial Y_i \partial Y_j}\left(P(Y,t) \dfrac{\Gamma}{2\omega_n^2} \left[B^T B\right](Y)_{ij} \right)
\end{align}
This multi-variable Fokker-Planck equation is equivalent to a system of stochastic differential equations \cite{gardiner04}:
\begin{align}
\dot{Y} = F_\text{slow}(Y) + \dfrac{\Gamma}{2}H(Y) + B(Y)N(t)
\end{align}
where $N(t)=[\zeta_A, \zeta_{\chi}, \zeta_{\theta}, \zeta_{\phi} ]$ is a four component vector of uncorrelated gaussian white noises of intensities $\Gamma/2\omega_n^2$.
The function $F_\text{slow}$ is equal to the function describing the slow-flow without stochastic forcing. The modification of the deterministic part of the system through the addition of noise is included in the function $H$. $B$ accounts for the random fluctuations.

\subsection{From standing modes to spinning modes}
\begin{figure}
	\centering
	\begin{subfigure}{.5\textwidth}
		\centering
		\includegraphics[width=\textwidth]{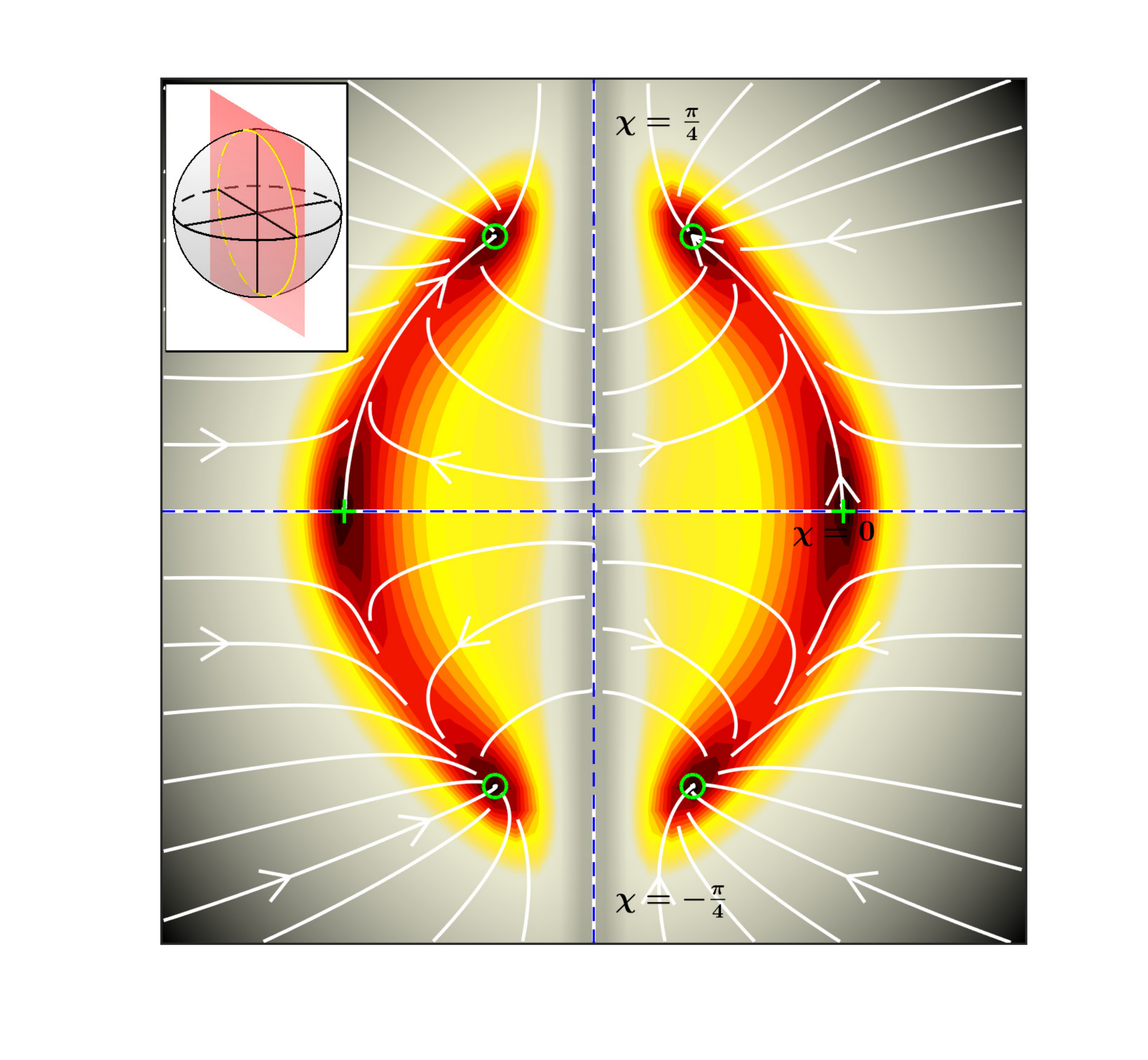}
		\caption{$\Gamma$ = 1.6$.10^9$ Pa$^2$.s$^{-3}$ $< \Gamma_\text{thr}$}
		\label{fig:stochMix}
	\end{subfigure}%
	\begin{subfigure}{.5\textwidth}
		\centering
		\includegraphics[width=\textwidth]{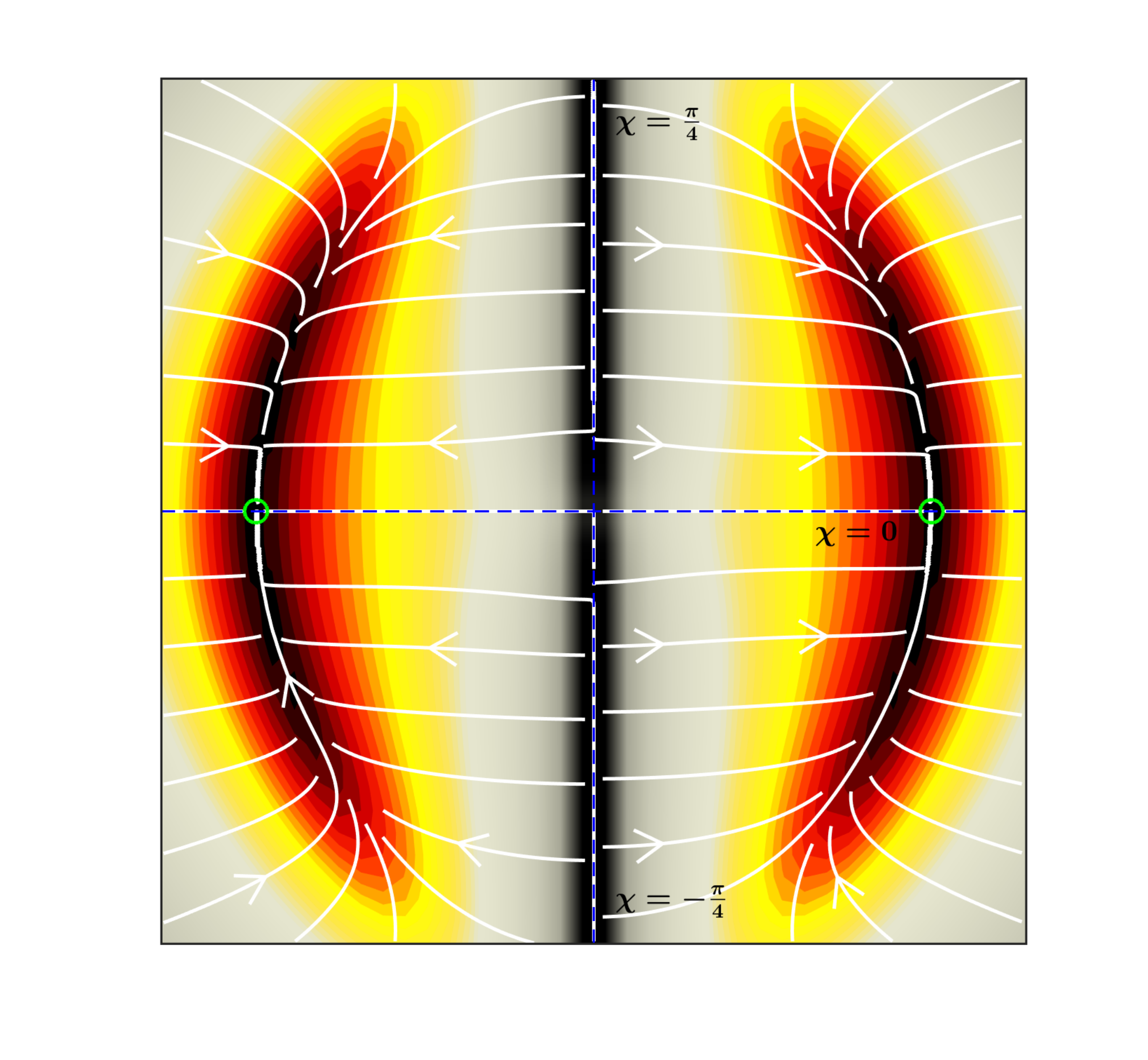}
		\caption{$\Gamma$ = 2.5$.10^{10}$ Pa$^2$.s$^{-3}$  $> \Gamma_\text{thr}$}
		\label{fig:stochStand}
	\end{subfigure}%
	\caption{Streamlines in the plane $\theta = 0 \mod\pi$ for the system \eqref{eq:sysSymStochDet}. Parameters: $\beta_0$=160 rad.s$^{-1}$, $\alpha$=100 rad.s$^{-1}$, $\kappa$=0.5 rad.s$^{-1}$Pa$^{-2}$, $\omega_n$=1167 rad.s$^{-1}$. $\Gamma_\text{thr}$ = 2.3$.10^10$ Pa$^2$.s$^{-3}$. Symbols: $\circ$: attractors, +: saddles. The color scale is given in fig. (\ref{fig:color})}
	\label{fig:basicStoch}
\end{figure}

For the configuration without mean flow, time delay and spatial asymmetries, the system of Langevin equations describing the slow-flow dynamics in presence of turbulent forcing is:
\begin{align}\label{eq:sysSymStoch}
\begin{cases}
\;\dot{A} \:=\: \nu A \,-\, \dfrac{3\kappa}{64}\left[5+\cos(4\chi)\right]A^3 \,+\, \dfrac{ 3\Gamma }{ 4A \omega_n^2} \,+\, \zeta_A
\\[10pt]
\; \dot{\chi} \:=\: \dfrac{3\kappa}{64}A^2\sin(4\chi) \,-\, \dfrac{ \Gamma \tan(2\chi) }{ 2 A^2 \omega_n^2 } \,+\, \dfrac{1}{A}\zeta_{\chi}
\\[10pt]
\; \dot{\theta} \:=\: \dfrac{1}{A n \cos(2\chi)}\zeta_{\theta} - \dfrac{\tan(2\chi)}{A}\zeta_{\varphi}
\\[10pt]
\; \dot{\varphi}  \:=\: \dfrac{1}{A}\zeta_{\varphi}
\end{cases}
\end{align}
In this perfectly symmetric configuration, $\dot{\varphi}$ and $\dot{\theta}$ are governed by pure multiplicative noise. The Langevin equations for $A$ and $\chi$ exhibit deterministic and stochastic contributions, and they are independent of $\varphi$ and $\theta$.  The deterministic part of the equations for $A$ and $\chi$ is:
\begin{align}\label{eq:sysSymStochDet}
\begin{cases}
\;\dot{A} \:=\: \nu A \,-\, \dfrac{3\kappa}{64}\left[5+\cos(4\chi)\right]A^3 \,+\, \dfrac{ 3\Gamma }{ 4 A \omega_n^2}
\\[10pt]
\; \dot{\chi} \:=\: \dfrac{3\kappa}{64}A^2\sin(4\chi) \,-\, \dfrac{  \Gamma \tan(2\chi) }{ 2 A^2 \omega_n^2 }
\end{cases}
\end{align}
The behavior of this system depends on the random forcing amplitude: if $\Gamma < 256\nu^2\omega_n^2/(27\kappa)\equiv \Gamma_\text{thr}$, the equilibrium points of the system are the standing mode
\begin{align}\label{eq:solStandStoch}
A=\dfrac{4}{3}\sqrt{\dfrac{\nu+\sqrt{\nu^2+\dfrac{27 \kappa \Gamma}{32\omega_n^2}}}{\kappa}} \equiv A^\text{sto}_\text{stand}
\quad\text{and}\quad
\chi=0
\end{align}
which is a saddle point, and the mixed modes
\begin{align}
\begin{cases}
A=\sqrt{ \dfrac{8}{3\kappa} \left( \nu + \sqrt{ \nu^2 + \dfrac{3\kappa\Gamma}{16\omega_n^2} } \right) } \equiv A^\text{sto}_\text{mix}
\\[15pt] \chi=\pm\dfrac{1}{2}\arccos\left(\dfrac{\sqrt{3\kappa\Gamma/(4\omega_n^2)}}{\nu+\sqrt{\nu^2+\dfrac{3\kappa\Gamma}{16\omega_n^2}}}\right)
\end{cases}
\end{align}
which are stable. The noise changes therefore the preferential dynamics of the system, which is not anymore attracted to pure spinning modes. Increasing the noise intensity has the effect to move the attractors of the associated deterministic system away from the poles of the Bloch sphere and towards the equatorial plane. This conclusion has been recently drawn in \cite{ghirardo19}, where a similar equation for $\chi$ was derived. The standing modes are also equilibrium points, but they are unstable. The streamlines for this case are shown in Fig. \ref{fig:stochMix}.
When $\Gamma \geq \Gamma_{\text{thr}}$, the mixed mode merges in the equatorial plane and the standing modes become stable. The threshold value $ \Gamma_{\text{thr}}$ grows proportionally with $\nu^2$.
It implies that i) for a given noise intensity $\Gamma$ and ii) for a fixed saturation constant $\kappa$, if the linear growth rate is increased from the Hopf point, standing modes will be the  attractor, until the growth rate reaches $\nu_\text{thr} \equiv (3/16)\times\sqrt{3\kappa\Gamma}/\omega_n$. Beyond this critical growth rate, the attractor of the deterministic component of the system will be a mixed mode which will move toward the poles of the Bloch sphere as the linear growth rate is further increased. $(\Gamma/(\omega_n^2\nu)^1/2$ gives an order of magnitude of the amplitude of the spatial average of the random fluctuations induced by the noise $\Xi_n(t)$. The local perturbation term $\Xi(t,\Theta)$ can possibly reach significantly higher levels.
It is important to emphasize that the fields shown in Fig. \ref{fig:stochMix} only indicate the dynamics associated with deterministic part of the system of coupled Langevin equations. Additive and multiplicative noise will of course lead to random trajectories in the phase space.
Considering again cases with linear growth rates that are low compared to the critical growth rate $\nu_\text{thr}$ (for which the equator of the Bloch sphere, which corresponds to standing modes, is the attractor) one can assume that $\chi\approx0$, and $A\approx A^\text{sto}_\text{stand}$ whose expression is given in \eqref{eq:solStandStoch}. Under these assumptions, it is interesting to consider the equation for preferential direction of the mode: $$\dot{\theta} \approx 1/(nA^\text{sto}_\text{stand})\,\zeta_\theta.$$ This is the equation of a Wiener process \cite{gardiner04,einstein05}. The mean square displacement of $\theta$ will therefore follow the law
\begin{align}
\left< \left[\theta(\Delta t+t) - \theta(t)\right]^2\right> = \dfrac{\Gamma}{2\omega_n^2 n^2(A^\text{sto}_\text{stand})^2}\Delta t \:.
\end{align}
The direction of the oscillation follows a random walk around the annulus. These conclusions can be used to shed light on unexplained experimental observations of the dynamics of azimuthal thermoacoustic modes in turbulent annular combustors with uniform  distribution of identical flames and without mean flow. 

\section{Symmetry breaking induced by a non-uniform thermoacoustic distribution}
\label{sec8}

\begin{figure}
	\centering
	\begin{subfigure}{.5\linewidth}
		\centering
		\includegraphics[width=\linewidth]{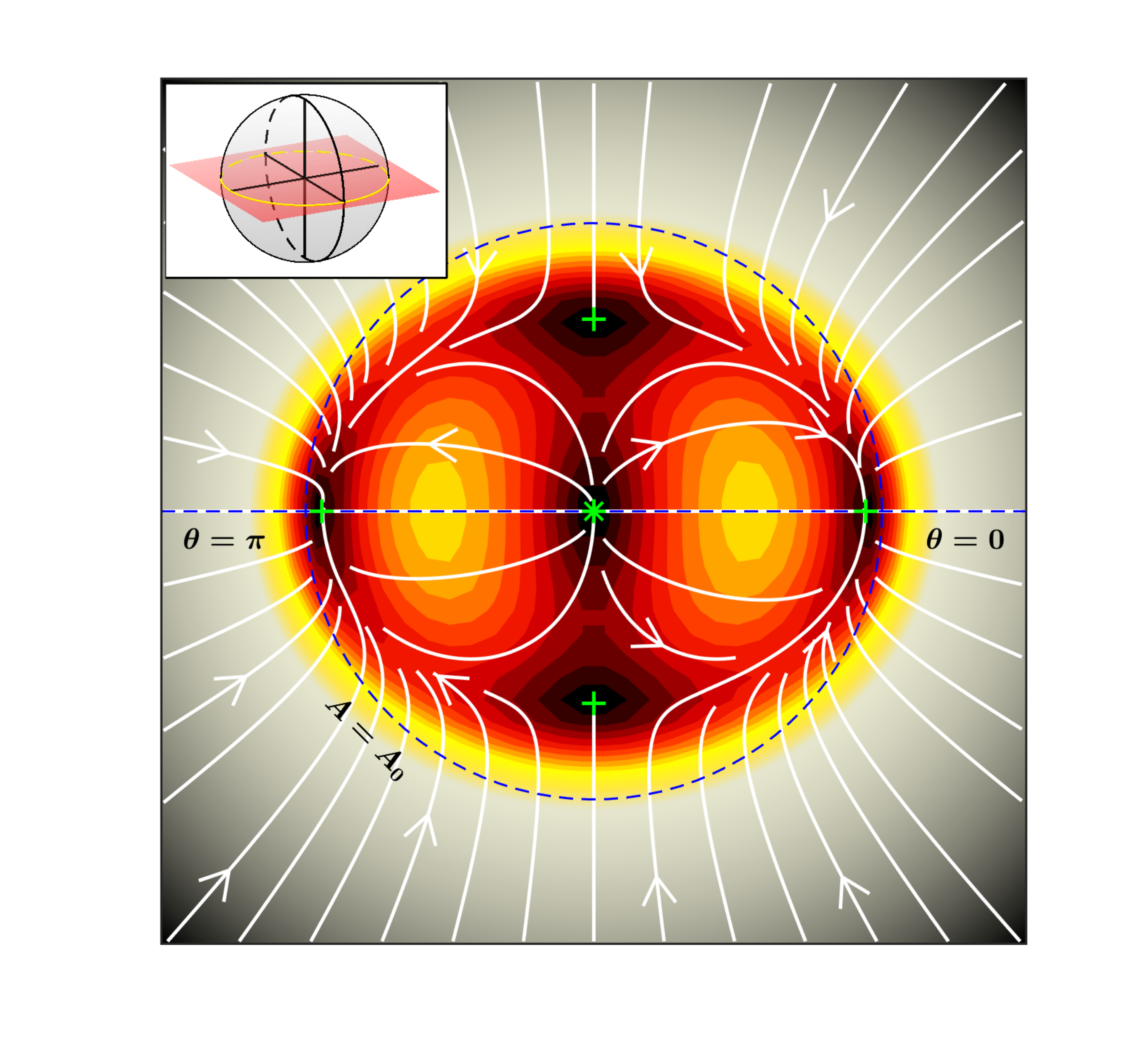}
		\caption{plane $\chi = 0$}
	\end{subfigure}%
	\begin{subfigure}{.5\linewidth}
		\centering
		\includegraphics[width=\linewidth]{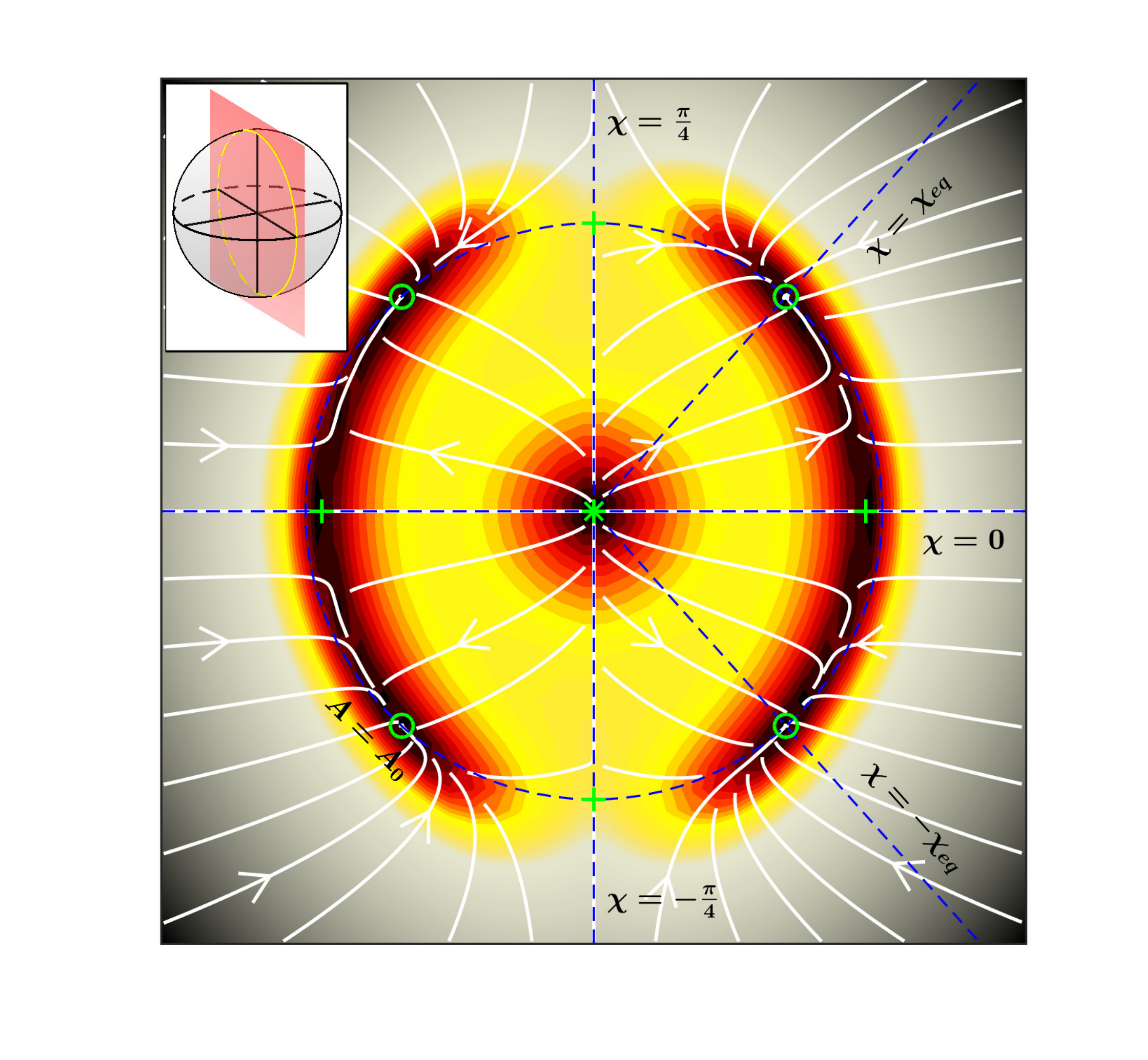}
		\caption{plane $\theta=0\mod\pi$}
	\end{subfigure}%
	
	\begin{subfigure}{.5\linewidth}
		\centering
		\includegraphics[width=\linewidth]{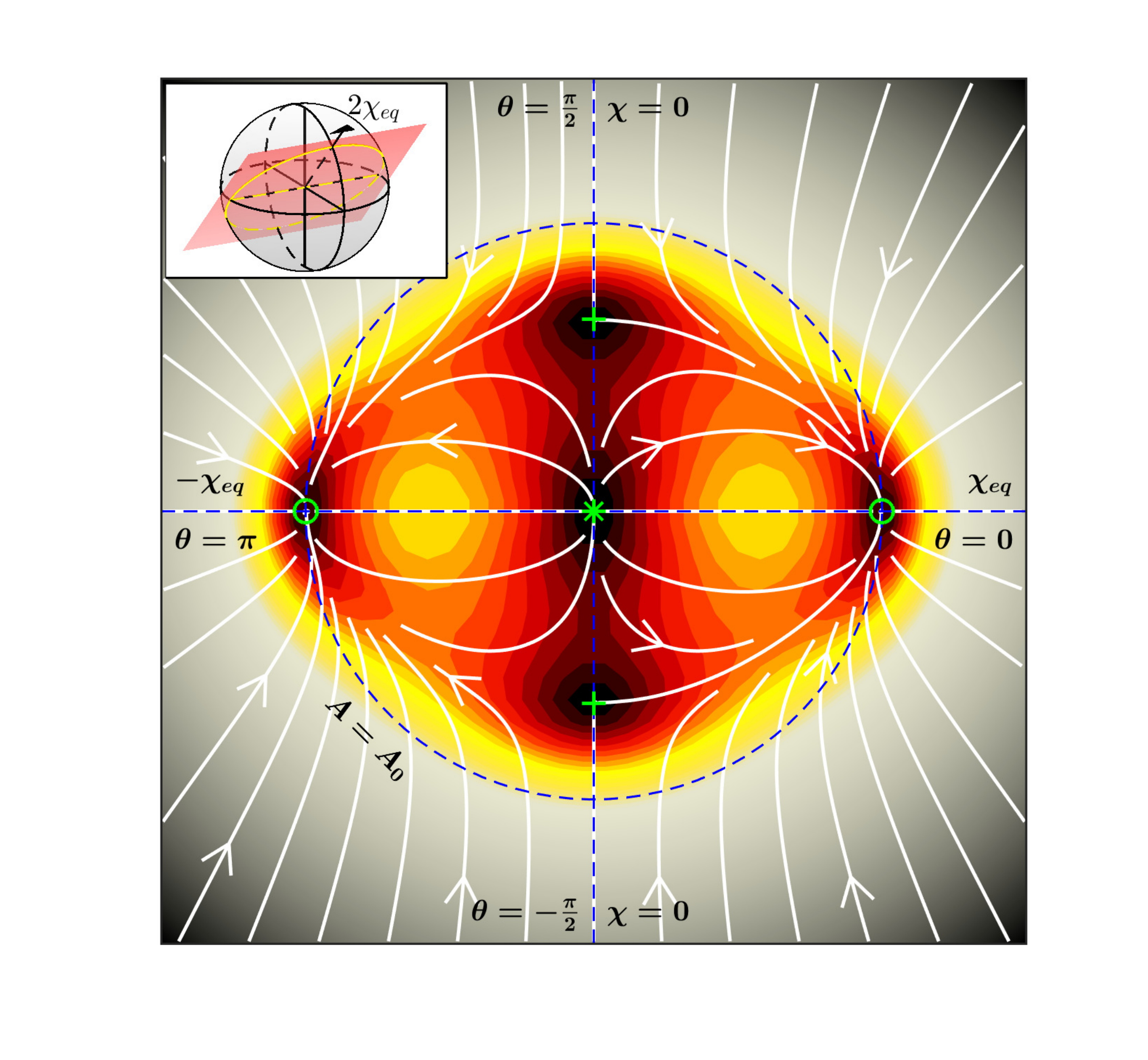}
		\captionsetup{width=.8\textwidth}
		\caption{plane inclinated of $\chi_\text{eq}$, showing the two attractors $\chi=\chi_\text{eq}$  and $\chi=-\chi_\text{eq}$.}
	\end{subfigure}%
	\begin{subfigure}{.5\linewidth}
		\centering
		\includegraphics[width=\linewidth]{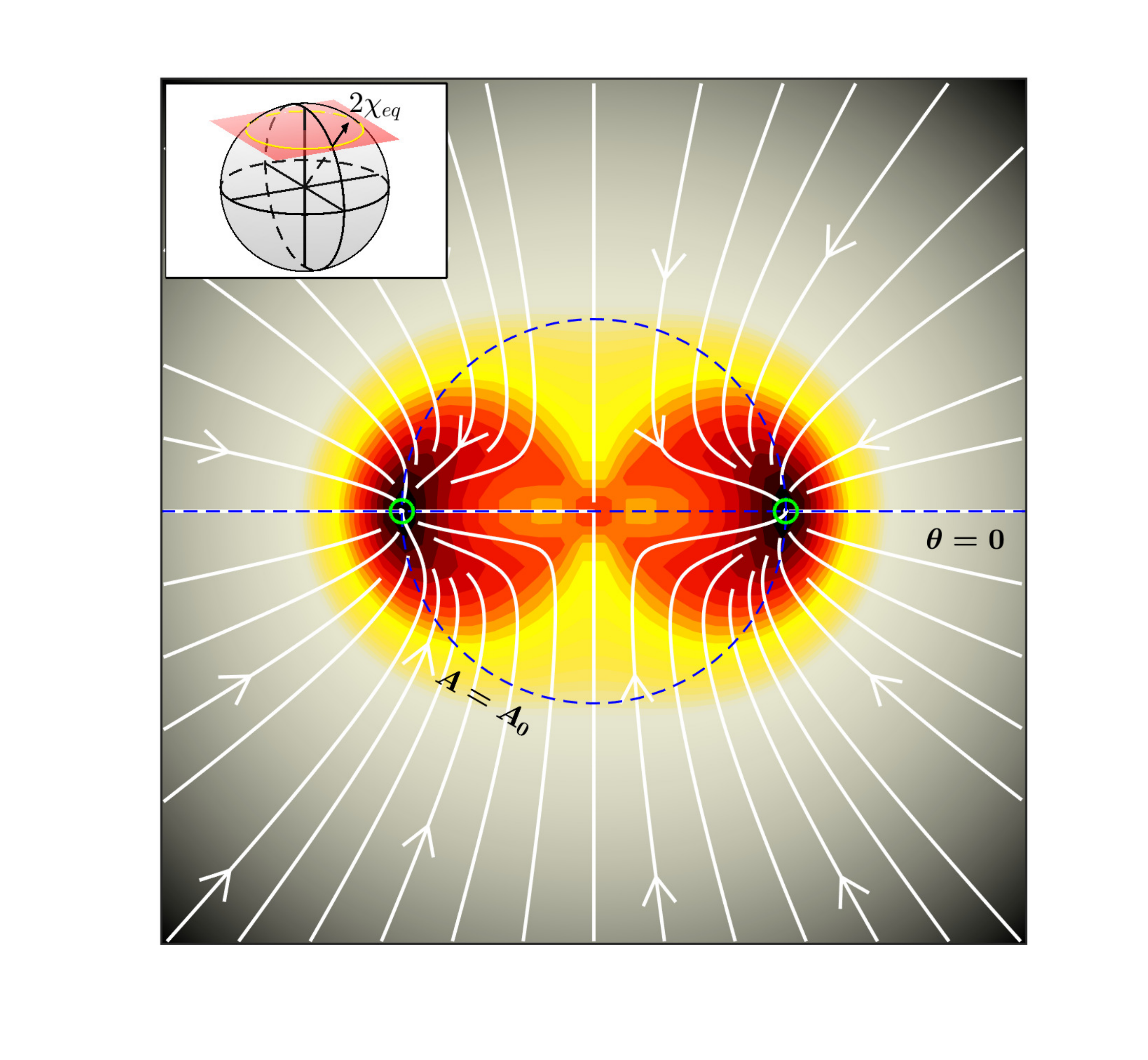}
		\captionsetup{width=.8\textwidth}
		\caption{Plane z=cst. including the points $\{\chi=\chi_\text{eq},\theta=0\}$ and $\{\chi=\chi_\text{eq},\theta=\pi\}$, corresponding to the same state. }
	\end{subfigure}%
	\caption{ Stream lines describing the evolution of the state of the linearly-unstable azimuthal-thermoacoustic-mode on the Bloch Sphere, for a non-uniform distribution of the thermoacoustic feedback:   $c_{2n}$=0.25 $<$ $C_{2n}$=0.375. $\beta_0$=160 rad.s$^{-1}$, $\alpha$=100 rad.s$^{-1}$, $\kappa$=0.5 rad.s$^{-1}$Pa$^{-2}$, $\omega_n$=1167 rad.s$^{-1}$. Symbols: $\circ$: attractors $*$: repellers, +: saddles. The color scale is given in fig. (\ref{fig:color})}
	\label{fig:asym}
\end{figure}

We  consider now the effect of azimuthal asymmetries of the thermoacoustic coupling without random forcing from turbulence. All the results obtained with the present quaternion formalism are in perfect agreement with the findings in reference \cite{noiray11}, which were obtained with a simpler approach based on modal projection. It will be shown in the next section that the latter  approach is less general and cannot be used to predict the system dynamics when symmetry-breaking results from the presence of mean swirl. 
When the Fourier coefficient $c_{2n}$ is different from 0, the system of equations governing the dynamics of the slow-flow variables that define the state of the azimuthal thermoacoustic mode is:
\begin{align}
\begin{cases}
\;\dot{A} \:=\: \left(\nu + \dfrac{c_{2n}\beta_0}{4}\cos(2n\theta)\cos(2\chi)\right)A \,-\, \dfrac{3\kappa}{64}\left(5+\cos(4\chi)\right)A^3
\\[10pt]
\; \dot{\chi} \:=\: \dfrac{3\kappa}{64}A^2\sin(4\chi) \,-\, \dfrac{c_{2n}\beta_0}{4}\cos(2n\theta)\sin(2\chi)
\\[10pt]
\; n\dot{\theta} \:=\:  -\dfrac{c_{2n}\beta_0}{4}\dfrac{\sin(2n\theta)}{\cos(2\chi)} 
\\[10pt]
\; \dot{\varphi} \:=\: \dfrac{c_{2n}\beta_0}{4}\sin(2n\theta)\tan(2\chi)
\end{cases}
\end{align}
If $0< c_{2n}<(\beta_0-\alpha)/\beta_0$, the equilibrium points of the system are:
\begingroup
\allowdisplaybreaks
\begin{align*}
&A=0&\quad &\mbox{(repeller)}&
\\[10pt]
A=\dfrac{4}{3}\sqrt{\dfrac{\beta_0\left(1+c_{2n}/2\right)-\alpha}{\kappa}}\quad \mbox{and} \quad &\chi=0& &&
\\[4pt]
\mbox{and} \quad &\theta=0 \mod (\pi/n)& \quad &\mbox{(saddle)}&
\\[10pt]
A=\dfrac{4}{3}\sqrt{\dfrac{\beta_0\left(1-c_{2n}/2\right)-\alpha}{\kappa}} \quad \mbox{and} \quad &\chi=0& &&
\\[4pt]
\mbox{and} \quad &\theta=\dfrac{\pi}{2n} \mod (\pi/n)& \quad &\mbox{(saddle)}&
\\[10pt]
A=\sqrt{\dfrac{8(\beta_0-\alpha)}{3\kappa}}\quad \mbox{and} \quad &\chi=\pm\dfrac{\pi}{4}&  &\mbox{(saddles)}&
\\[10pt]
A=\sqrt{\dfrac{8(\beta_0-\alpha)}{3\kappa}}=A_0 \quad \mbox{and} \quad &\chi=\pm\dfrac{1}{2}\arccos\left(\dfrac{c_{2n}\beta_0}{\beta_0-\alpha}\right)&&\equiv \pm\chi_\text{eq} &
\\[4pt]
\mbox{and} \quad &\theta=0 \mod (\pi/n)&  &\mbox{(attractors)}&
\end{align*}
\endgroup
The stable solutions are mixed mode, rotating either in counterclockwise or clockwise direction. The equilibrium points in the $\chi=0$ plane are two different kind of saddle points: The ones located at angular positions $n\theta = 0 \mod \pi$ are attractive in the $A$ and $\theta$ directions and the ones located on $n\theta = \pi/2 \mod \pi$ are only attractive in the $A$ direction, the two other eigendirections being repulsive. The preferential direction is therefore attracted towards $\theta=0 (\mod\pi/n)$, which was defined in the section \ref{sec43} as the angle where the spatial Fourier term of order $2n$ of the linear gain $\beta$ reaches its maximum. The temporal slow phase $\varphi$ stabilizes to a constant value depending on the initial conditions. The streamlines plots in fig. (\ref{fig:asym}) show that the system is no longer attracted by the poles of the Bloch sphere, which became saddles at the saddle-node bifurcations. Indeed, when $c_{2n}$ is decreased and ultimately vanishes for the uniform distribution of thermoacoustic feedback,  the saddles at the poles coalesce with the attractors. On the other end, increasing the value of $c_{2n}$ pushes $\chi_\text{eq}$ toward zero and therefore makes\\
 the standing part of the mode more and more prominent. 
\begin{figure}
	\centering
	\begin{subfigure}{.5\textwidth}
		\centering
		\includegraphics[width=\linewidth]{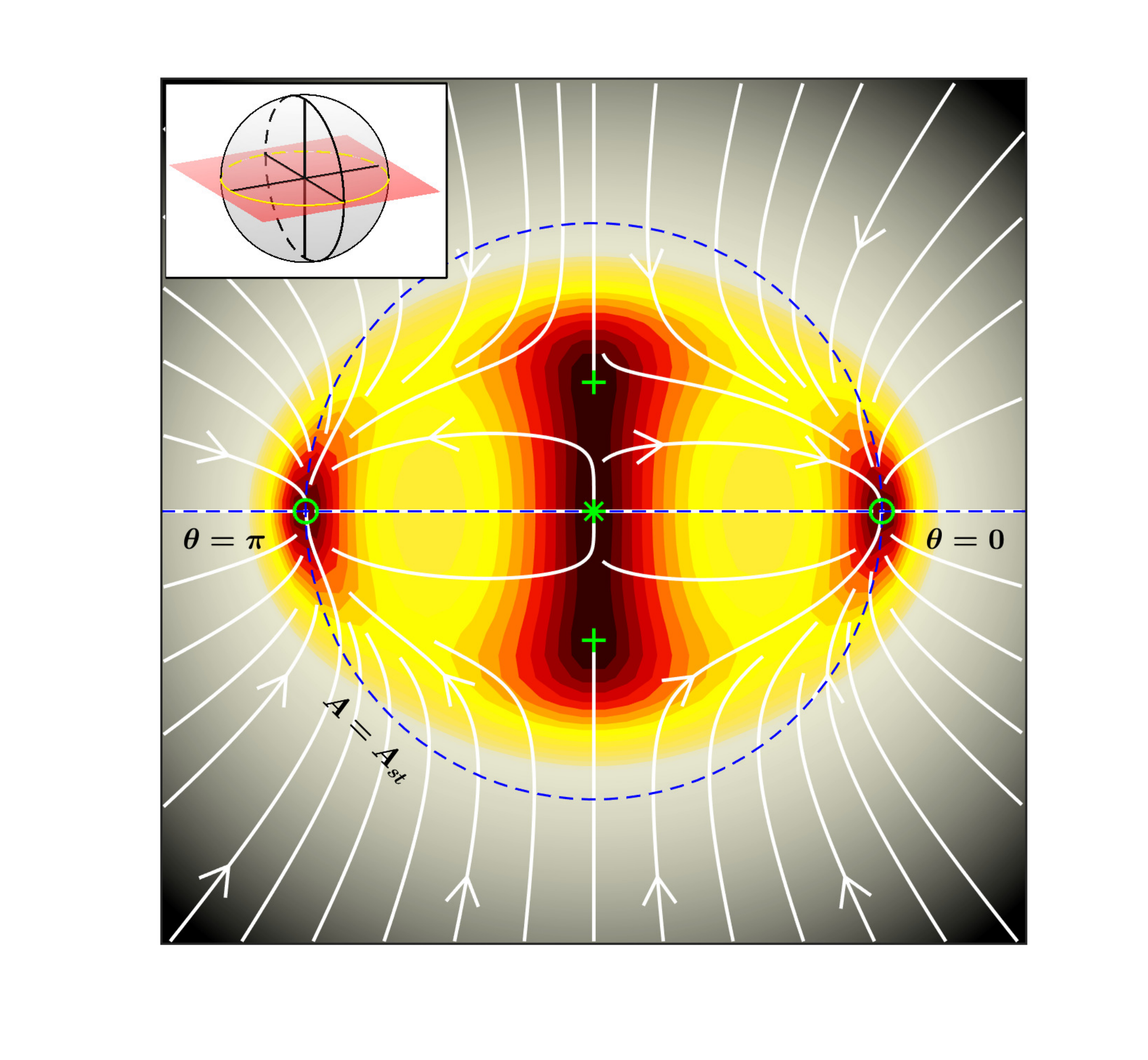}
		\caption{plane $\chi=0$}
	\end{subfigure}%
	\begin{subfigure}{.5\linewidth}
		\centering
		\includegraphics[width=\textwidth]{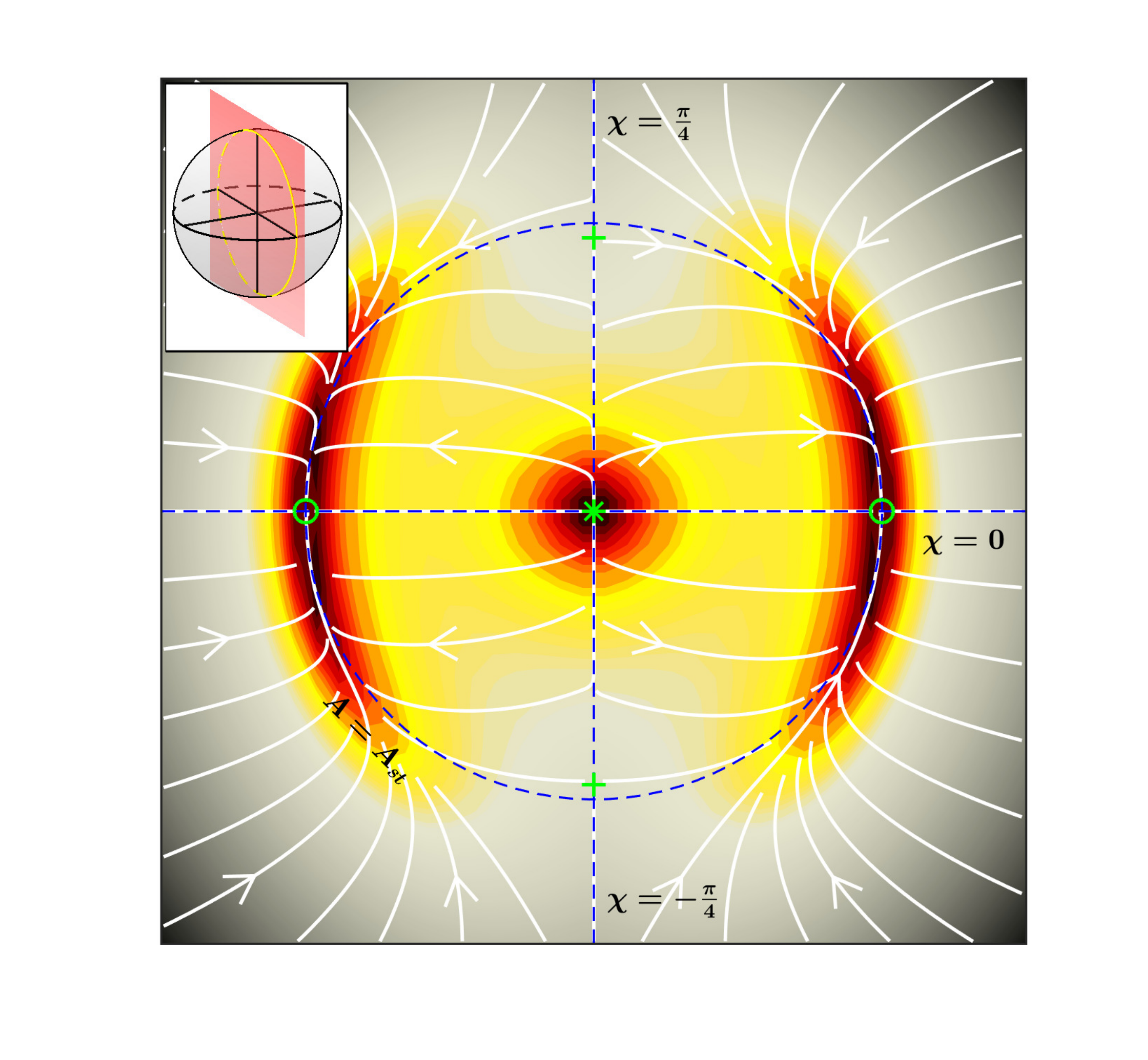}
		\caption{plane $\theta=0\mod\pi$}
	\end{subfigure}%
	\caption{Stream lines for a large azimuthal asymmetry: $c_{2n}$=0.5 $>$ $C_{2n}$ = 0.375. $\beta_0$=160 rad.s$^{-1}$, $\alpha$=100 rad.s$^{-1}$, $\kappa$=0.5 rad.s$^{-1}$Pa$^{-2}$, $\omega_n$=1167 rad.s$^{-1}$. Symbols: $\circ$: attractors $*$: repellers, +: saddles. The color scale is given in fig. (\ref{fig:color})}
	\label{fig:asym_stand}
\end{figure}
If the asymmetries of the thermoacoustic feedback along the annular combustor circumference are important enough to get $c_{2n} > C_{2n}=(\beta_0-\alpha)/\beta_0$, another bifurcation takes place where the attractor merge with the saddle of the equatorial plane at $\theta=0$. It leads to the following  equilibrium points:
\begingroup
\allowdisplaybreaks
\begin{align*}
&A=0&\quad &\mbox{(repeller)}&
\\[10pt]
A=\dfrac{4}{3}\sqrt{\dfrac{\beta_0\left(1+c_{2n}/2\right)-\alpha}{\kappa}}\equiv A_\text{st}\quad \mbox{and} \quad &\chi=0&   &&
\\[4pt]
\mbox{and} \quad &\theta=0 \mod (\pi/n)& \quad &\mbox{(attractor)}&
\\[10pt]
A=\dfrac{4}{3}\sqrt{\dfrac{\beta_0\left(1-c_{2n}/2\right)-\alpha}{\kappa}}\quad \mbox{and} \quad &\chi=0&  &&
\\[4pt]
\mbox{and} \quad &\theta=\dfrac{\pi}{2n} \mod (\pi/n)&  &\mbox{(saddle)}&
\\[10pt]
A=\sqrt{\dfrac{8(\beta_0-\alpha)}{3\kappa}}\quad \mbox{and} \quad &\chi=\pm\dfrac{\pi}{4}&  \quad &\mbox{(saddle)}&
\end{align*}
\endgroup
The only stable solution is a standing mode locked on the maxima of the $2n$-th order Fourier component of $\beta(\Theta)$ . Again, $\varphi$ stabilizes on a constant value depending on the initial conditions. Figure \ref{fig:asym_stand} shows that the attractors moved to the plane $\chi=0$ and merged in two points $\{\chi=0,\:\theta=0\}$ and  $\{\chi=0,\:\theta=\pi\}$ corresponding to the same attractor. When $c_{2n}>2 C_{2n}$, the saddle point located in $\chi=0, \theta=\pi/2 \mod \pi$ merges with the repeller in $A=0$  and the origin becomes a saddle point attracting in the direction $\theta=\pi/2\mod\pi$ and repelling to the direction $\theta=0 \mod\pi$ and to the poles.
In conclusion, symmetry breaking induced by a non-uniform distribution of the heat release rate feedback, first induces a switch from pure spinning to mixed spinning-standing solutions and that for a sufficiently large asymmetry of the 2$n$ Fourier component of the distribution, the linearly unstable thermoacoustic system has a pure standing wave as solution.
\section{Symmetry breaking induced by the presence of mean swirl}\label{sec9}
The effect of a non-zero azimuthal velocity of the mean flow is now scrutinized. This configuration with mean swirl is still axisymmetric but does no longer exhibit reflectional symmetry. \\
In their theoretical study about symmetry breaking in symmetric systems undergoing bifurcations, Crawford and Knobloch \cite{crawford91} show that removing the reflectional symmetry will no longer ensure that both spinning solutions will have the same amplitudes and eigenvalues. This symmetry breaking is accompanied with a split of the degenerate eigenvalues of the associated linear problem, as observed for instance in disk-shaped micrometer-sized ferromagnetic elements, in which the polarization of a core mode breaks the symmetry between the two azimuthal spin-wave modes \cite{guslienko08} -- Hoffmann \textit{et al.} \cite{hoffmann07} shows that removing the core mode allows to restore the degeneracy and suppress the frequency splitting. In the field of thermoacoustics, Bauerheim \textit{et al.} \cite{bauerheim15}  showed that a mean swirl in annular combustors affects the solutions of the linear eigenvalue problem and that it leads to counter-rotating modes with different linear growth rates and frequencies. \\
In this section, we consider the nonlinear thermoacoustic problem of a thin annular combustor with a low-Mach mean swirl (azimuthal velocity $U<0.1c$), for a uniformly distributed heat release rate feedback. To our  knowledge, this problem has never been addressed, and we show in what follows that this type of symmetry breaking bifurcation can be conveniently investigated with the present quaternion-based acoustic field projection.
Considering the case of $c_{2n}=0$, which is less restrictive than the uniform distribution $\beta=\beta_0$, we apply the slow flow averaging on the wave equation (\ref{eq:thEq}) to obtain the following system:
\begin{align}\label{eq:sw}
\begin{cases}
\dot{A} \:=\: \dfrac{1}{2}\bigg(\beta_0\left[1-M\sin(2\chi)\right]-\alpha\bigg)A
\:+\: \dfrac{3\kappa}{64}\bigg(-5-\cos(4\chi) + 4M\sin(2\chi) \bigg)A^3,
\\[15pt]
\dot{\chi} \:=\: \dfrac{3\kappa}{64}A^2\bigg(\sin(4\chi)+6M\cos(2\chi)\bigg) \,-\, \dfrac{M\beta_0}{2}\cos(2\chi),
\\[12pt]
n\dot{\theta} \:=\: M\omega_n,
\\[10pt]
\dot{\varphi}  \:=\: 0.
\end{cases}
\end{align}
The linear growth rate in the amplitude equation is $(\beta_0\left[1-M\sin(2\chi)\right] -\alpha)/2$. It gets smaller (resp. larger) when $M\sin(2\chi)$ is positive (resp. negative), in other words the amplitude grows faster when the wave propagates against the swirl ($M>0$ and $\chi < 0$, or $M<0$ and $\chi>0$) than when it propagates in the same direction ($M>0$ and $\chi>0$, or $M<0$ and $\chi<0$) . It shows that the swirl breaks the symmetry between the two spinning waves in the vicinity of the origin. This symmetry breaking effect comes from the combined effect of the mean azimuthal flow and the linear response of the flame. The term responsible of this effect in the thermoacoustic wave equation \eqref{eq:thEq} is : $(U/\mathcal{R})\partial(\beta p)/\partial\Theta$. This term is a product of terms with slow time scales associated to the mean azimuthal flow, and to the flame response. As specified at the end of the section  \ref{sec4}.\ref{sec42}, keeping this term  is meaningful only if $\beta_0\gg\beta_0-\alpha$. If this condition is not fulfilled, the splitting of the counter-spinning waves growth rates is a second order phenomenon and should not be considered in the present first order analysis. Moreover, such second order effects would be negligible in a non ideal chamber, where several other physical phenomena would have a stronger influence on the growth rate splitting, e.g. the presence of a radial mean velocity gradient \cite{rouwenhorst17}.\\
Coming back to the present model, we can go one step further and consider the consequences of the symmetry breaking for states with non-vanishing amplitudes, in particular the saddles and attractors which govern this intriguing nonlinear dynamics.\\
From now on, we will always assume that $M\geq0$, corresponding to a counterclockwise swirl. The clockwise direction will sometimes be referred to as the counter swirl direction. The analysis for $M\leq0$ will give exactly the same results in a phase space where $-\chi$ replaces $\chi$.
The equilibrium solutions for $A$ and $\chi$ are:
\begin{align*}\label{eq:solSw}
&\hspace{3cm}  A=0&; \quad&   \chi=\pm\dfrac{\pi}{4} &
\\[10pt]
&A=\dfrac{4}{3}\sqrt{\dfrac{\beta_0(1-M^2)-\alpha}{\kappa(1-M^2)}}\equiv A_m& ; \quad &\chi=\dfrac{1}{2}\arcsin\left(\dfrac{3M\alpha}{\beta_0(1-M^2)-\alpha}\right)\equiv\chi_m &
\\[10pt]
&A=\sqrt{\dfrac{8(\beta_0(1-M)-\alpha)}{3\kappa(1-M)}}\equiv A_{ccw} & ; \quad & \chi=\dfrac{\pi}{4}&
\\[10pt]
&A=\sqrt{\dfrac{8(\beta_0(1+M)-\alpha)}{3\kappa(1+M)}}\equiv A_{cw} & ; \quad& \chi=-\dfrac{\pi}{4} &
\end{align*}
The first solution is the trivial 0 solution. The second solution corresponds to a mixed mode of amplitude $A_m$ and nature angle $\chi_m$ increasing with the mach. The two last solutions are pure spinning modes, respectively in the counterclockwise and the clockwise direction. The amplitude of the clockwise spinning mode is larger than the equilibrium amplitude $A_0 = 4\sqrt{\nu/3\kappa}$ without swirl, and the amplitude of the counterclockwise mode is smaller. Some of these solutions are not always defined depending on the value of $M$. The equilibrium point $\{A_m,\:\chi_m\}$ exists only under the condition:
\begin{align}
\dfrac{\beta_0}{\alpha} > \dfrac{1+3\lvert M\rvert}{1-M^2}
\end{align}
in the case of a positive mach small compared to 1, this condition becomes:
\begin{align}
\dfrac{\beta_0}{\alpha} > \dfrac{1+3M}{1-M^2}
\end{align}
The condition on $M$ is a 2$^{\text{nd}}$ order polynomial equation with only one positive root. Its value is:
\begin{align}
M_1 = \sqrt{\left(\dfrac{3\alpha}{2\beta_0}\right)^2 + \left(1-\dfrac{\alpha}{\beta_0}\right)} - \dfrac{3\alpha}{2\beta_0}
\end{align}
The counterclockwise mode is defined only if the following condition is satisfied:
\begin{align}
M < 1-\dfrac{\alpha}{\beta_0} \equiv M_2
\end{align}
Using $\alpha<\beta_0$, we have:
\begin{align}
0\leq M_1< M_2
\end{align}

\begin{figure}
	\centering
	\begin{subfigure}{.5\textwidth}
		\centering
		\includegraphics[width=\textwidth]{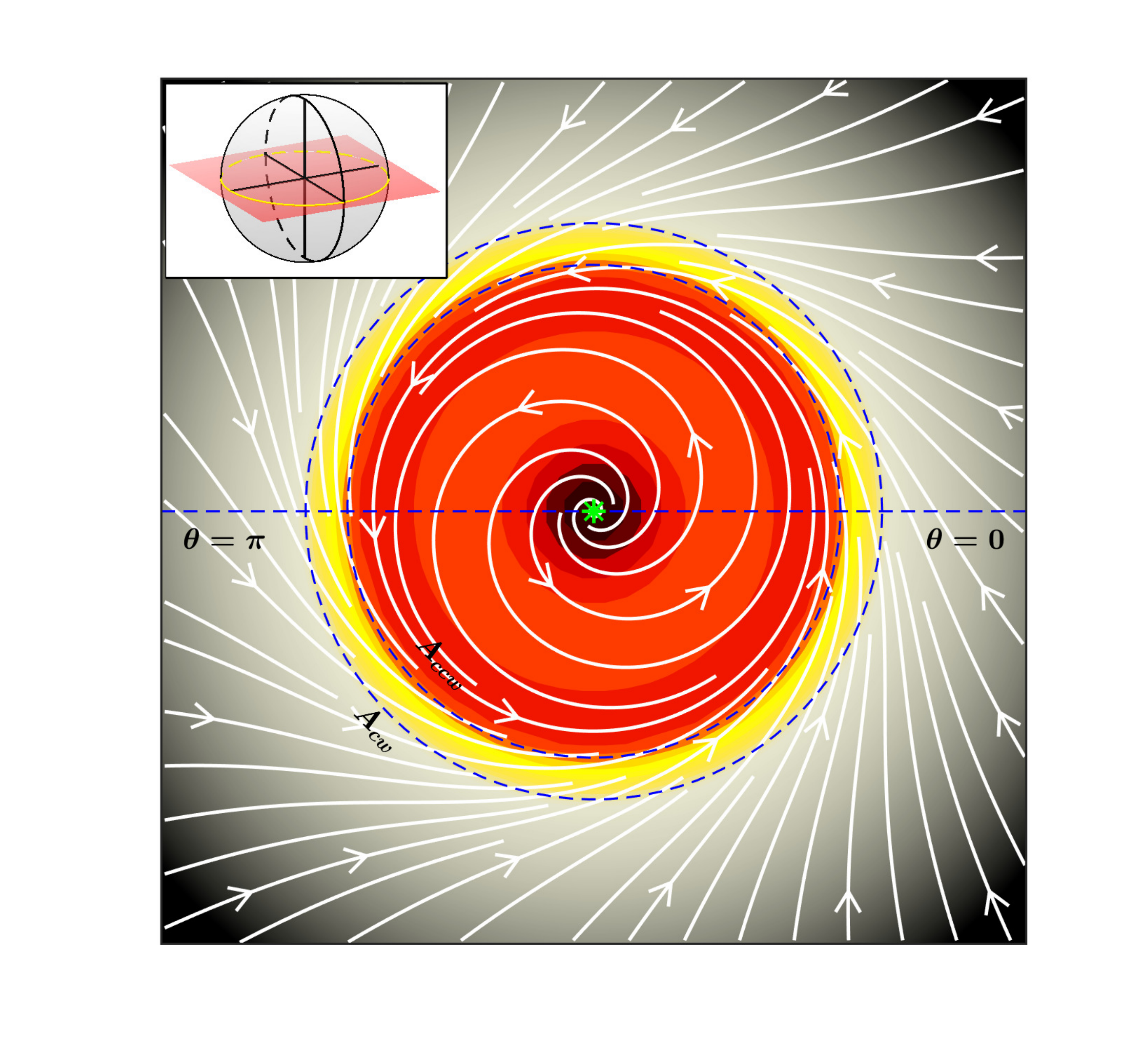}
		\caption{plane $\chi = 0$}
	\end{subfigure}%
	\begin{subfigure}{.5\textwidth}
		\centering
		\includegraphics[width=\textwidth]{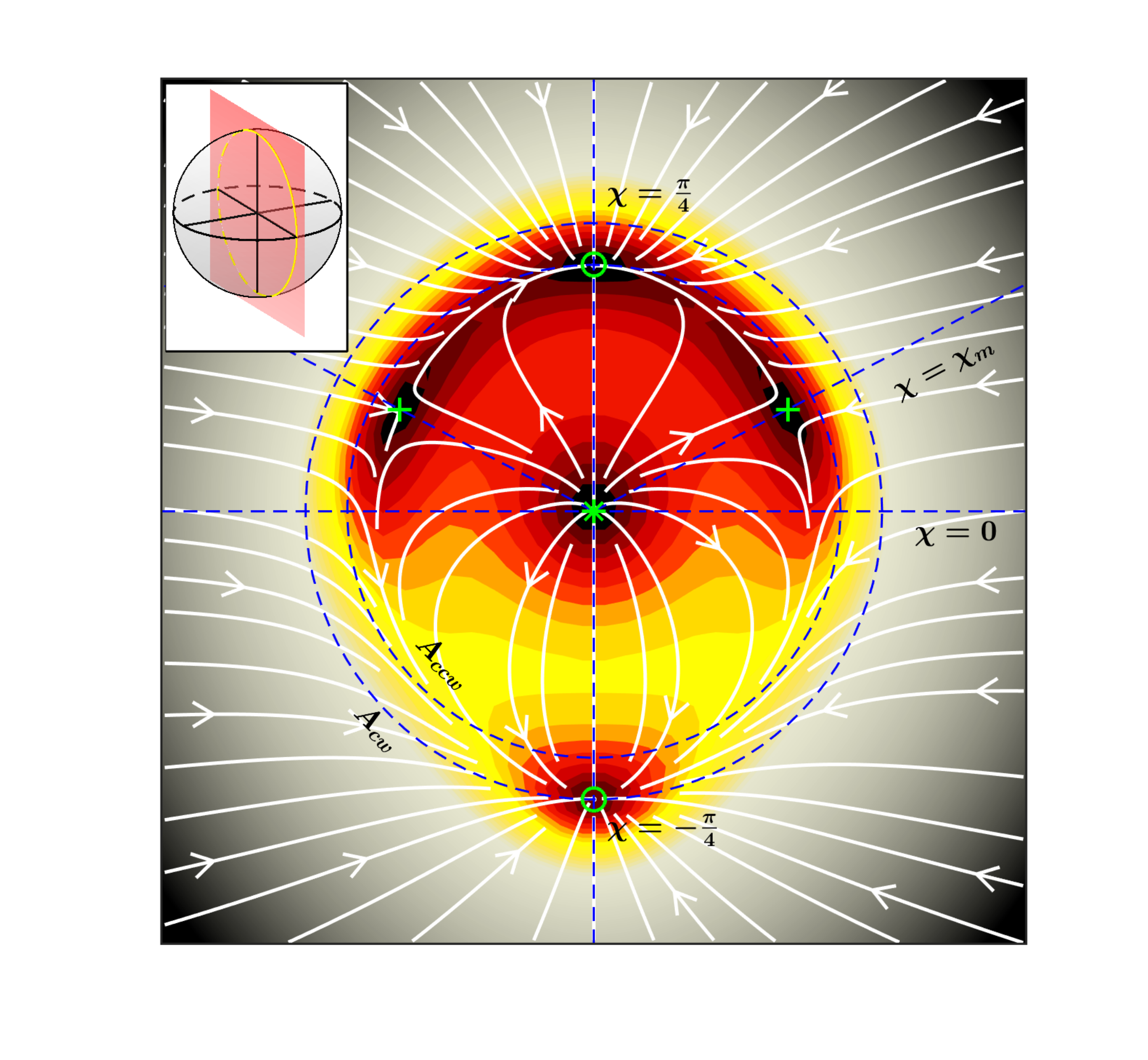}
		\caption{plane $\theta=0\mod\pi$}
	\end{subfigure}%
	\caption{Streamlines for $M=0.005$. $M_1=0.0107 > M$.The green symbols are the equilibrium points: $\circ$: attractors, $*$: repellers, +: saddles. Parameters: $\beta_0$=160 rad.s$^{-1}$, $\alpha$=155 rad.s$^{-1}$, $\kappa$=0.5 rad.s$^{-1}$Pa$^{-2}$, $\omega_n$=1167 rad.s$^{-1}$. The maximum value of the color scale is not the same as shown in Fig. (\ref{fig:color}).}
	\label{fig:sw0_005}
\end{figure}
\begin{figure}
	\centering
	\begin{subfigure}{.5\textwidth}
		\centering
		\includegraphics[width=\textwidth]{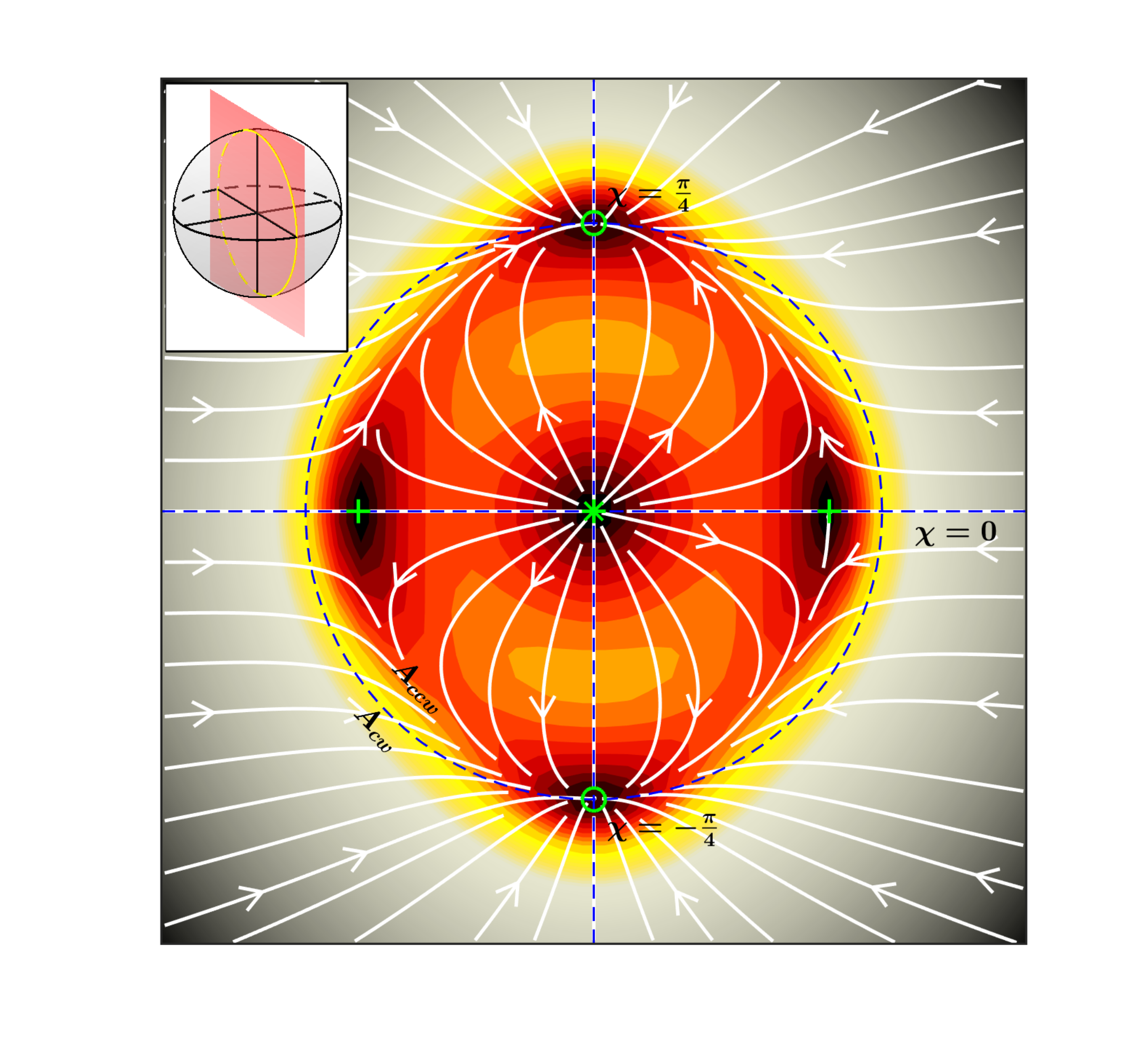}
		\caption{$M=0$}
		\label{fig:swsta}
	\end{subfigure}%
	\begin{subfigure}{.5\textwidth}
		\centering
		\includegraphics[width=\textwidth]{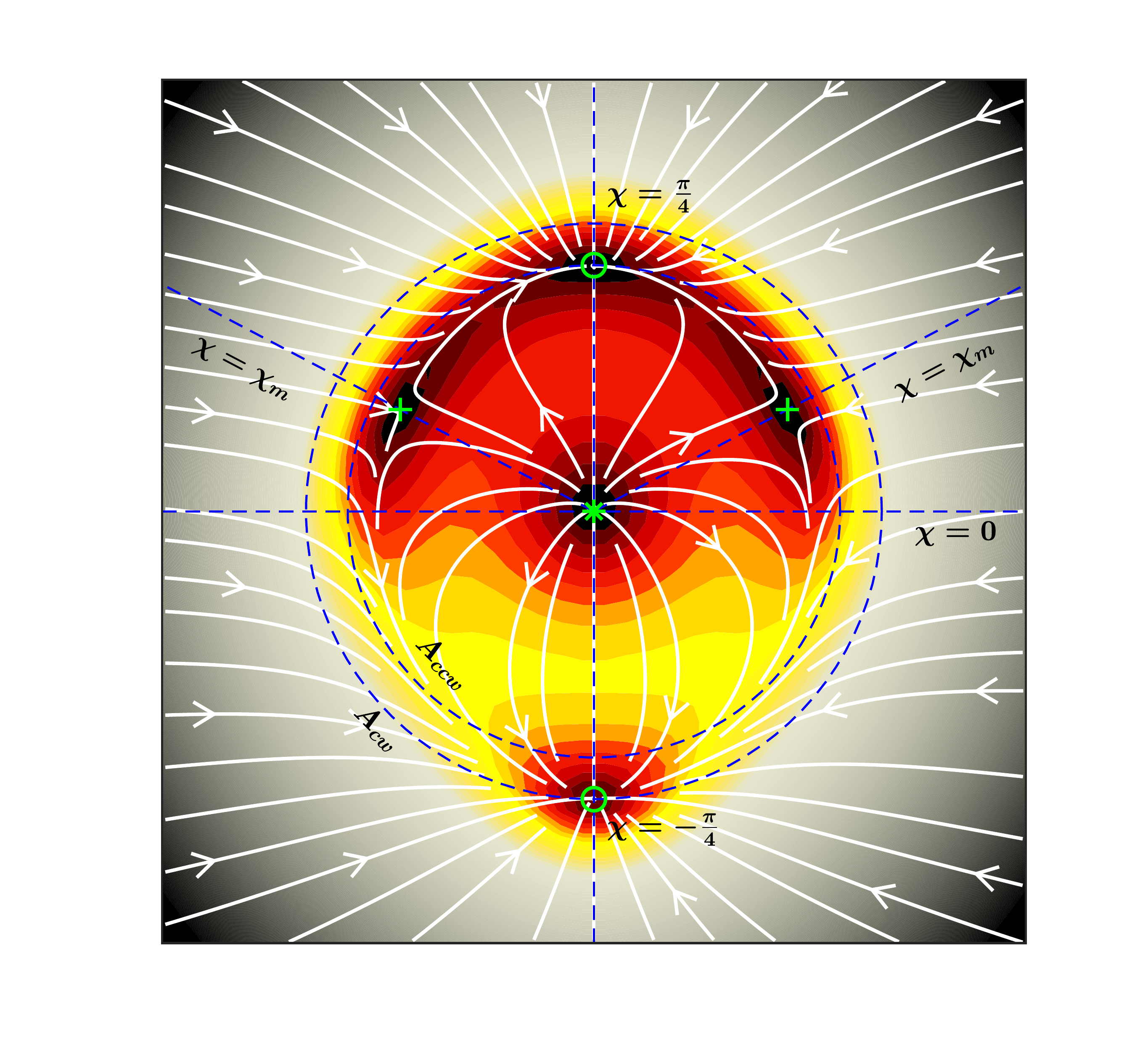}
		\caption{$M=0.005<M_1$}
		\label{fig:swstb}
	\end{subfigure}%
	
	\begin{subfigure}{.5\textwidth}
		\centering
		\includegraphics[width=\textwidth]{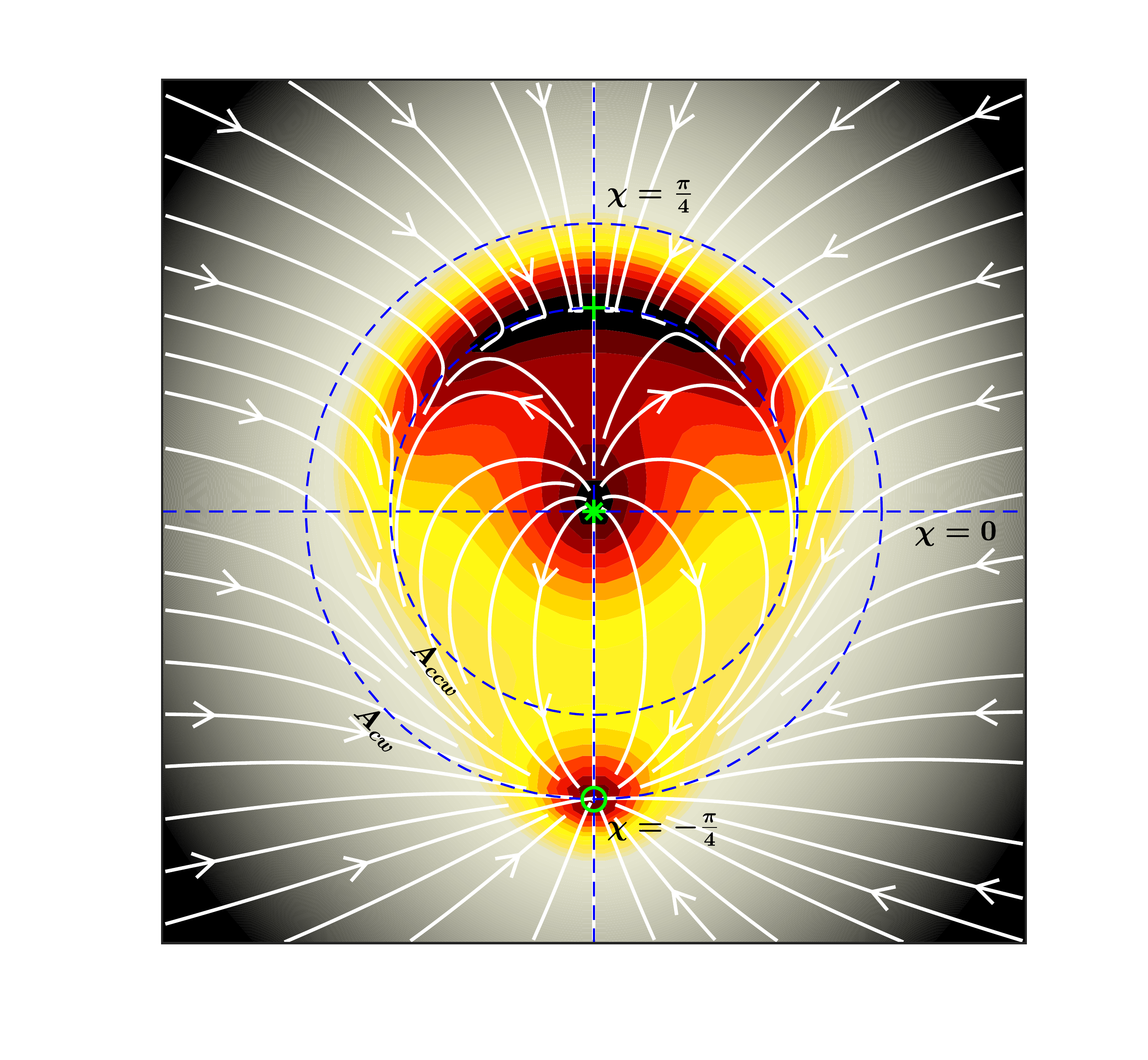}
		\caption{$M=M_1$}
		\label{fig:swstc}
	\end{subfigure}%
	\begin{subfigure}{.5\textwidth}
		\centering
		\includegraphics[width=\textwidth]{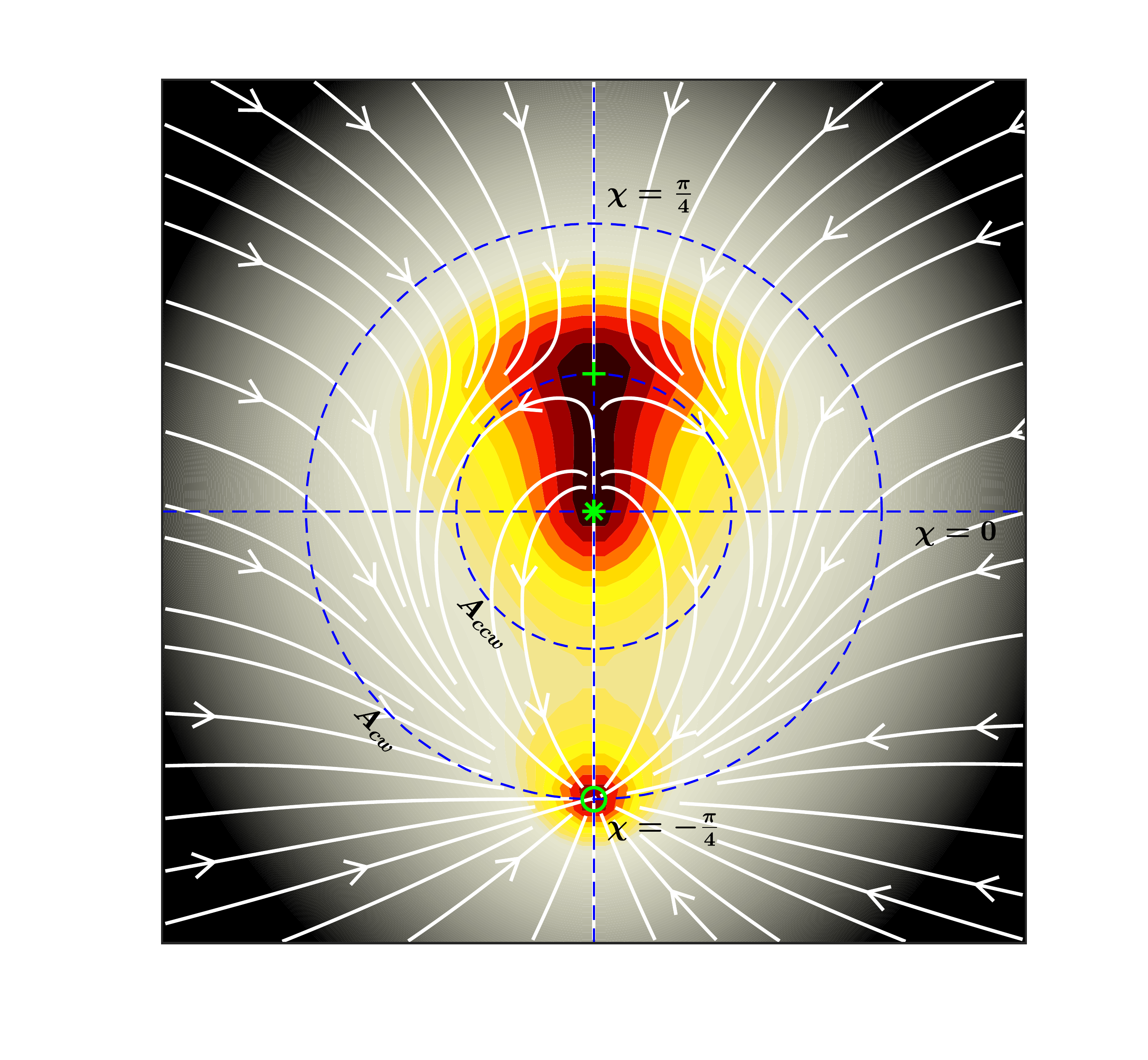}
		\caption{$M=0.02$ ($M_1<M<M_2$)}
		\label{fig:swstd}
	\end{subfigure}%
	
	\begin{subfigure}{.5\textwidth}
		\centering
		\includegraphics[width=\textwidth]{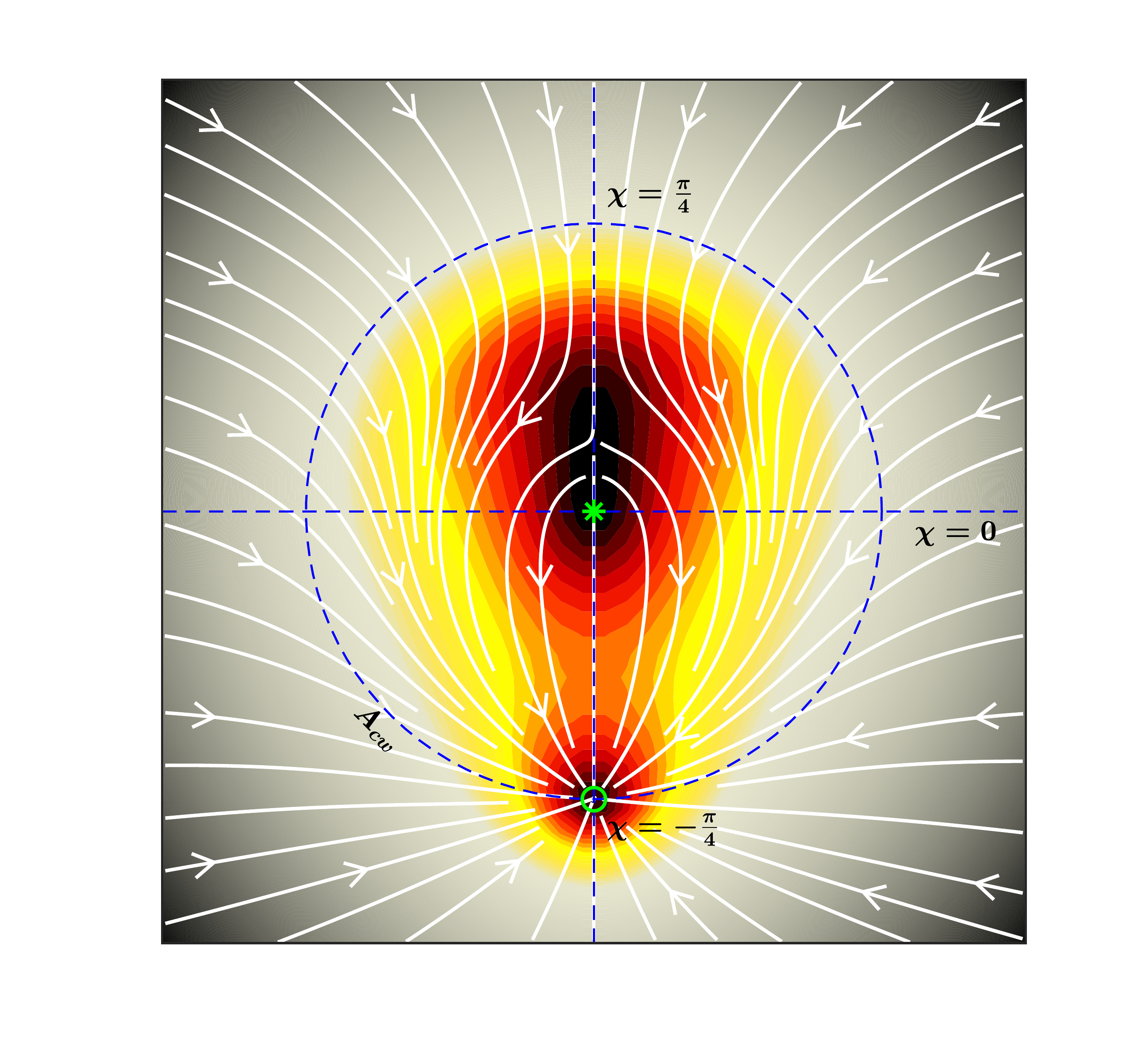}
		\caption{$M=M_2$}
		\label{fig:swste}
	\end{subfigure}%
	\begin{subfigure}{.5\textwidth}
		\centering
		\includegraphics[width=\textwidth]{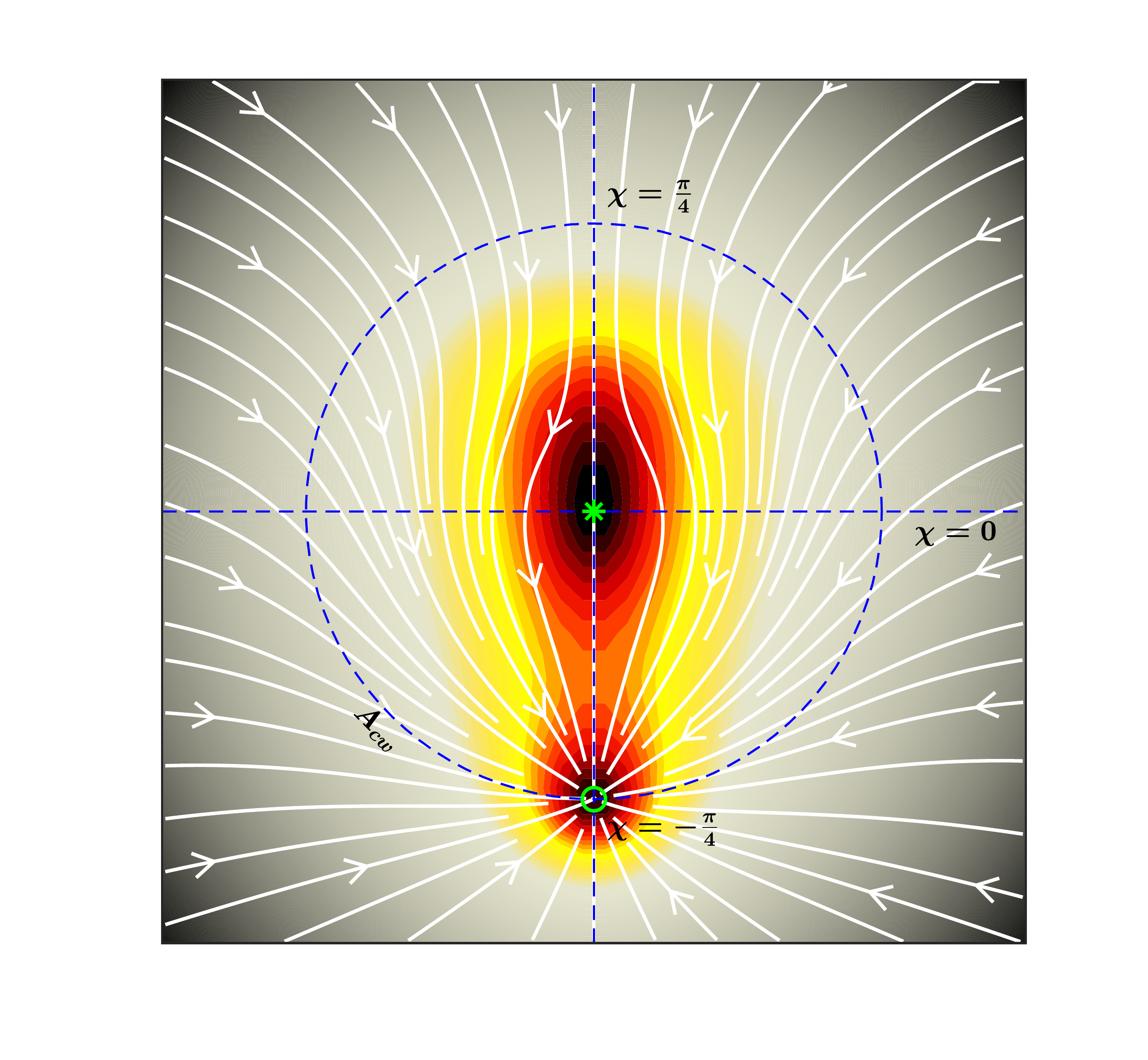}
		\caption{$M=0.1>M_2$}
		\label{fig:swstf}
	\end{subfigure}%
	\caption{Streamlines in the plane $\theta=0$ for different values of $M$. $M_1= 0.0107$, $M_2=0.0313$. Symbols: $\circ$: attractors $*$: repellers, +: saddles. The parameters are the same as in Fig. (\ref{fig:sw0_005}). The same color scale is used for the the four first plots.}
	\label{fig:swst}
\end{figure}
\begin{figure}
	\centering
	\begin{subfigure}{.5\textwidth}
		\centering
		\includegraphics[width=\textwidth]{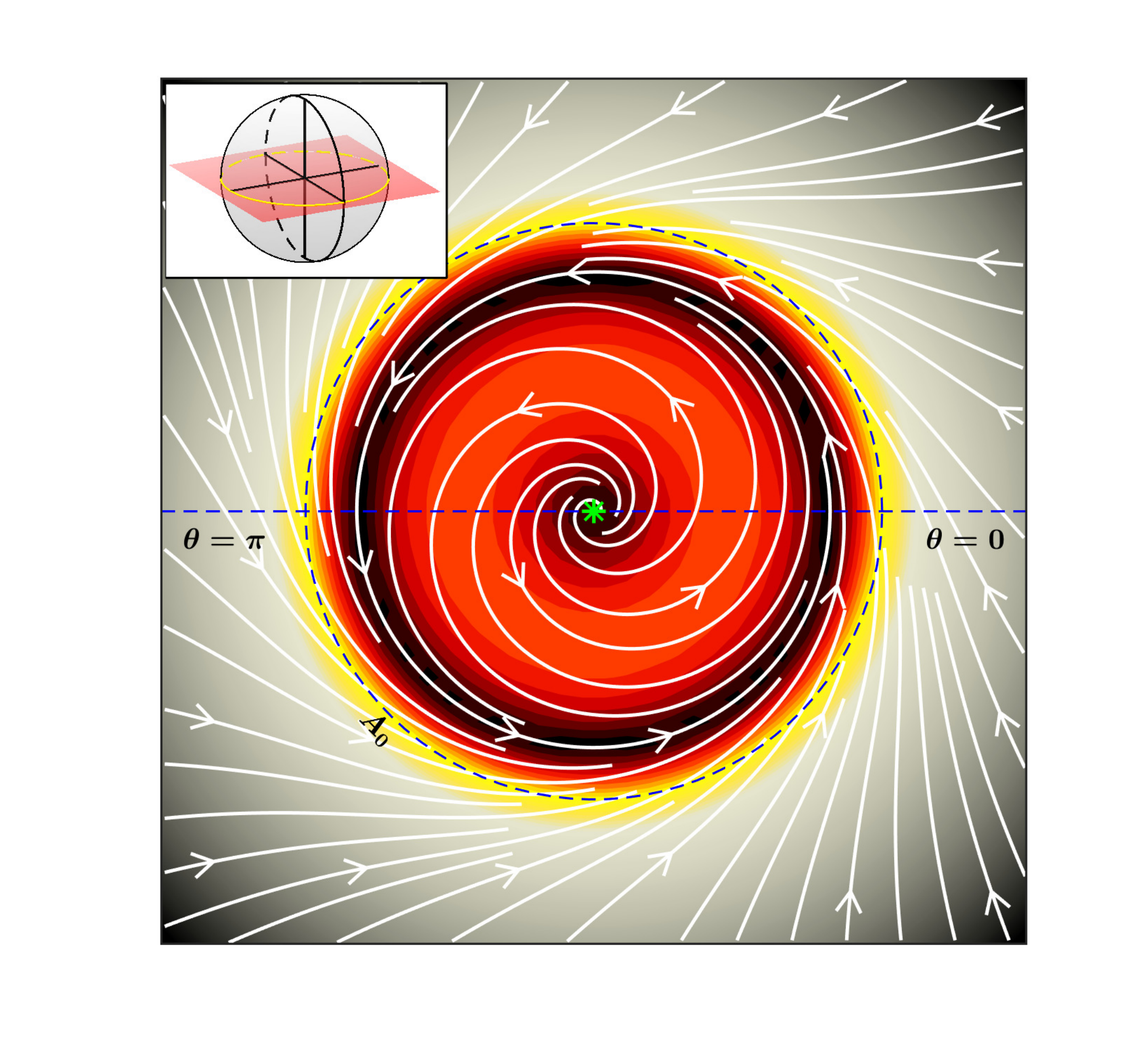}
		\caption{plane $\chi = 0$}
	\end{subfigure}%
	\begin{subfigure}{.5\textwidth}
		\centering
		\includegraphics[width=\textwidth]{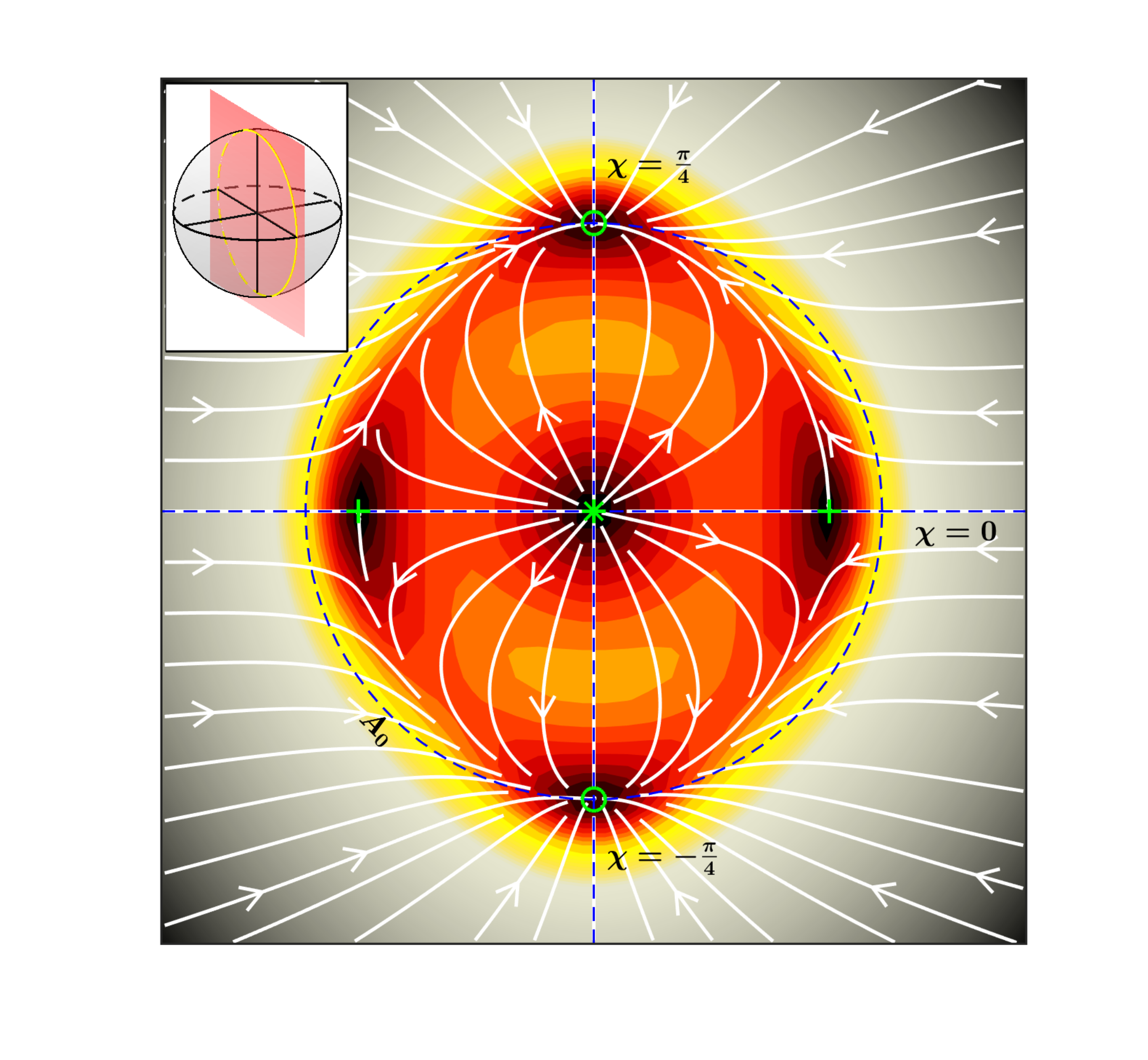}
		\caption{plane $\theta=0\mod\pi$}
	\end{subfigure}%
	\caption{Streamlines for $M=0.005$, for the model without convective terms in the thermoacoustic source term.The green symbols are the equilibrium points: $\circ$: attractors, $*$: repellers, +: saddles. Parameters: $\beta_0$=160 rad.s$^{-1}$, $\alpha$=155 rad.s$^{-1}$, $\kappa$=0.5 rad.s$^{-1}$Pa$^{-2}$, $\omega_n$=1167 rad.s$^{-1}$.}
	\label{fig:swNoConvection}
\end{figure}

For each of the cases $0<M<M_1$, $M_1<M<M_2$ and $M_2<M$, a stability analysis of the fixed points was performed:
\begin{itemize}
	\item For the lowest range of mach $0<M<M_1$, the two spinning modes are stable solutions. The mixed mode is a saddle circle, and the solutions $A=0, \chi=\pm\pi/4$ are repellers.
	The amplitude of the counter-swirl mode is higher than the amplitude of the mode spinning in the swirl direction. Figure \ref{fig:sw0_005} shows the streamlines of the system. The streamlines approaching the amplitude $A_m$ are either attracted to the lower pole if they are in the sector $\chi<\chi_m$, either to the upper pole if they are in the sector $\chi>\chi_m$. When the mach increases, the angular sector $\chi>\chi_m$ gets smaller and it is therefore less likely to end up in the swirl direction. 
	\item When $M$ reaches $M_1$, the saddle circle and the counterclockwise mode attractor at the north pole merge together.\\
	\item For the range $M_1<M<M_2$, the saddle  circle $\{A_m,\:\chi_m\}$ is no longer defined. The spinning modes are still equilibrium points of the system, but a stability analysis shows that the counterclockwise mode has become a saddle point. Figure \ref{fig:swstd}  shows that the streamlines are first pointing towards the counterclockwise mode and  are subsequently bending toward the clockwise mode. When the mach number increases, the amplitude of the counterclockwise mode decreases and eventually reaches 0 when $M=M_2$.\\
	\item When $M>M_2$, the only stable solution is the counter-swirl mode. The point $A=0, \chi=+\pi/4$ becomes a saddle, while the point $A=0, \chi=-\pi/4$ remains a repeller. This behaviour is obtained in our example for a Mach number $M_2=0.03$ which is relevant for practical applications.
\end{itemize}
It has to be noted that all these different regimes are due to the presence of the nonlinear term
$-3\kappa p^2(U/\mathcal{R})\partial(\beta p)/\partial\Theta$ in the wave equation \eqref{eq:thEq}. This behaviour is therefore sensitive to the model used for the flame's nonlinear response.\\
Let us now discuss the evolution of $\theta$ and $\varphi$, and the associated consequences on the oscillation frequency at the stable fixed points when the Mach number is increased. The time integration of the $\theta$ and $\varphi$ equations in the system (\ref{eq:sw}) gives:
\begin{align}\label{eq:linDrift}
\begin{cases}
\theta = \dfrac{M\omega_n}{n}t + \theta_i
\\[10pt]
\varphi = \varphi_i
\end{cases}
\end{align}
with $\theta_i$ and $\varphi_i$ the initial values for $\theta$ and $\varphi$.
These expression are introduced in the expression of $p$ as a sum of two counter propagating waves, Eq. \eqref{eq:pReSpin}, which gives
\begin{align}\label{eq:pReSw}
p \;=\; A^{+} \cos(-n(\Theta-\theta_i) + t\omega^+ + \varphi_i) \:+\: A^{-} \cos(n(\Theta-\theta_i) + t\omega^- + \varphi_i),
\end{align}
where we defined
\begin{align}
\omega^+\:=\:\omega_n(1+M)
\end{align}
and
\begin{align}
\omega^-\:=\:\omega_n(1-M)\,.
\end{align}
The counterclockwise and clockwise spinning components of the pressure have therefore different frequencies. The counter-swirl component has the lower frequency. The frequency gap between the two components is $2M\omega_n$, which is small compared to the pure acoustic frequency $\omega_n$. This frequency splitting due to the azimuthal flow has been previously observed experimentally in \cite{hummel16} and predicted from a linear stability analysis of a lumped-element description of the thermoacoustics of an annular combustor \cite{bauerheim15}. It has to be noted that the phenomenon of frequency splitting does not depend on the model adopted for the flame response. In the wave equation \eqref{eq:thEq}, the frequency splitting comes from the term $-2(U/\mathcal{R})(\partial^2 p / \partial \Theta \partial t)$, which is present in the convected acoustic wave operator, regardless on the presence of source terms. Furthermore, the expression of $\theta$ in (\ref{eq:linDrift}) shows that the preferential direction angle will drift at the velocity of the mean flow during the transient phases where the mode is not purely spinning. However, in real gas turbines, the turbulent combustion noise adds a constant excitation in the system, which prevents the azimuthal thermoacoustic mode to be purely spinning as demonstrated in section \ref{sec7}. The combined effects of stochastic forcing from turbulence and symmetry breaking can be modelled by combining the results of sections \ref{sec7},  \ref{sec8} and \ref{sec9}.\\\\
In the previous case, the splitting of the counter-spinning-waves growth rates and the phase space evolution are governed by the presence of the convective derivative of $\dot{Q}'$ in Eq. \eqref{eq:wavepSimp}. Indeed, it results from $(U/\mathcal{R})(\partial\dot{Q}'/\partial\Theta)$ in 
$\bar{D}\dot{Q}'/D t$ when the azimuthal mean flow is considered uniform in the whole chamber. However, if the flames are located in a region where there is no mean azimuthal bulk velocity, the thermoacoustic source term would result from the spatial averaging of  $\partial \dot{Q}'/\partial t$ only. Then, the slow flow system becomes simpler:
\begin{align}\label{eq:sw2}
\begin{cases}
\dot{A} \:=\: \dfrac{1}{2}(\beta_0-\alpha)A
\:+\: \dfrac{3\kappa}{64}(-5-\cos(4\chi))A^3,
\\[15pt]
\dot{\chi} \:=\: \dfrac{3\kappa}{64}A^2\sin(4\chi) ,
\\[12pt]
n\dot{\theta} \:=\: M\omega_n,
\\[10pt]
\dot{\varphi}  \:=\: 0.
\end{cases}
\end{align}
The equations for $\dot{A}$ and $\dot{\chi}$ are exactly the same as the equations for the case without swirl presented in section \ref{sec5}: the independent system for $A$ and $\chi$ is not affected by the swirling motion and  leads to stable spinning modes with amplitude $A_0$ given in Eq. \eqref{eq:solSym}. However, the equation for $\dot{\theta}$ involves $M$, which induces splitting of the frequencies of the counter-spinning modes. Figure \ref{fig:swNoConvection} shows the streamlines of this new system. In conclusion, the phase space evolution is very sensitive to the presence or not of mean azimuthal flow in the heat release rate regions.
\section{Conclusion}
In this paper, the problem of thermoacoustic instabilities in annular combustion chambers has been considered with a new theoretical approach. Starting with a low-Mach convected wave equation for the acoustic pressure with forcing from heat release rate and velocity field fluctuations, we derived a 1D approximation for thin annulus with mean swirling flows. Then, we inserted a quaternion-based ansatz for the acoustic field in order to find the systems of differential equations describing the dynamics of the slow-flow variables (amplitude, nature angle, preferential direction and temporal phase drift of the azimuthal thermoacoustic modes) for different scenarios: i) thermoacoustic feedback with or without time delay, ii) in presence or not of stochastic forcing by turbulence, iii) with uniform or non-uniform thermoacoustic distribution along the annulus circumference, iv) in presence or not of a mean swirl in the annulus. This quaternion-based ansatz was initially proposed for bi-variate time-series analysis and applied for seismic data processing \cite{flamant17}, and subsequently taken up for analysing acoustic pressure data recorded in annular combustors \cite{ghirardo18}.  Classical deterministic and stochastic  methods of slow-flow averaging were adapted to this quaternion formalism to obtain nonlinear dynamical systems describing the dynamics of azimuthal modes over time scales that are large compared to the acoustic period.
The quaternion representation allows us to simultaneously model, symmetry breaking induced by non-uniform distribution of the thermoacoustic sources and by the presence of a mean swirl in the annulus. This complete description cannot be achieved with the alternative models available in the literature for this problem. Indeed, the approach consisting in projecting the acoustic field onto the orthogonal acoustic eigenmodes of the annulus \cite{noiray11,noiray13,moeck18} captures rotational symmetry breaking induced by spatial thermoacoustic non-uniformities, while it is not appropriate to describe the reflectional symmetry breaking induced by a mean swirl. In contrast, the approach that is based on  a representation of the acoustic field using clockwise and counterclockwise spinning waves \cite{hummel17b} cannot capture the former symmetry breaking, while it is very suited for describing the latter. As shown already in \cite{noiray11} with the model based on Helmholtz eigenmode pairs, the quaternion-based framework predicts i) spinning waves limit-cycles for a uniform distribution of thermoacoustic sources without mean flow, where the direction of the spinning wave depending on the initial conditions,  ii) mixed-modes for non-zero 2$n$ component $c_{2n}$ of the Fourier series for the thermoacoustic source distribution and ultimately standing modes when $c_{2n}$ exceeds $(\beta_0-\alpha)/\beta_0$, where the damping coefficient $\alpha$ characterizes  uniformly distributed acoustic losses around the annular combustor, and where $\beta_0$ is the zero order component of the Fourier series. Notice that the present modelling approach depicts nonlinear flame response that depends on acoustic pressure or, via the time delay, on acoustic axial velocity. Although it is generally acknowledged that the amplitude increase of the latter acoustic perturbation is the most important driver of the saturation of the heat release rate response, another model which accounts for azimuthal acoustic velocity has been proposed some years ago \cite{ghirardo13} and leads to stable standing modes even for situations where $c_{2n}=0$. The main new conclusions of the present work are now listed: 
\begin{itemize}
	\item For annular combustors without low-Mach mean swirl ($M=0$) and with  $c_{2n}=0$, a delayed thermoacoustic feedback induces a slight shift of the eigenfrequency of the spinning mode limit-cycle; this shift depends on the time-delay $\tau$, which also applies to longitudinal modes in can-combustors. It does not depend on the spinning direction.
	\item For annular combustors with $M=0$, $c_{2n}=0$, $\tau=0$, and with stochastic forcing from turbulence, whose spatially averaged contribution is modelled as a bivariate white noise of intensity $\Gamma$, pure spinning waves are not anymore the fixed points of the phase space defined by the deterministic part of the Langevin-equations-system for the slow-flow variables. When the thermoacoustic system is linearly unstable and when the noise intensity is increased from zero, the attracting fixed points are on two circles in the Bloch sphere representation, which corresponds to mixed modes in every direction. For a given linear growth rate $\nu$ and saturation constant $\kappa$, there exists  a critical noise intensity beyond which standing modes (equator on the Bloch sphere), with nodal direction evolving as a random walk, are attracting the thermoacoustic system. In other words, for a given noise intensity $\Gamma$, standing modes  statistically prevail when the linear growth rate becomes positive, and beyond a critical growth rate $\nu_\text{thr} \equiv (3/16)\times\sqrt{3\kappa\Gamma}/\omega_n$, mixed modes govern the system dynamics. For further increase of the linear growth rate, these mixed-modes asymptotically tend to pure spinning modes. 
	\item For annular combustors with mean swirl ($M\neq0$) and with $c_{2n}=0$, $\tau=0$ and $\Gamma=0$, the counter-spinning modes have no longer the same frequency $\omega_n(1\pm M)$. Additionally, if the heat release occurs in a region where there is a uniform azimuthal mean flow, the spinning mode rotating against the swirl direction attracts the system with more strength than the co-rotating spinning mode. Beyond a critical azimuthal Mach number, a saddle-node bifurcation leads to a situation where the only attractor of the system is the spinning mode travelling against the mean swirl. Conversely, when the flames are not located in the convection zone, both counter-spinning modes are equally strong attractors.
\end{itemize}
We conclude this study with the two following remarks: First, these  results are quantitative for an \textit{idealized} annular combustion chamber and they can only be used for qualitatively interpreting observations from a \textit{real-world} annular combustor. Indeed, in practical systems, there are additional effects that can have a very significant impact on the thermoacoustic dynamics and that are not accounted for in this model, e.g. 
the acoustic coupling between the chamber and the plenum. Second, several combinations of these effects were not considered in the present paper and will be the topic of future work. For instance, one can expect that other intriguing nonlinear dynamics would result from $\left\{M\neq0, c_{2n}=0, \tau\neq0, \sigma\neq0\right\}$, i.e. combined effects of mean swirl, delayed flame response and stochastic forcing. Third, these results show that there can be different root causes for statistically more frequent standing modes or spinning modes. Consequently, when interpreting experimental and numerical observations of azimuthal thermoacoustic modes, it is important to quantify i) the rotational and reflectional symmetry of the annular turbulent combustor considered, which is not easy in the case of experimental or practical setups, and ii) the statistics of the observed slow-flow thermoacoustic dynamics, which is not easy in the case of computationally expensive Large Eddy Simulations.

\section*{Acknowledgements}

This project has received funding from the European Union's Horizon 2020 research and innovation programme under Grant Agreement No 765998.

\begin{appendices}
\section{Derivation of a wave equation for the pressure with thermoacoustic and aeroacoustic source terms}\label{apC}
In this appendix, we build upon the works of Howe \cite{howe75,howe03} to derive a wave equation for the acoustic pressure in annular chambers in presence of a uniform mean flow and of aeroacoustic and thermoacoustic sources. \\
~\\
The pressure, velocity, density, temperature, specific entropy and vorticity at point $\xx$ and time $t$ are respectively denoted by $p(\xx,t)$, $\uu(\xx,t)$, $\rho(\xx,t)$, $T(\xx,t)$, $S(\xx,t)$, $\ww(\xx,t)$. The considered fluid fulfils the ideal gas law $p=\rho r T$, with $r$ the specific gas constant. The sound speed and the specific heat capacity for constant pressure are respectively denoted by $c=\sqrt{\gamma r T}$ and $c_p=\gamma r/(\gamma-1)$, with $\gamma$ the heat capacity ratio. The Euler momentum equation for a non-viscous flow is:
\begin{align}
\dfrac{D\uu}{Dt}+\dfrac{1}{\rho}\nabla p  \,=\, 0\, ,
\end{align}
where $D/Dt = \partial /\partial t + \uu\cdot\nabla$ is the material derivative.
This equation can be rewritten in Crocco's form:
\begin{align}\label{eq:crocco}
\dfrac{\partial \uu}{\partial t} + \nabla B = -\ww\times \uu + T \nabla S\,,
\end{align}
where the total enthalpy per unit mass $B=\uu^2/2+\int \rho^{-1}dp + \int\,TdS$  and the Lamb vector  $\ww\times \uu$  have been introduced.
The mass balance equation is given by
\begin{align}\label{eq:massBalance}
\dfrac{1}{\rho}\dfrac{D\rho}{Dt}\,+\,\nabla \cdot \uu = 0.
\end{align}
When heat is released in a fluid, the density $\rho$ can be written as a function of the two thermodynamic variables $p$ and $S$, and one can write the differential equality
\begin{align}
d\rho\:=\:\left.\dfrac{\partial \rho}{\partial p}\right|_S dp\:+\: \left.\dfrac{\partial \rho}{\partial S}\right|_p dS\,.
\end{align}
Considering the ideal gas law and the facts that $c_p dT = T dS$ at constant pressure and that  $dp=c^2 d\rho$ at constant entropy, one obtains
\begin{align}
d\rho\:=\: \dfrac{1}{c^2}dp \:-\: \dfrac{\rho}{c_p}dS,
\end{align}
and therefore
\begin{align}\label{eq:stateEqS0}
\dfrac{D\rho}{D t} = \dfrac{1}{c^2}\dfrac{D p}{D t} - \dfrac{\gamma-1}{c^2}\rho T \dfrac{D S}{D t}.\end{align}
In this model, one considers entropy changes resulting from the combustion heat release and from heat diffusion, but one neglects the effects of viscous heating and species diffusion. Under these assumptions, eq. \eqref{eq:stateEqS0} becomes
\begin{align}\label{eq:heatrelease}
\dfrac{D\rho}{D t} = \dfrac{1}{c^2}\dfrac{D p}{D t} - \dfrac{\gamma-1}{c^2} \left(\dfrac{DQ}{Dt} + \nabla\cdot\lambda\nabla T\right)\, ,
\end{align}
where $Q$ is the heat released by the combustion reaction and $\lambda$ the thermal conductivity of the fluid.
Taking the divergence of eq. \eqref{eq:crocco}, the time derivative of the continuity equation (\ref{eq:massBalance}),  using eq. \eqref{eq:heatrelease}, and $$\frac{DB}{Dt}=\frac{1}{\rho}\frac{\partial p}{\partial t}+T\frac{DS}{Dt},$$  one can derive the following equation for the total enthalpy: 
\begin{multline}\label{eq:waveB}
\dfrac{D}{Dt}\left(\dfrac{1}{c^2}\dfrac{DB}{Dt}\right)+\dfrac{1}{c^2}\frac{D \uu}{D t}\cdot\nabla B-\nabla^2 B 
\\
=\nabla\cdot\left(\ww\times\uu-T\nabla S\right) -\dfrac{1}{c^2}\dfrac{D\uu}{Dt}\cdot\left(\ww\times\uu -  T \nabla S\right)+ \dfrac{D}{Dt}\left(\dfrac{T}{c^2}\dfrac{DS}{Dt}\right)
\\
+\dfrac{\partial}{\partial t}\left\{\dfrac{\gamma-1}{\rho c^2}\left( \dfrac{DQ}{Dt} + \nabla\cdot\lambda\nabla T\right)\right\}.
\end{multline}
This equation is eq. (4.14) in the famous paper from Howe \cite{howe75} and we just have explicitly expressed the contribution from combustion heat release and heat diffusion in the last right hand side term. The other forcing terms are acoustic sources due to the unsteady heat release rate, the vorticity and entropy inhomogeneities. This equation is exact and will be linearised in the next paragraph to obtain a wave equation for the  pressure fluctuations.\\
~\\
From now on, we consider small perturbations around a stationary low Mach mean flow at thermal equilibrium and work in the frame of linear acoustics. The flow variables are decomposed as the sum of a steady  and a fluctuating component, e.g. $p(\xx,t)=\overline{p}(\xx)+p'(\xx,t)$. This decomposition is applied to the variables of eq. \eqref{eq:waveB} in order to derive an equation for the fluctuations where the terms involving products of the fluctuating quantities are neglected. The resulting equation, which contains many terms and which is not presented here, will be further simplified by considering incompressible mean flows with uniform density field, and by discarding coherent vorticity and entropy disturbances.
In addition, only the  first order terms in Mach number are retained.
\\
To get the perturbations of the last right-hand-side term of eq. \eqref{eq:waveB}, one has to subtract its time-averaged component. Under low Mach approximation, this time-averaged contribution satisfies: $\overline{D S/D t}=0$.
With the other assumptions made to derive eq.  (\ref{eq:heatrelease}), it yields
\begin{align}\label{eq:stateEqAv}
\overline{\dot{Q}} + \nabla\cdot\lambda\nabla \overline{T} = 0 ,
\end{align}
which just means that the mean heat release rate $\overline{\dot{Q}}=\overline{DQ/Dt}$ is locally balanced by heat diffusion. 
Consequently, the linearized energy equation for the perturbations around the incompressible mean flow is
\begin{align}\label{eq:stateEqSLin}
\dfrac{\Dbar\rho'}{D t} = \dfrac{1}{c^2}\dfrac{\Dbar  p'}{D t} - \dfrac{\gamma-1}{c^2}\left(\dot{Q}'+ \nabla\cdot\lambda\nabla T'\right),
\end{align}
where we introduced the notation $\Dbar / D t=\partial/\partial t + \overline{\uu}\cdot\nabla$.  In what follows, one assumes that the term $\nabla\cdot\lambda\nabla T'$ corresponding to unsteady conduction phenomena is negligible compared to the one associated with the heat release rate fluctuations from the unsteady flames.
With the assumption of negligible temperature and entropy gradients in the mean flow and uniform speed of sound $c$, the linearisation of \eqref{eq:waveB} leads to the following convected wave equation:
\begin{align}\label{eq:waveBLinM1}
\left(\dfrac{1}{c^2}\dfrac{\Dbar^2}{D t^2}-\nabla^2\right)(B'-\overline{T}S') = \dfrac{\gamma-1}{\overline{\rho} c^2}\dfrac{\partial \dot{Q}'}{\partial t}
+ \nabla\cdot (\ww\times\uu)'
\end{align}
The objective is now to obtain a convected wave equation for the acoustic pressure $p'$.
With the definition of the total enthalpy, one can write
$B'- \overline{T}S'=p'/\overline{\rho}+\overline{\uu}\cdot\uu'$ and therefore,  the left-hand-side of eq. \eqref{eq:waveBLinM1}  can be decomposed as \begin{align}\label{eq:waveBpu}
\Box_{\bar{\uu}}(B'-\overline{T}S') = \dfrac{1}{\overline{\rho}}\Box_{\bar{\uu}}(p')+ \Box_{\bar{\uu}}(\overline{\uu}\cdot\uu'),
\end{align}
where we introduced a convected wave operator, $\Box_{\bar{\uu}}=c^{-2}\Dbar ^2/Dt^2 - \nabla^2$. The term $\Box_{\bar{\uu}}(\overline{\uu}\cdot\uu')$ comes from the interaction of the mean flow with the velocity fluctuations and vanishes for zero mean flow velocity.\\
In this paragraph, the term $\Box_{\bar{\uu}}(\overline{\uu}\cdot\uu')$ is rewritten as a source term  of the wave equation or the fluctuating pressure. To do so, one writes a convected wave equation for the velocity fluctuations $\uu'$ by considering, with the assumption of an incompressible mean flow, the linearised mass balance equation
\begin{align}\label{eq:linMassEq}
\dfrac{\Dbar \rho'}{D  t} + \overline{\rho}\nabla\cdot\uu' = 0,
\end{align}
the linearized energy equation
\begin{align}
\dfrac{\Dbar \rho'}{D t} = \dfrac{1}{c^2}\dfrac{\Dbar p'}{Dt} - \dfrac{\gamma - 1}{c^2}\dot{Q}',
\end{align}
and the linearized Euler-Crocco momentum equation
\begin{align}\label{eq:linMomentumEq}
\dfrac{\partial \uu'}{\partial t} + (\ww\times\uu)' + (\uu'\cdot\nabla)\overline{\uu} + (\overline{\uu}\cdot\nabla)\uu' = - \dfrac{1}{\overline{\rho}} \nabla p'.
\end{align}
The material derivative of (\ref{eq:linMomentumEq}) and the gradient of (\ref{eq:linMassEq}) are combined with the assumption of a uniform mean density in the domain of interest. Using the identity $\nabla\nabla\cdot =  \nabla^2 + \nabla\times\nabla\times$, one obtains a convected wave equation for the fluctuating velocity with source terms. Taking the dot product of this equation with $\overline{\uu}$ and neglecting the second order terms in Mach, one obtains:
\begin{align}\label{eq_Luu}
\Box_{\bar{\uu}}(\uu')\cdot\overline{\uu}=-\dfrac{\gamma-1}{\overline{\rho}\,c^2}\overline{\uu}\cdot\nabla \dot{Q}' + (\nabla\times\ww')\cdot\overline{\uu}.
\end{align}
Tensor calculus identities give us: \footnote{The definition used here for the tensor double dot product is: $(a_{ij}):(b_{ij}) = \sum_{ij} a_{ij}b_{ij}$. An other definition exists with the transpose of $b$.  }
\begin{align}\label{eq_Luu2}
\Box_{\bar{\uu}}(\overline{\uu}\cdot\uu') \:=\: \Box_{\bar{\uu}}(\uu')\cdot\overline{\uu} \:-\: (\nabla^2\overline{\uu})\cdot \uu' 
\:-\: 2(\nabla\overline{\uu}):(\nabla \uu').
\end{align}
The last step consists in combining the wave equation (\ref{eq:waveBLinM1}), the decomposition (\ref{eq:waveBpu}), and Eqs. \eqref{eq_Luu} and  \eqref{eq_Luu2} in order to obtain a convected wave equation for $p'$: 
\begin{multline}\label{eq:wavep0}
\Box_{\bar{\uu}}\left(\frac{p'}{\overline{\rho}}\right) \:=\: \dfrac{\gamma - 1}{\overline{\rho}c^2}\dfrac{\Dbar \dot{Q}'}{D t} \:+\: \nabla\cdot(\overline{\ww}\times\uu')\:+\: \nabla\cdot(\ww'\times\overline{\uu})
\\[10pt]
\:-\: (\nabla\times\ww')\cdot\overline{\uu} \:+\: (\nabla^2\overline{\uu})\cdot \uu' 
\:+\: 2(\nabla\overline{\uu}):(\nabla \uu').
\end{multline}
This equation can be rewritten
\begin{align}\label{eq:wavep} \Box_{\bar{\uu}}\left(p'/\overline{\rho}\right) \:=\: (\gamma - 1)/(\overline{\rho}c^2)(\Dbar \dot{Q}'/D t)	\:+\: 2(\nabla\overline{\uu}):(\nabla \uu')^T.
\end{align}
and is in agreement with the one derived in \cite{phillips60}.
This is a convected wave equation for the fluctuating pressure, which is applicable to low-Mach mean flows with uniform density and sound speed distributions, which results from a first-order approximation in Mach, and which has source terms resulting from fluctuations of heat release rate, velocity and vorticity.\\
~\\
Equation \eqref{eq:wavep} will now be further simplified for the case of thin annular chambers with unsteady heat release rate. The thin annulus has a mean radius $\mathcal{R}$, a thickness $\delta \mathcal{R}$, with $\mathcal{R}\gg\delta \mathcal{R}$, and a length $\mathcal{Z}$. Figure \ref{fig:chamber} presents this simplified geometry and the associated cylindrical coordinate system. We assume $\overline{\uu}=U\ee_\Theta+V\ee_z$ to model a simple swirling motion. Indeed, for some practical combustors, a mean low-mach azimuthal velocity $U$  can reach the same order of magnitude as the mean axial velocity component $V$ due to the design of the burners and their arrangement along the circumference of the annulus. 
One also considers situations where the velocity fluctuations are quasi-irrotational, with negligible axial and radial components and gradients, i.e.  $\uu'\approx u'_\Theta(\Theta)\,\ee_\Theta$, where $\ee_\Theta$ is the unit vector in the azimuthal direction. The heat release rate fluctuations are also assumed to depend on $\Theta$ and $t$ only. These assumptions lead to the cancellation of the aeroacoustic source term $2(\nabla\overline{\uu}):(\nabla \uu')^T$ and the wave equation becomes:
\begin{align}\label{eq:wavepSimp0}
\Box_{\bar{\uu}}\,p' \:=\: \dfrac{\gamma - 1}{c^2}\left(\dfrac{\partial \dot{Q}'}{\partial t} +\frac{U}{\mathcal{R}}\frac{\partial \dot{Q}'}{\partial\Theta}\right).
\end{align}
\section{General system of slow-flow equations including the effects of swirl, time delay,  azimuthal asymmetry and stochastic forcing}\label{apA}

\textit{Amplitude equation}  
\begin{align*}
\;\dot{A}\:=\:& \dfrac{1}{2}\left(\beta_0\cos(\omega_n\tau)\bigg(1-M\sin(2\chi)\bigg) -\alpha\right)A
\\[6pt]
&\:+\: \dfrac{c_{2n}\beta_0}{4}\bigg(\cos(2n\theta)\cos(\omega_n\tau)-M\sin(2n\theta)\sin(\omega_n\tau)\bigg)\cos(2\chi)A
\\[6pt]
&\:+\: \dfrac{3\kappa}{64}\cos(\omega_n\tau)\Big(-5-\cos(4\chi) + 4M\sin(2\chi) \Big)A^3
\\[6pt]
& \:+\: \dfrac{ 3\Gamma }{ 4\omega_n^2 A } \,+\, \zeta_A
\end{align*}
\\
\textit{Nature angle equation}     
\begin{align*}
\; \dot{\chi} \:=\:& \dfrac{3\kappa}{64}A^2\cos(\omega_n\tau)\bigg(\sin(4\chi)+6M\cos(2\chi)\bigg) \,-\, \dfrac{M\beta_0}{2}\cos(2\chi)\cos(\omega_n\tau)
\\[6pt]
&\,-\, \dfrac{c_{2n}\beta_0}{4}\bigg(\cos(2n\theta)\sin(2\chi)\cos(\omega_n\tau) + \sin(2n\theta)\sin(\omega_n\tau)\bigg)
\\[6pt]
&\,-\, \dfrac{c_{2n}\beta_0 M}{4}\bigg(\cos(2n\theta)\cos(\omega_n\tau) + \sin(2n\theta)\sin(2\chi)\sin(\omega_n\tau)\bigg)
\\[6pt]
&\,-\, \dfrac{ \Gamma \tan(2\chi) }{ 2 \omega_n^2 A^2 } \,+\, \dfrac{1}{A}\zeta_{\chi}
\end{align*}
\\
\textit{Preferential angle equation}
\begin{align*}
\; n\dot{\theta} \:=\: &  M\left(\omega_n+\dfrac{\beta_0}{2}\sin(\omega_n\tau)\right)
\,-\, \dfrac{3\kappa}{32}A^2\bigg(3M+\sin(2\chi)\bigg)\sin(\omega_n\tau)
\\[6pt]
&\,+\, \dfrac{c_{2n}\beta_0}{4}\left(-\dfrac{\sin(2n\theta)\cos(\omega_n\tau)}{\cos(2\chi)} + \cos(2n\theta)\sin(\omega_n\tau)\tan(2\chi)\right)
\\[6pt]
&\,+\, \dfrac{c_{2n}\beta_0 M}{4}\left(\dfrac{\cos(2n\theta)\sin(\omega_n\tau)}{\cos(2\chi)} - \sin(2n\theta)\cos(\omega_n\tau)\tan(2\chi)\right)
\\[6pt]
&\,+\, \dfrac{1}{A \cos(2\chi)}\zeta_{\theta} - \dfrac{\tan(2\chi)}{A}\zeta_{\varphi}
\end{align*}
\\
\\
\textit{Phase equation}      
\begin{align*}
\; \dot{\varphi} \:=\:& \,-\, \dfrac{\beta_0}{2}\sin(\omega_n\tau)
\,+\, \dfrac{3\kappa}{32}A^2\bigg(3+M\sin(2\chi)\bigg)\sin(\omega_n\tau)
\\[6pt]
&\,+\, \dfrac{c_{2n}\beta_0}{4}\left(\sin(2n\theta)\cos(\omega_n\tau)\tan(2\chi) - \dfrac{\cos(2n\theta)\sin(\omega_n\tau)}{\cos(2\chi)}\right)
\\[6pt]
&\,+\, \dfrac{c_{2n}\beta_0 M}{4}\left(\dfrac{\sin(2n\theta)\cos(\omega_n\tau)}{\cos(2\chi)} - \cos(2n\theta)\sin(\omega_n\tau)\tan(2\chi)\right)
\\[6pt]
&\,+\, \dfrac{1}{A}\zeta_{\varphi}
\end{align*}
\section{Expression of the random fluctuation terms in the dynamic system before applying the stochastic averaging procedure}\label{apB}

Before the stochastic averaging (see section \ref{sec7}), the system describing the evolution of the variables $Y=(A,\chi,\theta,\varphi)^T$ can be written as
\begin{align*}
\dot{Y} = F_\text{slow}(Y) + S^{(1)}(Y,t)\zeta_1(t) + S^{(2)}(Y,t)\zeta_2(t)
\end{align*}
where $F_\text{slow}$ is the deterministic part (which can be obtained from the equations of the appendix \ref{apA} with $\sigma = 0$) and $S^{(1)}(Y,t)\zeta_1(t)$, $S^{(2)}(Y,t)\zeta_2(t)$ are the fluctuating terms. The components of $S^{(1)}$ and $S^{(2)}$ have the following expressions:
\begingroup
\allowdisplaybreaks
\begin{align*}
S^{(1)}_A &= - \dfrac{ \sin(\chi)\sin(n\theta)\cos(\varphi+\omega_nt) + \cos(\chi)\cos(n\theta)\sin(\varphi+\omega_nt) }{\omega_n}
\\[8pt]
S^{(1)}_\chi &= \dfrac{ \sin(\chi)\cos(n\theta)\sin(\varphi+\omega_nt) - \cos(\chi)\sin(n\theta)\cos(\varphi+\omega_nt)}{\omega_n A}
\\[8pt]
S^{(1)}_\theta &= 2\cos(2\chi) \dfrac{ \sin(\chi)\cos(n\theta)\cos(\varphi+\omega_nt) + \cos(\chi)\sin(n\theta)\sin(\varphi+\omega_nt)}{\omega_nA n (1+\cos(4\chi))}
\\[8pt]
S^{(1)}_\varphi &= -2\cos(2\chi) \dfrac{ \cos(\chi)\cos(n\theta)\cos(\varphi+\omega_nt) + \sin(\chi)\sin(n\theta)\sin(\varphi+\omega_nt)}{\omega_nA (1+\cos(4\chi))}
\\[10pt]
S^{(2)}_A &= \dfrac{ \sin(\chi)\cos(n\theta)\cos(\varphi+\omega_nt) - \cos(\chi)\sin(n\theta)\sin(\varphi+\omega_nt)}{\omega_n }
\\[8pt]
S^{(2)}_\chi &= \dfrac{ \cos(\chi)\cos(n\theta)\cos(\varphi+\omega_nt) + \sin(\chi)\sin(n\theta)\sin(\varphi+\omega_nt)}{\omega_n A}
\\[8pt]
S^{(2)}_\theta &= 2\cos(2\chi) \dfrac{ \sin(\chi)\sin(n\theta)\cos(\varphi+\omega_nt) - \cos(\chi)\cos(n\theta)\sin(\varphi+\omega_nt)}{\omega_n A n (1+\cos(4\chi))}
\\[8pt]
S^{(2)}_\varphi &= 2\cos(2\chi) \dfrac{ \sin(\chi)\cos(n\theta)\sin(\varphi+\omega_nt) - \cos(\chi)\sin(n\theta)\cos(\varphi+\omega_nt)}{\omega_n A (1+\cos(4\chi))}
\end{align*}
\endgroup
\end{appendices}


\end{document}